\documentclass[usenatbib]{mnras}
\pdfoutput=1
\usepackage{amssymb}
\usepackage{amsmath}
\usepackage{natbib}
\usepackage[pdftex]{graphicx}
\usepackage{subfigure}
\usepackage{booktabs}
\usepackage{tabularx}
\usepackage{float}
\usepackage{afterpage}
\usepackage{pdflscape}
\begin{document}
\label{firstpage}
\title[Tides in stars]
{Tidal dissipation in evolving low-mass and solar-type stars with predictions for planetary orbital decay}
 \author[A. J. Barker]{A. J. Barker\thanks{Email address: A.J.Barker@leeds.ac.uk} \\ Department of Applied Mathematics, School of Mathematics, University of Leeds, Leeds, LS2 9JT, UK}

\label{firstpage}
\pagerange{\pageref{firstpage}--\pageref{lastpage}}
\maketitle
%\date{Accepted . Received ; in original form }
\pubyear{2020}

\begin{abstract}
We study tidal dissipation in stars with masses in the range $0.1-1.6 M_\odot$ throughout their evolution, including turbulent effective viscosity acting on equilibrium tides and inertial waves in convection zones, and internal gravity waves in radiation zones. We consider a range of stellar evolutionary models and incorporate the frequency-dependent effective viscosity acting on equilibrium tides based on the latest simulations. We compare the tidal flow and dissipation obtained with the conventional equilibrium tide, which is strictly invalid in convection zones, finding that the latter typically over-predicts the dissipation by a factor of 2-3. Dissipation of inertial waves is computed using a frequency-averaged formalism accounting for realistic stellar structure for the first time, and is the dominant mechanism for binary circularization and synchronization on the main sequence. Dissipation of gravity waves in the radiation zone assumes these waves to be fully damped (e.g.~by wave breaking), and is the dominant mechanism for planetary orbital decay. We calculate the critical planetary mass required for wave breaking as a function of stellar mass and age, and show that this mechanism predicts destruction of many hot Jupiters but probably not Earth-mass planets on the main sequence. We apply our results to compute tidal quality factors following stellar evolution, and tidal evolutionary timescales, for the orbital decay of hot Jupiters, and the spin synchronization and circularization of binary stars. We also provide predictions for shifts in transit arrival times due to tidally-driven orbital decay of hot Jupiters that may be detected with NGTS, TESS or PLATO. %250 words
\end{abstract}


\section{Introduction}
\label{Intro}

Tidal interactions between planets and their host stars, and between the stars in close binaries, are important in driving long-term spin and orbital evolution in these systems \citep[e.g.][]{Maezh2008}. For solar-type stars, there is observational evidence for tidally-driven orbital circularization \citep{ZahnBouchet1989,Meibom2005,VanEylen2016,Nine2020} and spin synchronization \citep{Meibom2006,Lurie2017}. There is also evidence for tidal evolution of the eccentricities of binaries containing low mass \citep{Triaud2017} and evolved stars \citep{VerbuntPhinney1995,PriceWhelan2018,Beck2018}. Very recently, the first exciting indications of tidally-driven orbital decay of some of the shortest-period hot Jupiters \citep{Maciejewski2016,Patra2017,Bouma2019,Yee2020} have been revealed based on transit timing variations over decadal timescales \citep{Birkby2014,Wilkins2017}. These observations constrain the mechanisms of tidal dissipation in stars.

The mechanisms of tidal dissipation in low mass and solar-type stars with convective envelopes have been studied theoretically for several decades, but significant uncertainties remain \citep[e.g.][]{Zahn2008,Mathis2013,Ogilvie2014}. The tidal response in a star due to small-amplitude tidal forcing is usually decomposed into two components: an equilibrium and a dynamical tide, following the pioneering work of Jean-Paul Zahn \citep{Zahn1966,Zahn1970,Zahn1975,Zahn1977}, and the dissipation of each component is then studied separately. The equilibrium tide is a large-scale non-wavelike quasi-hydrostatic deformation of the star that represents the very low 
frequency response of the body to tidal forcing. The associated time-dependent flow is believed to be dissipated through its interaction with turbulent convection \citep{Zahn1966,Zahn1989}, or possibly through its own instabilities\footnote{These include nonlinear tidal effects such as the elliptical instability in convection zones \citep{BL2013,B2016} and similar instabilities in radiation zones \citep{Weinberg2012,Vidal2018}, which both require large tidal amplitudes to operate, and are potentially important for the very shortest orbital periods.} 
The dynamical tide is a wavelike component that exists in any region of a star that supports waves \citep[e.g.][]{Cowling1941,Zahn1975}. This is usually thought to primarily consist of internal gravity waves (restored by buoyancy and modified by rotation) in radiation zones and inertial waves (restored by Coriolis forces) in convection zones. This component of the tide is dissipated by non-adiabatic effects such as radiative damping \citep{Zahn1975}, nonlinear effects such as wave breaking \citep{BO2010,B2011}, or possibly through its interaction with turbulent convection \citep{Terquem1998,OL2007} or magnetic fields \citep{LO2018}.

It is commonly believed that the dominant tidal mechanism in solar-type stars is the action of convective turbulence in dissipating equilibrium tidal flows \citep{Zahn1966,Zahn1977,Zahn1989}. The turbulence is thought to act as an effective viscosity ($\nu_E$, that is much larger than the microscopic viscosity) in damping this large-scale tidal flow. However, an important objection is that tides are often much faster than the convection, i.e.~the tidal frequency $\omega$ often exceeds the dominant convective turnover frequency $\omega_c$, which is thought to reduce the efficiency of this mechanism in dissipating the tide \citep{Zahn1966,GN1977,GO1997,IvPap2004}. The magnitude of the high-frequency reduction which applies when $\omega/\omega_c\gg 1$ has however been disputed for several decades using phenomenological arguments, with \cite{Zahn1966} suggesting that $\nu_E \propto (\omega_c/\omega)$, and \cite{GN1977} suggesting $\nu_E\propto (\omega_c/\omega)^2$. This mechanism has remained very difficult to study theoretically however, and it is only recently that these scaling laws have been tested numerically \citep[e.g.][]{Penev2007,Penev2009,OL2012,braviner_stellar_2015,DBJ2020,DBJ2020a,VB2020,VB2020a}. Recent work has provided strong evidence in favour of $\nu_E\propto (\omega_c/\omega)^2$ for $\omega/\omega_c\gg 1$ \citep{OL2012,braviner_stellar_2015,DBJ2020,DBJ2020a,VB2020,VB2020a}, even if this may not be for the reasons originally proposed by \cite{GN1977}. This means that convective turbulence is probably much less efficient at dissipating large-scale tidal flows than is commonly believed \citep[cf.][]{ZahnBouchet1989,Zahn2008}. In addition, new regimes of frequency-dependence have also been uncovered for intermediate frequencies $10^{-2}\lesssim \omega/\omega_c\lesssim 1-5$, which remain to be fully explored \citep{DBJ2020a,VB2020,VB2020a}.

Furthermore, the conventional equilibrium tidal flow \citep{Zahn1966,Zahn1989,Remus2012} is only valid in radiation zones, which are stably-stratified, and does not correctly describe the tidal flow in convection zones, which are usually approximately neutrally (adiabatically) stratified \citep{GD1998,Terquem1998,Ogilvie2014}. The correct tidal flow in convective regions can be calculated using the conventional equilibrium tide together with a non-wavelike contribution to the dynamical tide \citep{GD1998}, or somewhat more cleanly, by defining an equilibrium tide that is formally valid in convection zones \citep{Terquem1998,Ogilvie2013}.
It is essential to determine how the damping of the conventional equilibrium tide may differ from that of the correct equilibrium tide in convection zones because the former has been applied in many works previously despite being strictly invalid. Indeed, the consequences of this distinction have not yet been fully explored. In this paper, we revisit this problem by comparing the flow and resulting dissipation from the conventional equilibrium tide \citep[e.g.][]{Zahn1989} with the correct equilibrium tide \citep{Terquem1998,Ogilvie2013}. To do so, we consider the action of convective turbulence in damping these large-scale tidal flows by applying the latest prescriptions for the effective viscosity \citep{DBJ2020a}. We will show that the conventional equilibrium tide over-predicts the dissipation by typically a factor of 2-3, and therefore we advocate against using it to infer the rates of astrophysical tidal evolution. In addition, using the frequency reduction of \cite{Zahn1966} together with the conventional equilibrium tide can (artificially) enhance the dissipation by several orders of magnitude, therefore potentially leading to the incorrect interpretation of observations.

The dynamical tide in convection zones consists of tidally-excited inertial waves, restored by the Coriolis force, if the forcing frequency is less than twice the spin frequency of the star \citep[e.g.][]{OL2007,PapIv2010,RV2010,Ivanov2013,Ogilvie2013,FBBO2014}. This mechanism can be important for stellar spin synchronization and circularization \citep{OL2007}, as well as for spin-orbit alignment \citep{BO2009,Lai2012,B2016a,LO2017,Damiani2018}, since certain tidal frequencies are then typically smaller than twice the rotation frequency. These waves are thought to be dissipated through their interaction with turbulent convection -- though it is unclear whether this mechanism should apply to very short-wavelength waves -- or magnetic fields \citep{LO2018,Wei2018,Astoul2019}, or by nonlinear effects \citep{GoodmanLackner2009,FBBO2014}. This mechanism does not currently operate in the slowly-rotating host stars of most short-period hot Jupiters, and so cannot explain their observationally-inferred orbital decay rates.

In this paper we compute the dissipation of inertial waves using a frequency-averaged formalism \citep{Ogilvie2013} which accounts for the realistic structure of the star for the first time. This builds upon prior work which has adopted a simplified piece-wise homogeneous two-layer stellar model \citep{Mathis2015,Bolmont2016,Gallet2017,Benbakoura2019}, which we compare with the more realistic model. The frequency-averaged dissipation gives an indication of the typical magnitude of the tidal dissipation due to inertial waves. We note that the actual dissipation may be strongly frequency dependent \citep{SavPap1997,OL2007,RV2010,Ogilvie2013} and could in principle differ from the frequency-averaged measure by up to 2-3 orders of magnitude for certain tidal frequencies (based on prior results). However, the frequency-averaged measure is indicative of the typical level of dissipation resulting from these waves, and is much more straightforward to compute.

In radiation zones, which are stably-stratified, internal gravity waves (or ``g-modes") are excited by tidal forcing. In stars with radiative cores, these waves are primarily excited at the radiative/convective interface and subsequently propagate to the centre of the star. Near the centre, these waves become geometrically focused and may achieve sufficiently large amplitudes to break or undergo weaker nonlinear interactions \citep{GD1998,Terquem1998,OL2007,BO2010,B2011,BO2011,Weinberg2012,EssickWeinberg2016,SunArrasWenberg2018}. Wave breaking occurs if the waves locally overturn the stratification, and when this occurs they are then efficiently absorbed \citep{BO2010,B2011}, resulting in efficient tidal dissipation. The critical planetary mass required for wave breaking is approximately 3 $M_J$ (Jupiter masses) in the current Sun, but it strongly depends on the mass and age of the star. In F-type stars (which have masses larger than about $1.1 M_\odot$) possessing convective cores, this mechanism may be unable to operate as effectively because the waves cannot reach the centre of the star \citep{BO2009,B2011}, and so they may not reach large enough amplitudes to break (this has been proposed to explain the survival of WASP-18 b; \citealt{Wilkins2017}). One of the goals of this paper is to determine the critical planetary mass required for wave breaking as a function of stellar mass and age, including in F-stars where these waves cannot reach the centre of the star. We will also study the variation in the resulting tidal dissipation, if these waves are fully damped, as a function of stellar mass and age.

Dissipation of gravity waves in radiative cores is probably responsible for driving orbital decay of hot Jupiters \citep{BO2010,B2011,Chernov2017}. Indeed, the resulting tidal dissipation can explain the observationally-inferred orbital decay rate of WASP-12 b \citep{Maciejewski2016,Patra2017,Maciejewski2018,Yee2020,Patra2020} if the waves are fully damped \citep{Chernov2017,Weinberg2017,2019Bailey}. Whether or not these waves should be fully damped likely depends strongly on the structure of the star (and therefore its mass and age). In particular, wave breaking may explain why these waves should be fully damped if WASP-12 is a subgiant with a radiative core \citep{Weinberg2017}, but this may not be compatible with the observed stellar properties \citep{2019Bailey}. We will revisit this issue later in this paper. We will also make predictions for shifts in transit arrival times due to tidally-driven planetary orbital decay as a result of this mechanism, as a function of stellar mass and age. Such predictions can be viewed as essential preparation for the PLATO mission \citep[e.g.][]{PLATO2014}, and they provide predictions which may be tested with e.g.~NGTS, TESS or PLATO.

Throughout this paper we will compute (modified) tidal quality factors $Q'$ as a way of representing the dissipation due to each mechanism, and we will only consider tides with spherical harmonic degrees and azimuthal wavenumbers $l=m=2$. We define this quantity formally by
\begin{eqnarray}
Q'=\frac{3}{2 k_2} \frac{2\pi E_0}{\oint D\,\mathrm{d} t},
\end{eqnarray}
where $E_0$ is the maximum energy stored in the tide, $D$ is the dissipation rate, $k_2$ is the second-order potential Love number ($k_2=3/2$ for a homogeneous fluid body), and the integral is over one tidal period. $Q'$ is an inverse measure of the dissipation, so that highly dissipative stars in a given application have relatively small $Q'$ values. The quantity $Q'$ is more useful than the tidal quality factor $Q=2k_2 Q'/3$ because the quantity $Q/k_2$ appears together in the tidal evolutionary equations, and in practice we do not wish to separately compute $k_2$. An alternative representation is to use the imaginary parts of tidal love numbers \citep[e.g.][]{Ogilvie2014}, but here we prefer to present $Q'$ due to its wider use in the community (where $\mathrm{Im}[k_2]=3/(2Q')$). It should be remembered that a given star does not possess a ``value of $Q'$", and in fact $Q'$ varies significantly depending on rotation, tidal period, tidal amplitude, as well as the stellar mass and age. The main aim of this paper is to present the variation in this quantity as a function of these input parameters to the best of our current theoretical understanding.

Attempts have been made to constrain the tidal dissipation occurring in planetary host stars and short-period binaries using population-wide statistical analysis \citep{CollierCameronJardine2018,Penev2018,HamerSchlaufman2019,Hwang2020}. These could provide important constraints on tidal dissipation, though we caution that we theoretically expect there to be significant variation in the rates of tidal dissipation as a function of stellar mass, age and rotation, in addition to the planetary mass and orbital period. This suggests that individual systems, rather than entire populations, would provide a more direct comparison of tidal theory with observations. However, the results of \cite{HamerSchlaufman2019} are particularly relevant for this paper. They found that hot Jupiter hosts have smaller Galactic velocity dispersion than stars without such planets, indicating that hot Jupiter hosts are on average younger than the field population. This strongly suggests that tidal interactions cause hot Jupiters to be destroyed while their host stars are on the main sequence, which requires $Q'\lesssim 10^7$ in hot Jupiter hosts. In this paper, we will demonstrate that the gravity wave breaking mechanism predicts such small $Q'$ for short orbital periods, and that the closest hot Jupiters should spiral into their stars as their stars evolve.

The key question addressed in this work is: \textit{how does tidal dissipation in the convective and radiative regions of low-mass and solar-type stars vary as a function of stellar mass, age, rotation and tidal period?} To answer this question, we compute the rates of tidal dissipation resulting from the equilibrium tide and inertial waves in convection zones, and internal gravity waves in radiation zones, following the evolution of low mass and solar-type stars. We consider masses in the range $0.1-1.6 M_\odot$ and use stellar models computed with the MESA stellar evolution code \citep{Paxton2011, Paxton2013, Paxton2015, Paxton2018, Paxton2019}.

The structure of this paper is as follows. In \S~\ref{NWLtides}, we describe how equilibrium tides are computed in convection and radiation zones, and our implementation of the dissipation of this component due to its interaction with turbulent convection. We then describe how we compute the dissipation of the dynamical/wavelike tide in \S~\ref{WLtides}, including inertial waves in convection zones in \S~\ref{IW} and internal gravity waves in radiation zones in \S~\ref{IGW}. We present our results for the dissipation of the correct equilibrium tide, and how it compares with the conventional equilibrium tide in \S~\ref{EQMCOMP}. In Fig.~\ref{QpNWL} we present the tidal dissipation due to this mechanism as a function of tidal period, stellar mass and age. We then present our results for the dissipation of inertial waves in convection zones in \S~\ref{IWresults}, also comparing our results with the simplified two-layer model. In \S~\ref{IGWresults}, we present our results for the dissipation of gravity waves in radiation zones, and the critical planetary masses required for wave breaking as a function of stellar mass and age. We apply our results to predict shifts in transit times due to tidally-driven orbital decay of hot Jupiters, and apply our results to interpret existing observations, and make some new predictions for these systems, in \S~\ref{HJimplications}. We briefly apply our results to binary stars in \S~\ref{binaryimplications}, and we finally conclude in \S~\ref{Conclusions}.

\section{Equilibrium (non-wavelike) tides}
\label{NWLtides}

We consider a slowly-rotating spherically-symmetric star of radius $R$ in hydrostatic equilibrium, such that
\begin{equation}
    \nabla p=\rho\boldsymbol{g}=-\rho\nabla\Phi,
\end{equation}
where $p(r)$ is the pressure, $\rho(r)$ is the density, $\boldsymbol{g}=-g(r)\boldsymbol{e}_r$ is the gravitational acceleration, and $\Phi(r)$ is the gravitational potential satisfying Poisson's equation
\begin{equation}
\nabla^2\Phi = 4\pi G\rho.
\end{equation}
We consider the star to rotate at the rate $\Omega=2\pi/P_\mathrm{rot}$, where $P_\mathrm{rot}$ is the rotation period, and define the dynamical frequency
\begin{eqnarray}
\omega_\mathrm{dyn}=\sqrt{\frac{GM}{R^3}}=\frac{2\pi}{P_\mathrm{dyn}},
\end{eqnarray}
where $P_\mathrm{dyn}$ is the dynamical timescale, as well as the parameter
\begin{eqnarray}
\epsilon_\Omega^2 = \frac{\Omega^2}{\omega_\mathrm{dyn}^2}=\frac{P_\mathrm{dyn}^2}{P_\mathrm{rot}^2},
\end{eqnarray}
Note that $\epsilon_\Omega^2\ll 1$ is implicitly assumed since we ignore centrifugal deformation of the star.
We adopt standard spherical polar coordinates $(r,\theta,\phi)$ centred on the star (with $\theta=0$ along the rotation axis) and we define the buoyancy frequency, or Brunt-V\"{a}is\"{a}l\"{a} frequency, $N$, by
\begin{equation}
    N^2 = g\frac{\mathrm{d}}{\mathrm{d} r} \ln\left(\frac{ p^{\frac{1}{\Gamma_1}}}{\rho}\right),
\end{equation}
where $\Gamma_1=(\partial \ln p/\partial \ln \rho)_s$, and the subscript $s$ refers to constant specific entropy. We consider stars with masses in the range $0.1M_\odot$ to $1.6M_\odot$ of spectral types MKGF, which are either fully convective or contain a mixture of radiative ($N^2>0$) and convective ($N^2\approx 0$) regions (cores and/or envelopes).

We perturb the star with a single tidal potential component
\begin{eqnarray}
\Psi(\boldsymbol{r},t)= \mathrm{Re}\left[\Psi_l (r) Y_l^m(\theta,\phi) \mathrm{e}^{-\mathrm{i}\omega t} \right],
\end{eqnarray}
which could be due to either an orbiting planet or another star in a close binary system, where $\Psi_l=A r^l$ and $A$ will be specified in a given application below, and $Y_l^m(\theta,\phi)$ is an orthonormalized spherical harmonic. Throughout this paper we will focus on the $l=m=2$ component of the tide, since this is typically the most important one. The tidal frequency is $\omega$, which will be specified when discussing specific applications in \S~\ref{HJimplications} and \ref{binaryimplications}. We also define the tidal period $P_\mathrm{tide}=2\pi/\omega$. The corresponding Eulerian gravitational potential perturbation is
\begin{eqnarray}
\Phi'(\boldsymbol{r},t) = \mathrm{Re}\left[\Phi_l(r) Y_l^m(\theta,\phi) \mathrm{e}^{-\mathrm{i}\omega t} \right],
\end{eqnarray}
and we use a similar form for other variables.

The equilibrium tide is computed by assuming the body remains in hydrostatic equilibrium, meaning that Eulerian perturbations to pressure, density and gravitational potential are determined by
\begin{align}
 p'&=-\rho(\Phi'+\Psi), \\
 \rho'&= -\frac{\mathrm{d}\rho}{\mathrm{d}p}\rho(\Phi'+\Psi),
\end{align}
and
\begin{equation}
    \nabla^2\Phi' = -4\pi G\frac{\mathrm{d}\rho}{\mathrm{d}p}\rho(\Phi'+\Psi),
\end{equation}
inside the star (and $\nabla^2\Phi' = 0$ outside). 
The latter can be written
\begin{eqnarray}
\label{PhiEqn}
\frac{1}{r^2}\frac{\mathrm{d}}{\mathrm{d}r}\left(r^2 \frac{\mathrm{d} \Phi'_l }{\mathrm{d}r}\right)-\frac{l(l+1)}{r^2}\Phi'_l+4\pi G\frac{\mathrm{d}\rho}{\mathrm{d}p} \rho(\Phi'_l+\Psi_l)=0,
\end{eqnarray}
where $\Phi'_l$ must satisfy the boundary conditions
\begin{eqnarray}
&&\frac{\mathrm{d}\ln \Phi'_l}{\mathrm{d} \ln r} = l \quad\text{at} \quad r=0, \\
&&\frac{\mathrm{d}\ln \Phi'_l}{\mathrm{d} \ln r} = -(l+1) \quad\text{at} \quad r=R.
\end{eqnarray}

The corresponding equilibrium tidal displacement field $\boldsymbol{\xi}_e=\xi_{e,r}\boldsymbol{e}_r+\boldsymbol{\xi}_{e,h}$ (and therefore tidal velocity field $\boldsymbol{u}_e=\partial \boldsymbol{\xi}_e/\partial t$) can be computed in radiative regions (in which $N^2>0$) by
\begin{equation}
\label{EQMtide}
    \xi_{e,r}=-\frac{\Phi'+\Psi}{g}, \quad\text{and}\text\quad \nabla\cdot\boldsymbol{\xi}_e=0,
\end{equation}
where $\boldsymbol{\xi}_{e,h}\cdot \boldsymbol{e}_r=0$. This is the conventional equilibrium tide \citep{Zahn1966,Zahn1989,Remus2012} and we will refer to it using a subscript $e$. This equilibrium tide solution does not apply in convective regions (with $N^2\approx 0$) however, and we must instead compute the displacement in a different manner \citep{Terquem1998,GD1998,Ogilvie2014}.

In a convective region with efficient convection such that it is adiabatically stratified ($N^2=0$), the low-frequency equilibrium tide is irrotational\footnote{We have implicitly neglected Coriolis forces in computing the equilibrium tide here, which is reasonable if $\omega^2\gg 4 \Omega^2$ \citep{Ogilvie2014}. When $\omega^2\sim 4\Omega^2$, inertial waves are also excited, which we will include in \S~\ref{IW}.} i.e. $\nabla \times \boldsymbol{\xi}=\boldsymbol{0}$ \citep{Terquem1998,GD1998}. This property is not satisfied by the conventional equilibrium tide. Instead, the tidal displacement for the correct equilibrium tide is determined by the solution (subject to appropriate boundary conditions) of
\begin{eqnarray}
\label{NWLtide}
\nabla \cdot \left(\rho \nabla X\right)=\frac{\mathrm{d}\rho}{\mathrm{d}p}\rho \left(\Phi'+\Psi\right),
\end{eqnarray}
where $\boldsymbol{\xi}_\mathrm{nw}=\nabla X$ \citep{Terquem1998,Ogilvie2013}, and we refer to this component of the equilibrium tide using the subscript nw (for ``non-wavelike"). After expanding $X$ in terms of spherical harmonics, this can be written as
\begin{eqnarray}
\label{Xeqn}
\frac{1}{r^2}\frac{\mathrm{d}}{\mathrm{d}r}\left(r^2\rho \frac{\mathrm{d} X_l }{\mathrm{d}r}\right)-\frac{l(l+1)}{r^2}\rho X_l=\rho\frac{\mathrm{d}\rho}{\mathrm{d}p}(\Phi'_l+\Psi_l).
\end{eqnarray}
If there is a convective core, the boundary condition at the centre is 
\begin{eqnarray}
\xi_{\mathrm{nw},r}=\frac{\mathrm{d}X_l}{\mathrm{d}r}=0 \quad \text{at} \quad r=0,
\end{eqnarray}
and for all other boundaries of any convection zone,
\begin{eqnarray}
\xi_{\mathrm{nw},r}=\frac{\mathrm{d}X_l}{\mathrm{d}r}=\xi_{e,r}=-\frac{\Phi'+\Psi}{g},
\end{eqnarray}
which also applies at $r=R$ for a convective envelope. We solve Eqs.~\ref{PhiEqn} and \ref{Xeqn} in a given stellar model using a Chebyshev collocation method, with a (large) number of points that is chosen to ensure that the solutions are accurately computed.

We wish to point out that $\boldsymbol{\xi}_e\ne\boldsymbol{\xi}_\mathrm{nw}$ in the interior of a convection zone, in general. One aim of this paper is to explore the consequences of adopting the correct form of the equilibrium tide ($\boldsymbol{\xi}_\mathrm{nw}$) in convection zones, compared with the conventional equilibrium tide ($\boldsymbol{\xi}_\mathrm{e}$). This is important because many papers have adopted $\boldsymbol{\xi}_e$ in convection zones instead of $\boldsymbol{\xi}_\mathrm{nw}$, which is strictly incorrect and may lead to misleading results. Note that we still ignore non-adiabatic effects, which may be important near the surface in modifying horizontal displacements there \citep{Bunting2019}.

\subsection{Viscous dissipation of the equilibrium tide}

The equilibrium tide is thought to be dissipated through its interaction with turbulent convective motions that act as an effective viscosity. To compute the viscous damping of this flow in a given stellar model we must specify the turbulent effective viscosity $\nu_E(r)$, which is accomplished by specifying the tidal frequency dependence of $\nu_E$. We assume an isotropic viscosity $\nu_E$ in this work, partly for simplicity and partly because existing simulations do not support a strongly anisotropic viscosity \citep{Penev2009}. The dynamic shear viscosity is then $\mu(r)=\rho(r)\nu_E(r)$, and we ignore any possible bulk viscosity. The viscous dissipation of the equilibrium tide is computed by
\begin{eqnarray}
D_\nu = \frac{1}{2} \omega^2\int r^2 \mu(r) D_l(r) \,\mathrm{d} r,
\end{eqnarray}
where the integral is carried out numerically over the entire radial extent of each convection zone in a given stellar model,
\begin{eqnarray}
\nonumber
&&D_l(r) = 3\left|\frac{\mathrm{d}\xi_r}{\mathrm{d} r} -\frac{\Delta_l}{3}\right|^2 +l(l+1)\left|\frac{\xi_r}{r}+r\frac{\mathrm{d}}{\mathrm{d} r}\left(\frac{\xi_h}{r}\right)\right|^2\\
&&\hspace{2.5cm} +(l-1)l(l+1)(l+2)\left|\frac{\xi_h}{r}\right|^2,
\end{eqnarray}
and we have defined
\begin{eqnarray}
\Delta_l=\frac{1}{r^2}\frac{\mathrm{d}}{\mathrm{d}r}\left(r^2\xi_r\right)-l(l+1)\frac{\xi_h}{r},
\end{eqnarray}
for the appropriate choice of $\boldsymbol{\xi}_e$ or $\boldsymbol{\xi}_\mathrm{nw}$, after each is expanded in terms of the appropriate $Y_l^m \mathrm{e}^{-\mathrm{i}\omega t}$ factors. The associated tidal quality factor is then \citep[e.g][]{Ogilvie2014}
\begin{eqnarray}
\frac{1}{Q'_\mathrm{eq}} = \frac{16\pi G}{3(2l+1)R^{2l+1}|A|^2}\frac{D_\nu}{|\omega|}.
\end{eqnarray}

\subsubsection{Turbulent viscosity prescriptions}
\label{turbvis}

We adopt various prescriptions for the turbulent effective viscosity acting on equilibrium tides in convection zones in this work. To apply these in a given stellar model we set $\nu_E$ to be equal to one of the prescriptions that we will define below. We ignore any dissipation of the equilibrium tide in radiation zones. First, we define
\begin{eqnarray}
\nu_{\mathrm{MLT}}=\frac{1}{3}u_c l_c,
\end{eqnarray}
as the frequency-independent prediction from mixing-length theory (MLT), where $u_c$ is the convective velocity and $l_c$ is the mixing-length (both of these quantities are directly output from MESA). We define $\omega_c=u_c/l_c$ to be the convective frequency (absent a factor of $2\pi$). Note that the factor of 1/3 is arbitrary but is adopted for consistency with \cite{Zahn1989} and many other works, which is based on a naive expectation using the analogy with a microscopic viscosity from kinetic theory. We also consider the following two conventional frequency-reductions (but smoothed) from \cite{Zahn1989}, which apply for high-frequency tidal forcing
\begin{eqnarray}
\label{Z}
\nu_\mathrm{Z} &=& \nu_{\mathrm{MLT}} \left(1+ \left(\frac{\omega}{\pi\omega_c}\right)^2\right)^{-\frac{1}{2}} \\
&\sim& \nu_{\mathrm{MLT}} \frac{\pi\omega_c}{\omega}
\quad \text{as} \quad \frac{\omega}{\omega_c}\rightarrow \infty
\nonumber
\end{eqnarray}
as the ``Zahn" reduction \citep{Zahn1966} and
\begin{eqnarray}
\label{GN}
\nu_{\mathrm{GN}} &=& \nu_{\mathrm{MLT}} \left(1+ \left(\frac{\omega}{\omega_c}\right)^2\right)^{-1} \\
&\sim& \nu_{\mathrm{MLT}} \left(\frac{\omega_c}{\omega}\right)^2
\quad \text{as} \quad \frac{\omega}{\omega_c}\rightarrow \infty,
\nonumber
\end{eqnarray}
as the ``Goldreich-Nicholson" reduction \citep{GN1977}. Note that $\nu_\mathrm{MLT}\gg \nu_\mathrm{Z}\gg \nu_{\mathrm{GN}}$ when $\omega\gg \omega_c$.

The latest numerical simulations exploring the interaction between tidal flows and convection provide strong evidence in favour of the quadratic ($
\nu_\mathrm{GN}$) reduction at high frequencies \citep{OL2012,braviner_stellar_2015,DBJ2020,DBJ2020a,VB2020,VB2020a}, albeit with a different constant of proportionality than 1/3, and are inconsistent with the linear ($\nu_Z$) reduction for $\omega/\omega_c\gg 1$. In addition, a new intermediate frequency regime has been uncovered \citep{DBJ2020a,VB2020,VB2020a} for $10^{-2} \lesssim \omega/\omega_c \lesssim 1-5$. To account for these latest results, we consider the continuous power-law fit to the simulations of \cite{DBJ2020a},
\begin{eqnarray}
 \nu_{FIT} =  u_{c} l_c \begin{cases}
 5 \quad & (\frac{|\omega|}{\omega_{c}}<10^{-2}), \\
 \frac{1}{2}\left(\frac{\omega_{c}}{|\omega|}\right)^{\frac{1}{2}} \quad &(\frac{|\omega|}{\omega_{c}} \in [10^{-2},5]), \\
\frac{25}{\sqrt{20}}\left(\frac{\omega_{c}}{|\omega|}\right)^2 \quad &(\frac{|\omega|}{\omega_{c}}>5). \\
 \end{cases}
 \label{nuEFIT1}
\end{eqnarray}
This is a fit to the upper envelope of simulation results for $\nu_E$ at high frequencies, which provides the maximum dissipation (for reasons that will become clear below). This appears to hold for a wide range of Rayleigh numbers in a local model of convection, but the simulations also indicate a possible tidal amplitude dependence for $\omega/\omega_c\gtrsim 1$, which we do not explore further here. We caution though that the intermediate regime may be somewhat model dependent \citep{VB2020a}.

Note that $\nu_\mathrm{FIT}$ matches the scaling of $\nu_\mathrm{GN}$ at high frequencies, but with a different proportionality constant that implies more efficient dissipation. This also predicts much more efficient dissipation at very low frequencies than any of the previous prescriptions, since the constant of proportionality is 5 rather than 1/3. However, we will show that $\nu_\mathrm{FIT}$ predicts much less efficient dissipation than $\nu_\mathrm{Z}$ at high frequencies. Note that when $\omega/\omega_c\gg 1$, the resulting $Q'_\mathrm{eq}\propto |\omega|^{-1}$ if we adopt $\nu_E=\nu_\mathrm{MLT}$, $Q'_\mathrm{eq}=$ const if we adopt $\nu_E=\nu_Z$ and $Q'_\mathrm{eq}\propto |\omega|$ if we adopt either $\nu_E=\nu_\mathrm{GN}$ or $\nu_E=\nu_\mathrm{FIT}$.

\begin{figure}
  \begin{center}
    \subfigure{\includegraphics[trim=3.5cm 0cm 4.5cm 1cm,clip=true,width=0.4\textwidth]{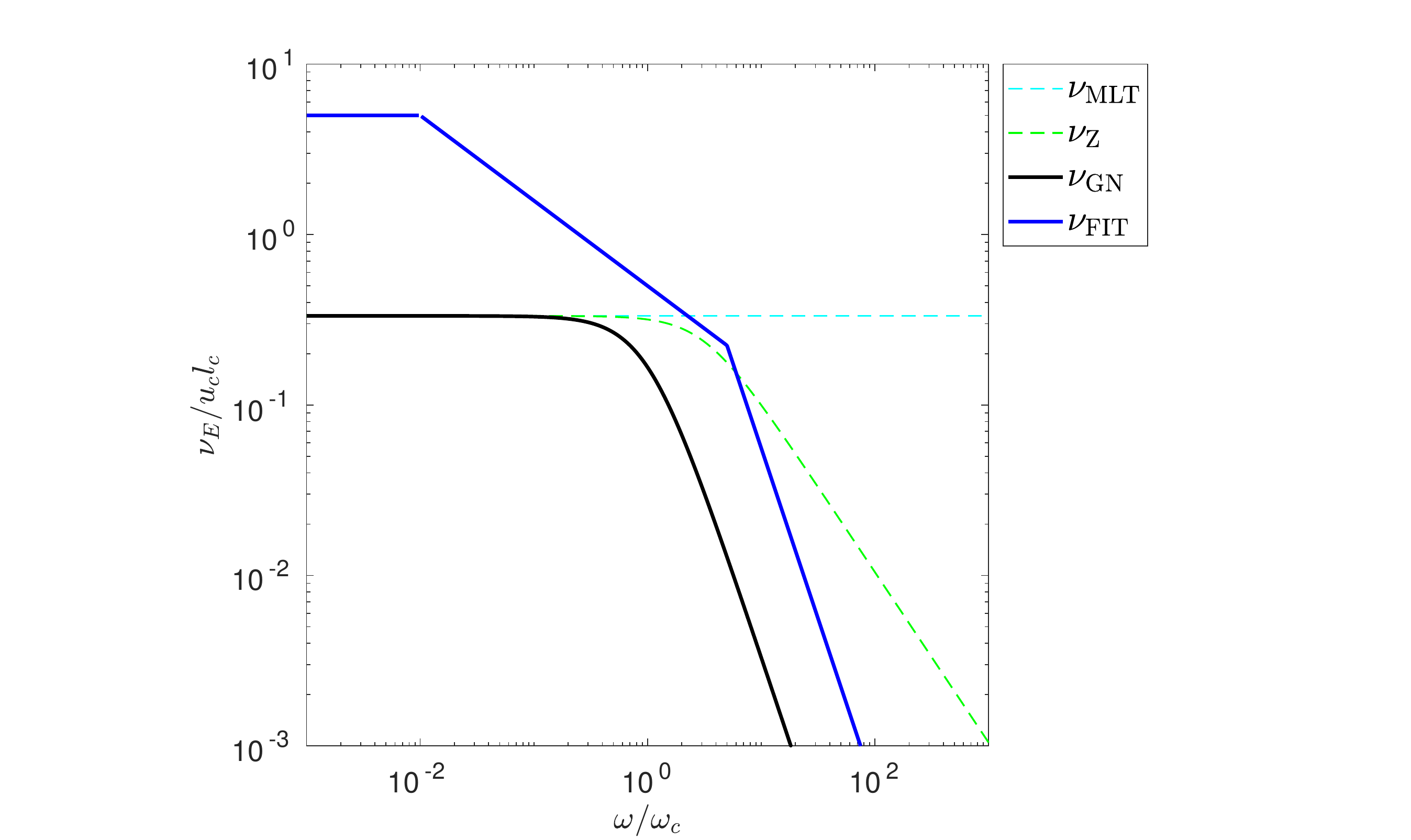}}
     \end{center}
  \caption{Effective viscosity prescription $\nu_E$ (normalised by $u_c l_c$) as a function of tidal frequency $\omega$ normalised by the convective frequency $\omega_c$. These prescriptions are evaluated at each radius in the convection zone of a given stellar model.}
  \label{nuEplot}
\end{figure}

These frequency-dependent prescriptions for $\nu_E$ are shown as a function of tidal frequency in Fig.~\ref{nuEplot}. Determining the correct viscosity prescription (and if such a prescription can indeed by applied at all) has long been considered as the Achilles' heel of tidal theory \citep{Zahn2008}. While significant uncertainties remain in our understanding of this mechanism, and much further work is required, the latest numerical simulations have shed some light on this matter, and we are now in a position to explore the consequences of these new results.

\section{Dynamical (wave-like) tides}
\label{WLtides} 

We consider the tidal excitation and dissipation of dynamical tides in both convective and radiative regions to quantify their importance for tidal dissipation.

\subsection{Inertial waves in convection zones}
\label{IW}

We compute the tidal excitation of inertial waves using a frequency-averaged formalism \citep{Ogilvie2013}. This builds upon prior work \citep[e.g.][]{Mathis2015,Bolmont2016,Gallet2017,Benbakoura2019} by accounting for the realistic structure of the star. The equilibrium tide in convection zones forces the dynamical tide through the inertial terms in the equation of motion. If we were to solve the linear tidal problem in the convection zone including rotation, we would obtain a strongly frequency-dependent response \citep[e.g.][]{SavPap1997,OL2007,RV2010}. Here we instead compute the frequency-averaged dissipation, which is a simplified way to quantify the typical dissipation resulting from inertial waves in convection zones. This is a crude measure which is nevertheless useful for population studies since it indicates the typical level of dissipation due to these waves. The actual dissipation in an individual system at a particular tidal frequency may differ by up to 2 to 3 orders of magnitude (based on prior results), but is subject to significant uncertainties, whereas the frequency-averaged measure is probably robust. The waves are calculated assuming an impulsive forcing and the specific dissipation mechanism (which may involve an uncertain combination of nonlinear effects, magnetic fields or the interactions of these waves with turbulent convection) is not explicitly studied.

The frequency-averaged dissipation due to these waves can be obtained by applying an impulsive forcing. Following \cite{Ogilvie2013}, the resulting pressure perturbation ($\propto W_l Y_l^m$) in an inhomogeneous body can be shown to satisfy
\begin{eqnarray}
\label{WL}
\frac{1}{r^2}\frac{\mathrm{d}}{\mathrm{d}r}\left(r^2\rho \frac{\mathrm{d} W_l }{\mathrm{d}r}\right)-\frac{l(l+1)}{r^2}\rho W_l=\frac{2\mathrm{i}m\Omega}{r}\frac{\mathrm{d}\rho}{\mathrm{d} r}X_l,
\end{eqnarray}
which must satisfy boundary conditions of vanishing radial velocity at the boundaries of each convection zone, i.e.
\begin{eqnarray}
\frac{\mathrm{d} W_l }{\mathrm{d}r}=\frac{2\mathrm{i}m\Omega X_l}{r}.
\end{eqnarray}
We solve Eq.~\ref{WL} in the same way as Eqs.~\ref{PhiEqn} and \ref{Xeqn}.
The associated tidal quality factor representing the frequency-averaged dissipation of inertial waves is then
\begin{eqnarray}
\frac{1}{\langle Q'_{\mathrm{IW}}\rangle} =
\frac{32\pi^2G}{3(2l+1)R^{2l+1}|A|^2}(E_l+E_{l-1}+E_{l+1}),
\label{QIW}
\end{eqnarray}
where
\begin{eqnarray}
E_l &=& \frac{1}{4}\int \rho r^2 \left(|a_l|^2 +l(l+1) r^2 |b_l|^2\right)\mathrm{d} r, \\
E_{l-1} &=& \frac{1}{4}\int \rho r^2 l(l-1) r^2 |c_{l-1}|^2\mathrm{d} r, \\
E_{l+1} &=& \frac{1}{4}\int \rho r^2 (l+1)(l+2) r^2 |c_{l+1}|^2\mathrm{d} r,
\end{eqnarray}
and
\begin{eqnarray}
a_l &=& \frac{2\mathrm{i}m\Omega}{r}X_l-\frac{\mathrm{d}W_l}{\mathrm{d}r}, \\
b_l &=& \frac{2\mathrm{i}m\Omega}{l(l+1) r^2}\left(r \frac{\mathrm{d}X_l}{\mathrm{d}r}+X_l\right)-\frac{W_l}{r^2}, \\
c_{l-1} &=& \frac{2\Omega q_l}{r^2}\left(r \frac{\mathrm{d}X_l}{\mathrm{d}r}+(l+1) X_l\right), \\
c_{l+1} &=& -\frac{2\Omega q_{l+1}}{r^2}\left(r \frac{\mathrm{d}X_l}{\mathrm{d}r}-l X_l\right), \\
q_l &=& \frac{1}{l}\left(\frac{l^2-m^2}{4l^2-1}\right)^{\frac{1}{2}}.
\end{eqnarray}
These quantities are straightforward to evaluate numerically in a given stellar model once we have solved Eq.~\ref{WL} (together with Eq.~\ref{PhiEqn} and Eq.~\ref{Xeqn}). Note that according to Eq.~\ref{QIW}, we find $\langle Q'_\mathrm{IW} \rangle \propto \epsilon_\Omega^{-2}\propto \Omega^{-2}$. This mechanism is therefore more efficient in (relatively) rapidly rotating stars. This mechanism does not operate at all if tidal frequencies are such that $|\omega|\not \leq 2\Omega$, since inertial waves cannot then be excited by tidal forcing. As a result, Eq.~\ref{QIW} should only be applied to infer the typical level of tidal dissipation due to inertial waves in circumstances when the relevant tidal frequencies satisfy $|\omega| \leq 2\Omega$.

In this paper, we will also compare the dissipation obtained in this way with the equivalent result from a simplified piece-wise homogeneous two-layer model \citep{Ogilvie2013}, which has been adopted in several previous papers \citep[e.g][]{Mathis2015,Bolmont2016,Gallet2017}. In this simplified two-layer model, which is meant to be applied only to stars with convective envelopes and radiative cores, we can show that
\begin{eqnarray}
\label{IWTwoLayer}
\nonumber
&& \frac{1}{\epsilon_\Omega^2\langle Q'_{\mathrm{IW}}\rangle}=\frac{200\pi}{189}\left(\frac{\alpha^5}{1-\alpha^5}\right)\left(1-\gamma\right)^2(1-\alpha)^4 \\
&& \hspace{0.3cm} \times
\left(1+2\alpha+3\alpha^2+\frac{3}{2}\alpha^3\right)^2\left(1+\left(\frac{1-\gamma}{\gamma}\right)\alpha^3\right)\\
&& \hspace{0.3cm} \times
\nonumber
\left(1+\frac{3}{2}\gamma+\frac{5}{2\gamma}
\left(1+\frac{1}{2}\gamma-\frac{3}{2}\gamma^2\right)\alpha^3-\frac{9}{4}(1-\gamma)\alpha^5\right)^{-2}.
\end{eqnarray}
We have followed \cite{Mathis2015} in defining $\alpha=r_c/R$, where $r_c$ is the radius of the radiative/convective interface, $\beta=M_c/M$, where $M_c$ is the mass contained in the radiative core, and  $\gamma=\alpha^3(1-\beta)/(\beta(1-\alpha^3))$, which is the ratio of mean densities of these two zones. The form of Eq~\ref{IWTwoLayer} clearly demonstrates the strong dependence of $\langle Q'_\mathrm{IW}\rangle$ on interior structure even in the two-layer model.

\subsection{Internal gravity waves in radiation zones}
\label{IGW}

We also calculate the tidal dissipation due to internal gravity waves in radiation zones. To do so, we follow e.g.~\cite{GD1998} (who applied the ideas of \citealt{Zahn1975}, \citealt{Zahn1977} and \citealt{GN1989} to solar-type stars), \cite{OL2007} and \cite{BO2010} by assuming that these waves are launched from the convective/radiative (envelope) interface and are then fully damped in the radiation zone, before they can reflect from the centre of the star or another boundary of the radiation zone. This ``travelling wave" regime gives the simplest estimate of the dissipation due to these waves, and it can be viewed as giving the most efficient dissipation in between resonances. This regime is expected to apply when the tidal amplitude is sufficiently large that the waves break \citep{BO2010,B2011}, but it is also relevant if the waves are otherwise efficiently damped (e.g. by radiative diffusion). If tidal forcing is resonant (or locked in resonance) with a global g-mode, enhanced dissipation over this estimate may occur. However, if the tidal amplitude in resonance is sufficiently large, wave breaking may be expected, and the fully damped regime may again be relevant.

To calculate the dissipation we must study the properties of the wave launching region near the radiative/convective interface in these stars. In stars with convective cores, we focus on the interface of the convective envelope rather than the core, since this is found to give the maximum dissipation in the mass range we consider. In the vicinity of a radiative/convective interface at $r=r_c$ in a stellar model, we fit the buoyancy frequency profile with the linear fit\footnote{For our purposes we find the energy flux and resulting tidal dissipation to be similar to fitting the profile with $N^2 \propto \sqrt{r_c-r}$ for $r\sim r_c$ in the convection zone \citep{B2011}, though this can give quantitatively different results in some cases for short-period tidal forcing \citep[see also][]{Ivanov2013}.}
\begin{eqnarray}
N^2(r) = \left(\frac{r}{r_c}-1\right) \left.\frac{\mathrm{d}N^2}{\mathrm{d}\ln r}\right|_{r=r_c}.
\end{eqnarray}
The dynamical tide in the vicinity of $r\sim r_c$ can then be shown to satisfy Airy's differential equation \citep{GD1998}, and the energy flux in gravity waves can be calculated. We will omit the details of this calculation, since they have been reported elsewhere \citep[e.g.][]{GD1998,B2011}, and just quote the resulting tidal quality factor due to gravity waves, which is determined from
\begin{eqnarray}
\label{QIGW}
\frac{1}{Q'_{\mathrm{IGW}}} = \frac{2\left[\Gamma\left(\frac{1}{3}\right)\right]^2}{3^{\frac{1}{3}}(2l+1) (l(l+1))^{\frac{4}{3}}}\frac{R}{GM^2} \mathcal{G} |\omega|^{\frac{8}{3}}.
\end{eqnarray}
The quantities that depend on the radiative/convective interface region in a particular stellar model are encapsulated in the quantity 
\begin{eqnarray}
\mathcal{G} = \sigma_c^2 \rho_c r_c^5 \left|\frac{\mathrm{d}N^2}{\mathrm{d}\ln r}\right|_{r=r_c}^{-\frac{1}{3}},
\end{eqnarray}
which takes the value $\mathcal{G}_\odot \approx 2\times 10^{47} \mathrm{kg}\mathrm{m}^2\mathrm{s}^{2/3}$ for the current Sun (in which $\sigma_c=-1.18$).
In this expression, $\rho_c=\rho (r_c)$, and the parameter
\begin{eqnarray}
\sigma_c=\frac{\omega_{dyn}^2}{A}\left.\frac{\partial \xi_{d,r}}{\partial r}\right|_{r=r_c},
\end{eqnarray}
where the derivative of the dynamical tide radial displacement $\xi_{d,r}$ is determined by integrating the linear differential equation given in Eq.3 in \cite{GD1998}. Note that $\sigma_c$ (and $Q'_\mathrm{IGW}$, for that matter) is defined such that it depends on the properties of the star but not on $A$.

A numerical evaluation of Eq.~\ref{QIGW} in a model of the current Sun gives \citep{BO2010}
\begin{eqnarray}
\label{QIGW1}
Q'_{\mathrm{IGW}}\approx 1.5\times 10^5 \left(\frac{\mathcal{G}_\odot}{\mathcal{G}}\right)\left(\frac{M}{M_\odot}\right)^2\left(\frac{R_\odot}{R}\right) \left(\frac{P_\mathrm{tide}}{0.5 \,\mathrm{d}}\right)^{\frac{8}{3}},
\end{eqnarray}
 for a tidal period relevant for a hot Jupiter on a 1 day orbit around a slowly rotating star. This is therefore an efficient mechanism of tidal dissipation for such short tidal periods. We will later evaluate how this estimate changes during stellar evolution for stars with radiation zones in the mass range 0.4 to 1.6 solar masses. Note that this estimate apparently agrees very closely with the value required to explain the inferred orbital decay rate of WASP-12 b \citep{Patra2017,Yee2020,Patra2020}. We will comment further on this issue later in the paper in \S~\ref{HJimplications}.

An important question is whether we would expect the gravity waves to be fully damped in the radiation zone. In stars with radiative cores, hydrodynamical simulations suggest that these waves may be fully damped if they have large enough amplitudes so that they overturn the stratification and break \citep{BO2010,B2011}. This requires the companion mass to exceed a particular threshold that depends on the structure of the star. For a circularly orbiting planet of mass $M_p$ orbiting a slowly rotating star with $\omega=2(n-\Omega)\approx 2n=4\pi/P_\mathrm{orb}$, where $n$ is the orbital mean motion (orbital angular frequency), $P_\mathrm{orb}$ is the orbital period, and the tidal amplitude in this case is
\begin{eqnarray}
A = \sqrt{\frac{6\pi}{5}} \omega_{\mathrm{dyn}}^2 \epsilon_T.
\end{eqnarray}
We have also defined the dimensionless tidal amplitude
\begin{eqnarray}
\epsilon_T=\frac{M_p}{M}\left(\frac{R}{a}\right)^3,
\end{eqnarray}
where
$a=\left(G(M_p+M)/n^2\right)^{\frac{1}{3}}$ is the semi-major axis. 

The criterion for the wave to overturn the stratification, and therefore be expected to break, at the centre of a star with a radiative core is $A_{nl}\gtrsim1$, where
\begin{eqnarray}
\label{Anl}
A_{nl}^2 = \frac{3^{\frac{2}{3}} 72\sqrt{6}\left[\Gamma\left(\frac{1}{3}\right)\right]^2}{40\pi^2(l(l+1))^{\frac{4}{3}}} \frac{\mathcal{G}\,  C^5}{\rho_0} \frac{|A|^2}{\omega_\mathrm{dyn}^4}|\omega|^{-\frac{13}{3}}.
\end{eqnarray}
where $\rho_0=\rho(r=0)$ is the central density and 
\begin{eqnarray}
N = C r
\end{eqnarray}
near the centre of the star. The quantity $C$ is obtained from a linear fit in a given  model, and for the current Sun, $C_\odot \approx 8\times 10^{-11} \mathrm{m}^{-1}\mathrm{s}^{-1}$.

Wave breaking is expected when $A_{nl}\gtrsim1$, but weaker wave-wave interactions could still be important even when $A_{nl}<1$, or even for $A_{nl}$ as small as 0.1 \citep{BO2011,Weinberg2012,EssickWeinberg2016}. A numerical evaluation of Eq.~\ref{Anl} for the current Sun gives \citep{BO2010}
\begin{eqnarray}
\label{Anl2}
A_{nl} \approx 0.3 \left(\frac{\mathcal{G}}{\mathcal{G}_\odot}\right)^{\frac{1}{2}}\left(\frac{C}{C_\odot}\right)^{\frac{5}{2}}\left(\frac{M_p}{M_J}\right)\left(\frac{P}{1 \mathrm{d}}\right)^{\frac{1}{6}}.
\end{eqnarray}
This can be rewritten as a criterion on the planetary mass $M_p\gtrsim M_{\mathrm{crit}}$ for wave breaking to occur, where
\begin{eqnarray}
M_{\mathrm{crit}} \approx 3.3 M_J \left(\frac{\mathcal{G}_\odot}{\mathcal{G}}\right)^{\frac{1}{2}}\left(\frac{C_\odot}{C}\right)^{\frac{5}{2}}\left(\frac{P}{1 \mathrm{d}}\right)^{-\frac{1}{6}}.
\end{eqnarray}
We will later evaluate Eq.~\ref{Anl} to determine how the critical mass $M_\mathrm{crit}$ depends on the stellar mass and age. This is necessary to indicate when we would expect waves to be fully damped in the radiation zone such that the efficiency of tidal dissipation is determined by Eq.~\ref{QIGW}.

In stars with convective cores (such as in F-stars), we cannot use Eq.~\ref{Anl} to determine whether wave breaking would be expected because the central regions of the star are no longer stably stratified and do not support gravity waves. Instead, we use the WKB expression
\begin{eqnarray}
\label{xidrkr}
|\xi_{d,r}k_r| \gtrsim 1,
\end{eqnarray}
to determine when wave breaking would be expected\footnote{This differs by a factor $\sqrt{4\pi}$ from the expression in \cite{GD1998} and \cite{SunArrasWenberg2018} to agree more closely with Eq.~\ref{Anl}.}, where $\xi_{d,r}$ is the coefficient of $Y_l^m \mathrm{e}^{-\mathrm{i}\omega t}$ and $k_r$ is the radial wavenumber. To compute this quantity, we first define the wave energy flux magnitude
\begin{eqnarray}
F = \frac{3^{\frac{2}{3}}}{8\pi}\frac{\left[\Gamma\left(\frac{1}{3}\right)\right]^2}{(l(l+1))^{\frac{4}{3}}}\mathcal{G}\frac{|A|^2}{\omega_\mathrm{dyn}^4} |\omega|^{\frac{11}{3}},
\end{eqnarray}
which is conserved in the radiation zone after the waves have been launched from the convective/radiative interface (neglecting damping processes).
To obtain $|\xi_{d,r}|^2$ in the interior of the radiation zone, we apply WKB theory for short-wavelength waves, which gives $F=r^2 \rho N^2 c_{g,r} |\xi_{d,r}|^2$. 
We employ the internal gravity wave dispersion relation $\omega^2=N^2 k_\perp^2/k^2$
for low tidal frequencies (that significantly exceed the rotation frequency), where $k_\perp=\sqrt{l(l+1)}/r$ is the horizontal wavenumber and $k^2=k_r^2+k_\perp^2\approx k_r^2$. The magnitude of the radial group velocity is then $c_{g,r} = \partial \omega/\partial k_r \approx \omega/k_r$. Eq.~\ref{xidrkr} then gives
\begin{eqnarray}
\label{xidrkr2} 
|\xi_{d,r} k_r| = \sqrt{\frac{[l(l+1)]^{\frac{3}{2}} N F}{\rho r^5 \omega^4}} \gtrsim 1
\end{eqnarray}
for wave breaking to occur. 
This formula applies in any radiation zone irrespective of whether it extends all the way to the centre of the star, and thus it allows us to predict whether wave breaking would be expected in F-stars with convective cores, for example. We will use the maximum value of Eq.~\ref{xidrkr2} (for which $\omega^2\leq N^2$) in the radiation zone later to estimate whether wave breaking would be expected whenever Eq.~\ref{Anl} cannot be applied. In stars with radiative cores, this estimate is found to agree quite well with Eq.~\ref{Anl} (though there is some variation depending on the precise radial grid points used for this estimate).

We neglect the effect of rotation on gravity waves, except insofar as it affects the tidal frequency $\omega$, since its inclusion would require a more complicated numerical evaluation (\citealt{PapSav1997,OL2007}, though also see \citealt{Ivanov2013}). Rotation can lead to a significant enhancement in tidal dissipation due to this mechanism if global r-modes are excited in the convection zone, but it can be ignored in applications where the tidal period is shorter than half the rotation frequency. Since most planetary hosts are slow rotators, the effects of rotation are unlikely to be important when we apply these predictions below.

\section{Dissipation of equilibrium tides in convection zones}
\label{EQMCOMP} \label{NWLdissipation}

We now present the results of our investigation. We begin by analysing the equilibrium tide and the resulting dissipation in the convection zones of a range of stellar models. The parameters that we use in MESA are given in Appendix~\ref{MESA} to allow our models to be reproduced. We have chosen the initial metallicity $Z_{init}=0.02$ except where otherwise specified, and different quantitative results could be obtained in stars with different $Z_{init}$, though the main results below are likely to be robust. We compare the dissipation obtained with the correct equilibrium tide (hereafter referred to as NWL), as defined by Eq.~\ref{NWLtide}, with that of the conventional equilibrium tide of \cite{Zahn1966} (hereafter referred to as EQ), as defined by the solutions of Eq.~\ref{EQMtide}. In Appendix~\ref{EQMCompsection}, we provide a complementary analysis by comparing the correct displacement with the conventional equilibrium tide.

\begin{table}
\begin{tabular}{cccc}
\hline
$M/M_\odot$ & $R/R_\odot$ & $r_g^2$ & $\mathrm{age}/\mathrm{Gyr}$ \\
\hline 
0.2 & 0.21 & 0.21 & 2.93 \\
0.5 & 0.45 & 0.18 & 3.34 \\
0.8 & 0.73 & 0.11 & 2.59 \\
1 & 1 & 0.074 & 4.7 \\
1.2 & 1.34 & 0.049 & 2.86 \\
1.4 & 1.57 & 0.039 & 1.29 \\
1.6 & 1.86 & 0.036 & 1.03 \\
\hline
\end{tabular}
\caption{Table of various stellar models plotted in Figs.~\ref{EQMNWLdisscomp} and \ref{EQMNWLcomp} and in \S~\ref{HJimplications} and \ref{binaryimplications}. The age quoted includes the pre-main sequence (PMS) phase and does not start at zero age main sequence (ZAMS). The model with $M=1M_\odot$ is ``solar-like" (albeit its slightly different age). All models have $Z_{init}=0.02$.}
\label{Table}
\end{table}

\begin{figure*}
  \begin{center}
    \subfigure[$M/M_{\odot}=0.2,\,\text{age/yr}=2.93\times10^9$]{\includegraphics[trim=1cm 0cm 3.5cm 0cm,clip=true,width=0.4\textwidth]{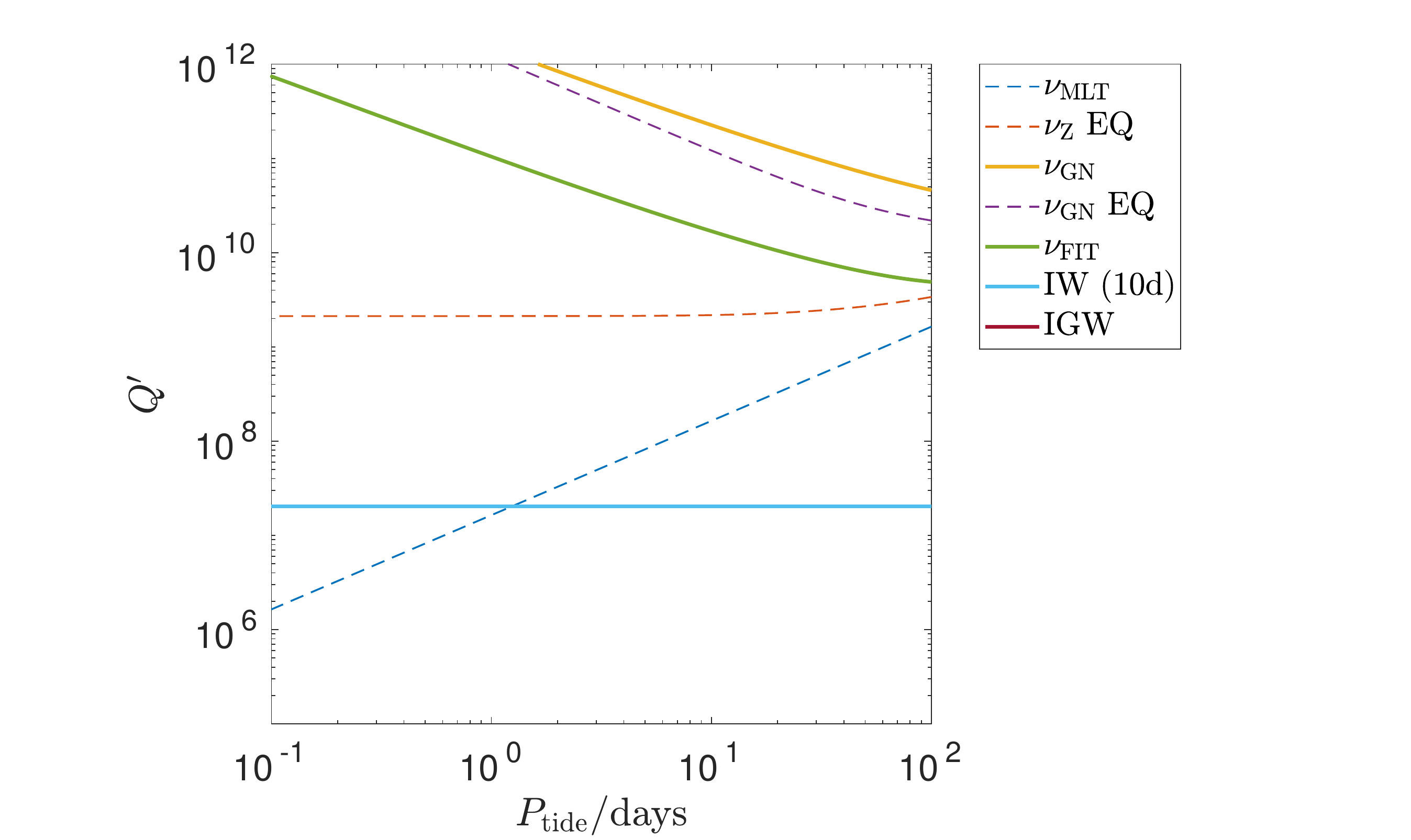}}
    \subfigure[$M/M_{\odot}=0.5,\,\text{age/yr}=3.34\times10^9$]{\includegraphics[trim=1cm 0cm 3.5cm 0cm,clip=true,width=0.4\textwidth]{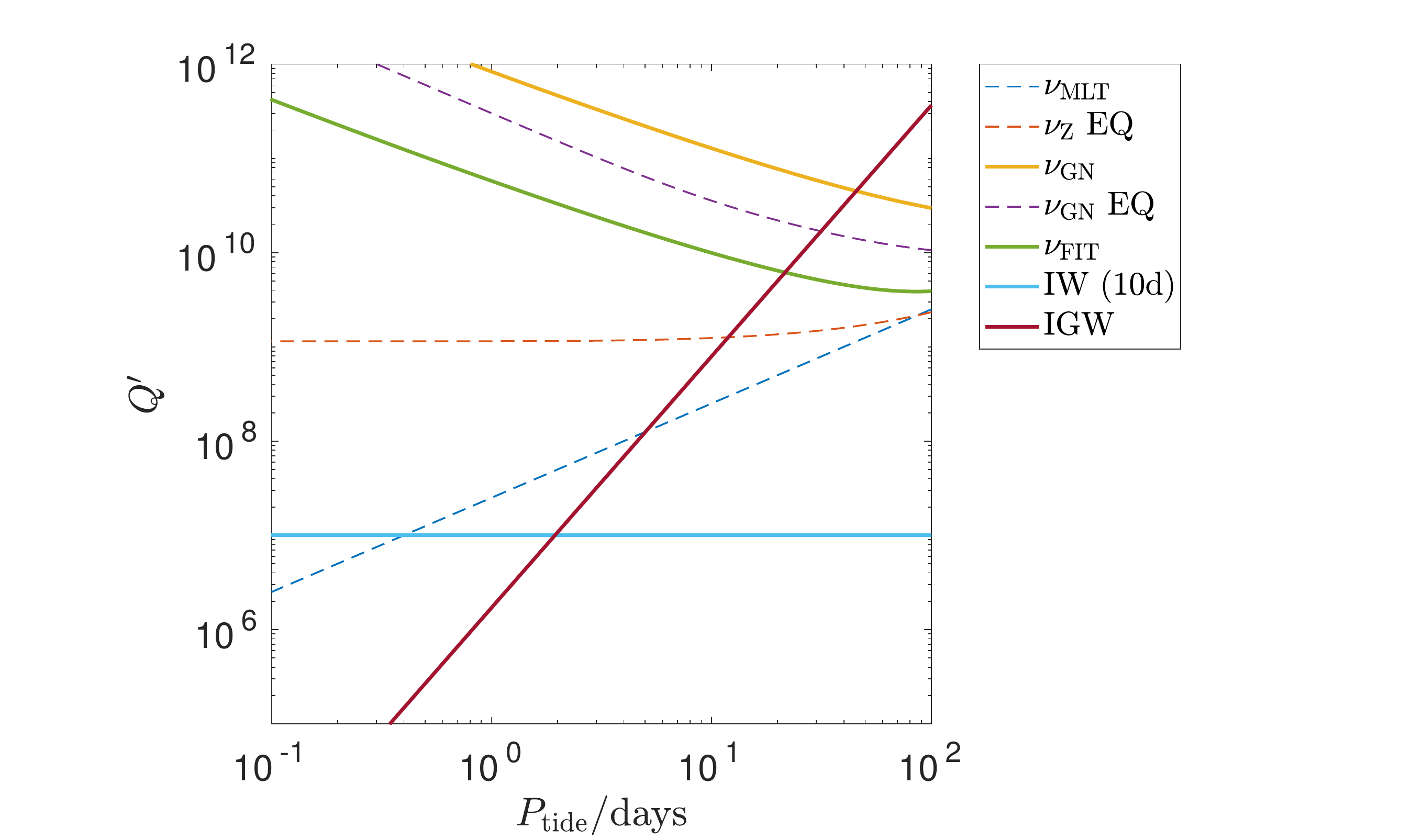}}
    \subfigure[$M/M_{\odot}=1,\,\text{age/yr}=4.70\times10^9$]{\includegraphics[trim=1cm 0cm 3.5cm 0cm,clip=true,width=0.4\textwidth]{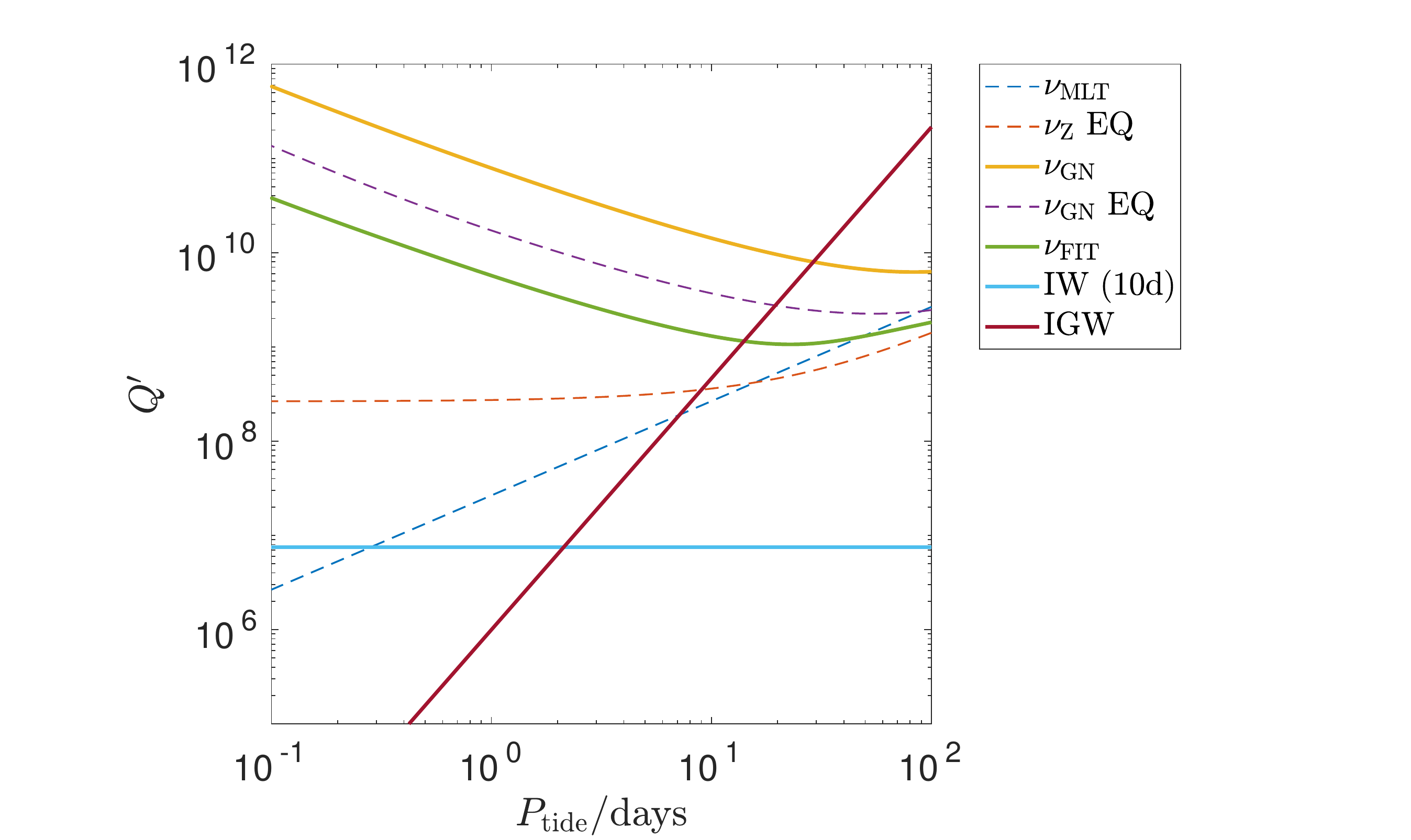}}
    \subfigure[$M/M_{\odot}=1.4,\,\text{age/yr}=1.29\times10^9$]{\includegraphics[trim=1cm 0cm 3.5cm 0cm,clip=true,width=0.4\textwidth]{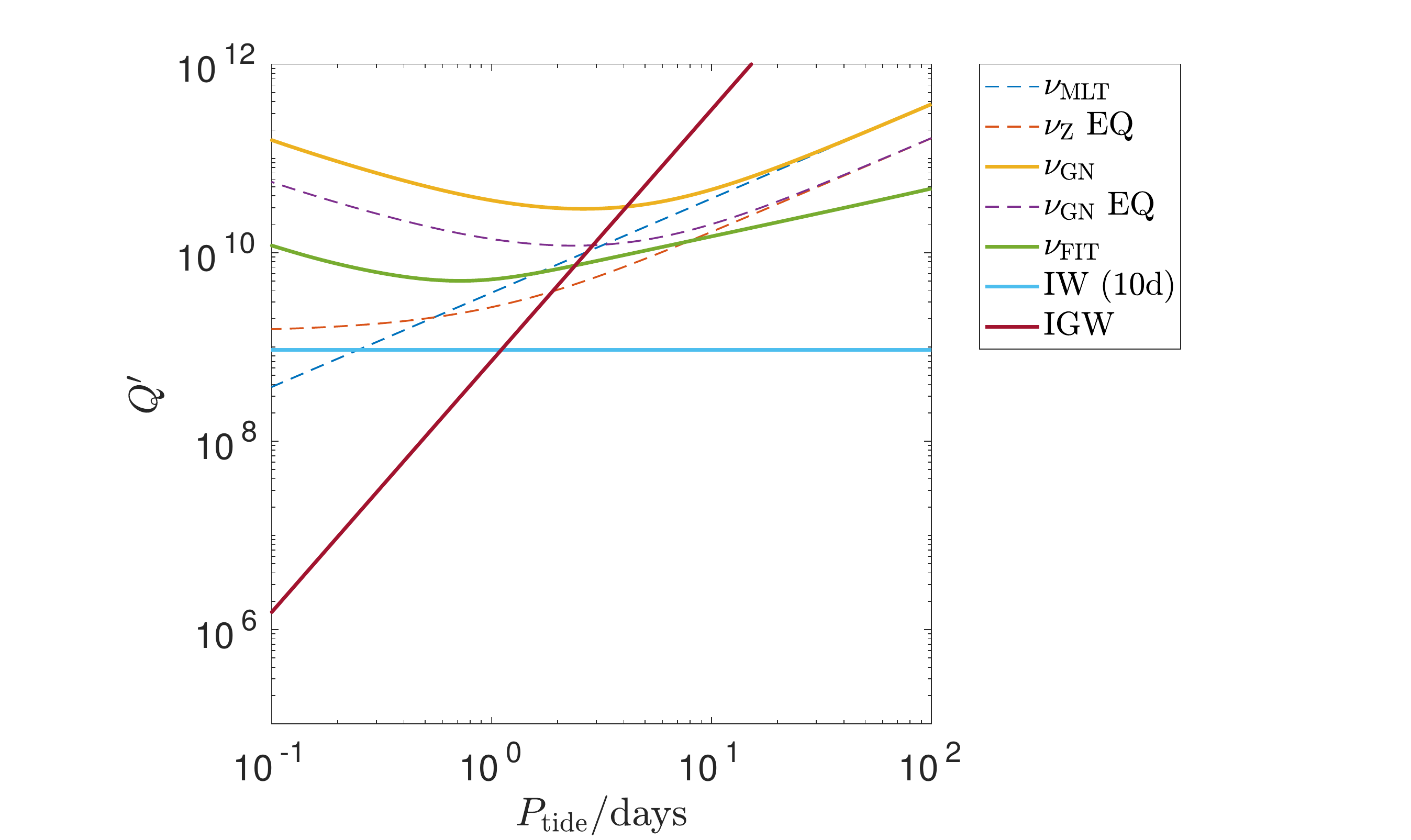}}
    \subfigure[$M/M_{\odot}=1.6,\,\text{age/yr}=1.03\times10^9$]{\includegraphics[trim=1cm 0cm 3.5cm 0cm,clip=true,width=0.4\textwidth]{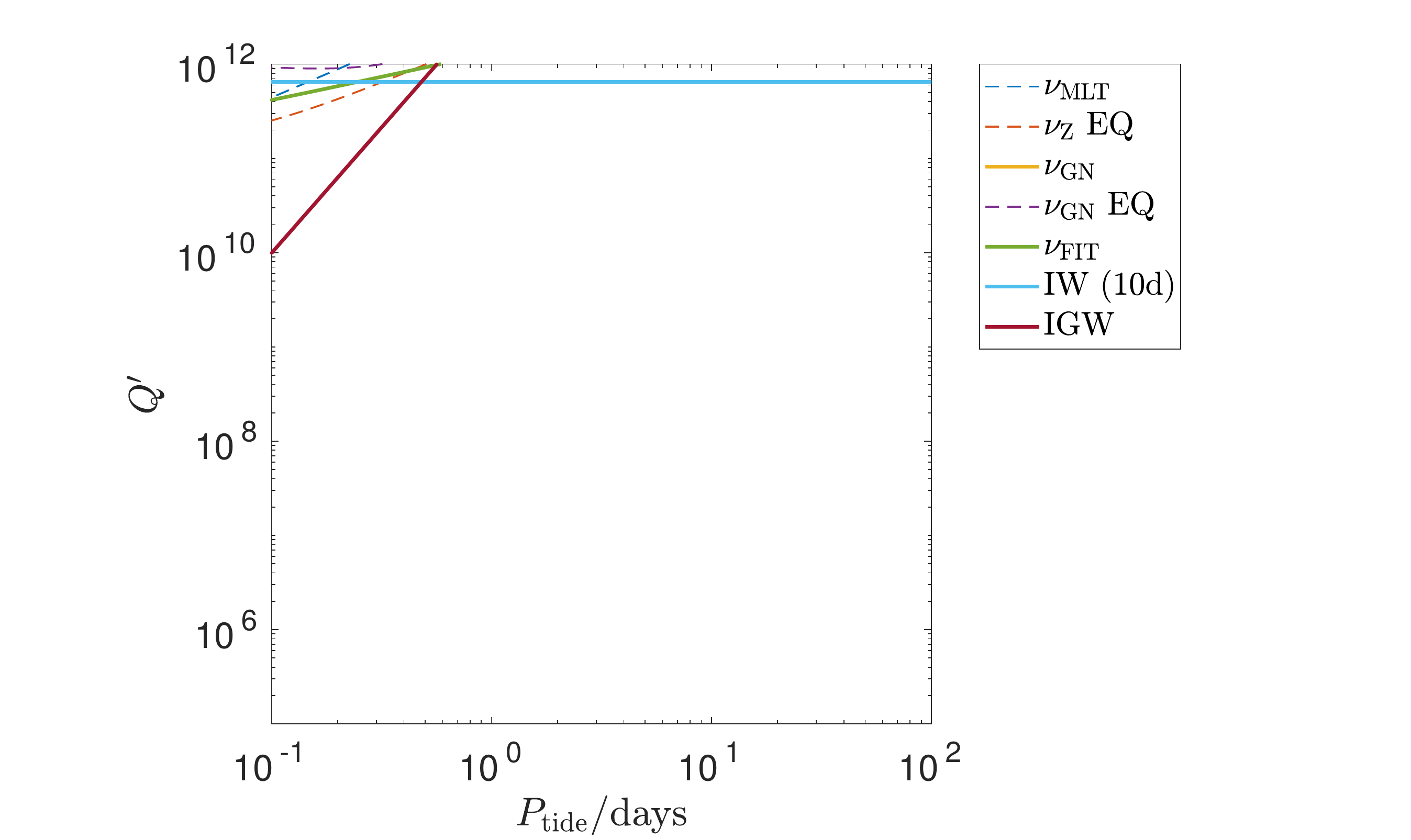}}
    \end{center}
  \caption{$Q^\prime_\mathrm{eq}$ as a function of tidal period (in days) for convective damping of the equilibrium tide. The associated tidal displacements are shown in Fig.~\ref{EQMNWLcomp}. We compute dissipation of the correct equilibrium tide (NWL) using various prescriptions for the turbulent viscosity according to \S~\ref{turbvis}. We compare the dissipation of NWL with that of the conventional equilibrium tide (EQ), where the latter is found to over-predict the dissipation by a factor of 2-3. For later reference, we have also plotted $\langle Q'_\mathrm{IW}\rangle$ for the frequency-averaged dissipation due to inertial waves in convection zones (IW) assuming a spin period of 10 days (which should only be applied when $P_\mathrm{tide}>5$ d; see \S~\ref{IWresults}), and $Q'_\mathrm{IGW}$ from the dissipation of gravity waves (IGW) in radiation zones (see \S~\ref{IGWresults}).}
  \label{EQMNWLdisscomp}
\end{figure*}

\subsection{Dissipation in a range of stellar models}

We calculate the dissipation of the equilibrium tide in a range of stellar models representing low-mass and solar-type stars on the main-sequence, and the results for the tidal quality factors $Q^\prime_\mathrm{eq}$ as a function of tidal period (measured in days) are shown in Fig.~\ref{EQMNWLdisscomp}. The masses and ages of each star are given in the panel captions and some details of these models are listed in Table~\ref{Table}. We show results computed using various prescriptions for the turbulent effective viscosity acting on this tidal component, as described in \S~\ref{turbvis}, and as identified in the legend. The tidal displacements in the same stellar models are analysed in Appendix~\ref{EQMCompsection} in Fig.~\ref{EQMNWLcomp} (except for the bottom panel, which shows an F-star with $M=1.6M_\odot$ that is omitted there). 

We find that $Q^\prime_\mathrm{eq}$ resulting from the dissipation of the equilibrium tide varies as a function of tidal period, and depends strongly on the adopted viscosity prescription (for both NWL and EQ). We also find that NWL and EQ give different predictions for $Q^\prime_\mathrm{eq}$ for each viscosity prescription, though we have only shown EQ for $\nu_\mathrm{Z}$ and $\nu_\mathrm{GN}$ in this figure. We have indicated models that we do not advocate using with dashed lines, which are shown to highlight how their predictions differ from those that are more compatible with simulations (identified by solid lines). As a test of our code, we have checked that we obtain the same results as Fig.~5 of \cite{Ogilvie2014} using a model similar to the current Sun.

We note that if there is no frequency-reduction applied to the turbulent viscosity (and we adopt $\nu_\mathrm{MLT}$), then convective damping of equilibrium tides would be expected to be relatively efficient ($Q'_\mathrm{eq}\sim 10^7$ for stars with $M\lesssim M_\odot$) for one-day tidal periods. However, this model is not consistent with any theoretical studies or simulations probing the interaction between tidal flows and convection, and is only shown for reference here. When the frequency-reduction is accounted for, we typically find $Q'_\mathrm{eq}\sim 10^{10}-10^{11}$ for one-day tidal periods due to convective damping of NWL in stars with $M\sim M_\odot$ using $\nu_\mathrm{FIT}$ (or $\nu_\mathrm{GN}$). Low-mass stars with $M\sim 0.2-0.5 M_\odot$ have weaker dissipation, with $Q'_\mathrm{eq}\sim 10^{11}-10^{12}$. F-type stars with $M\sim 1.3-1.6 M_\odot$ with thinner convective envelopes also have weaker dissipation, with $Q'_\mathrm{eq}\sim 10^{10}-10^{12}$. These results imply that convective damping of equilibrium tides in low mass and solar-type stars on the main sequence is generally an inefficient mechanism of tidal dissipation because of the strong reduction of $\nu_E$ for short tidal periods.

Solar-mass stars have (relatively) the most efficient convective damping of equilibrium tides because lower-mass stars typically have slower convective velocities, meaning smaller values of $\omega_c$. This means that the turbulent viscosity is expected to be smaller (even before considering any frequency-reduction), and secondly, the same tidal period typically has larger $\omega/\omega_c$, implying a larger frequency-reduction of the turbulent viscosity. This is why convective damping of equilibrium tides in low-mass stars is generally weaker than in solar-mass stars, even though the convection extends to greater depths (and higher density regions). On the other hand, in F-stars with masses above $1.2 M_\odot$, convection is faster than in the Sun, leading to larger turbulent viscosities and smaller frequency reductions for the same tidal period. However, the convective envelopes occupy a decreasing fraction of the stellar mass, and have lower densities, leading again to reduced dissipation compared with solar-mass models for one-day tidal periods, albeit for a different reason than for low-mass stars.

When using the same prescription we find that EQ typically over-predicts the dissipation relative to NWL by a factor of 2-3 in these stellar models. Here we have chosen to compare them with $\nu_\mathrm{GN}$, but similar results are obtained with the other viscosity prescriptions, and in other stellar models. This result is qualitatively consistent with what we might expect based on our analysis of the tidal displacements in Appendix~\ref{EQMCompsection}, as illustrated by Fig.~\ref{EQMNWLcomp}. As a result, we advocate the use of the correct equilibrium tide in convection zones (NWL) rather than the conventional equilibrium tide (EQ) of \cite{Zahn1966,Zahn1989}. One may argue that the remaining uncertainties in our understanding of the interaction between tidal flows and convection (i.e. in the correct prescription for $\nu_E$) is likely to make this small difference unimportant. This may well be true. However, we still advocate using the correct equilibrium tide since this is an error that is straightforward to eliminate.

This issue is compounded when using the linear frequency-reduction factor of $\nu_\mathrm{Z}$ rather than the quadratic reduction factors at high frequency of $\nu_\mathrm{GN}$ and $\nu_\mathrm{FIT}$. The associated $Q'_\mathrm{eq}$ using the conventional equilibrium tide and the linear reduction factor can be smaller by a large factor between 1-3 orders of magnitude for tidal periods of order 1 day in these stellar models. Consequently, the tidal dissipation resulting from convective damping of equilibrium tides may have been over-predicted in some previous studies by between 1-3 orders of magnitude. Note that adopting $\nu_\mathrm{Z}$ predicts a tidal period-independent (i.e. constant) $Q'_\mathrm{eq}$ for short periods, but this is not consistent with hydrodynamical simulations for such tidal periods \citep{OL2012,DBJ2020,DBJ2020a} (except perhaps for a narrow intermediate range of frequencies; \citealt{VB2020}).

Even if we adopt the correct equilibrium tide, some uncertainties remain in the viscosity prescription at high frequency, leading to uncertainties in $Q^\prime_\mathrm{eq}$ for tidal periods of 1 day by up to an order of magnitude, based on our current understanding \citep{DBJ2020a,VB2020a}. Our overall conclusions in this section agree with \cite{Penev2011}, though our numbers noticeably differ (they obtain $Q'_\mathrm{eq}\sim 10^8-3\times 10^9$ for $0.8-1.4 M_\odot$). This difference is partly because they used EQ, and partly because they used a different viscosity prescription based on their simulations, which spanned a narrower range of tidal periods.

We have also calculated $Q'_\mathrm{eq}$ due to convective damping of equilibrium tides in the convective cores of F-type stars, which exist on the main-sequence in stars with masses larger than about $1.1 M_\odot$. We find that $Q'_\mathrm{eq}$ is larger by many orders of magnitude than that due to dissipation in the envelope, so we omit showing this in any figures. This confirms prior theoretical expectations \citep[e.g.][]{Zahn1977}.

\subsection{Dissipation following stellar evolution}

\begin{figure*}
  \begin{center}
    \subfigure{\includegraphics[trim=3.5cm 0cm 5.5cm 0cm,clip=true,width=0.27\textwidth]{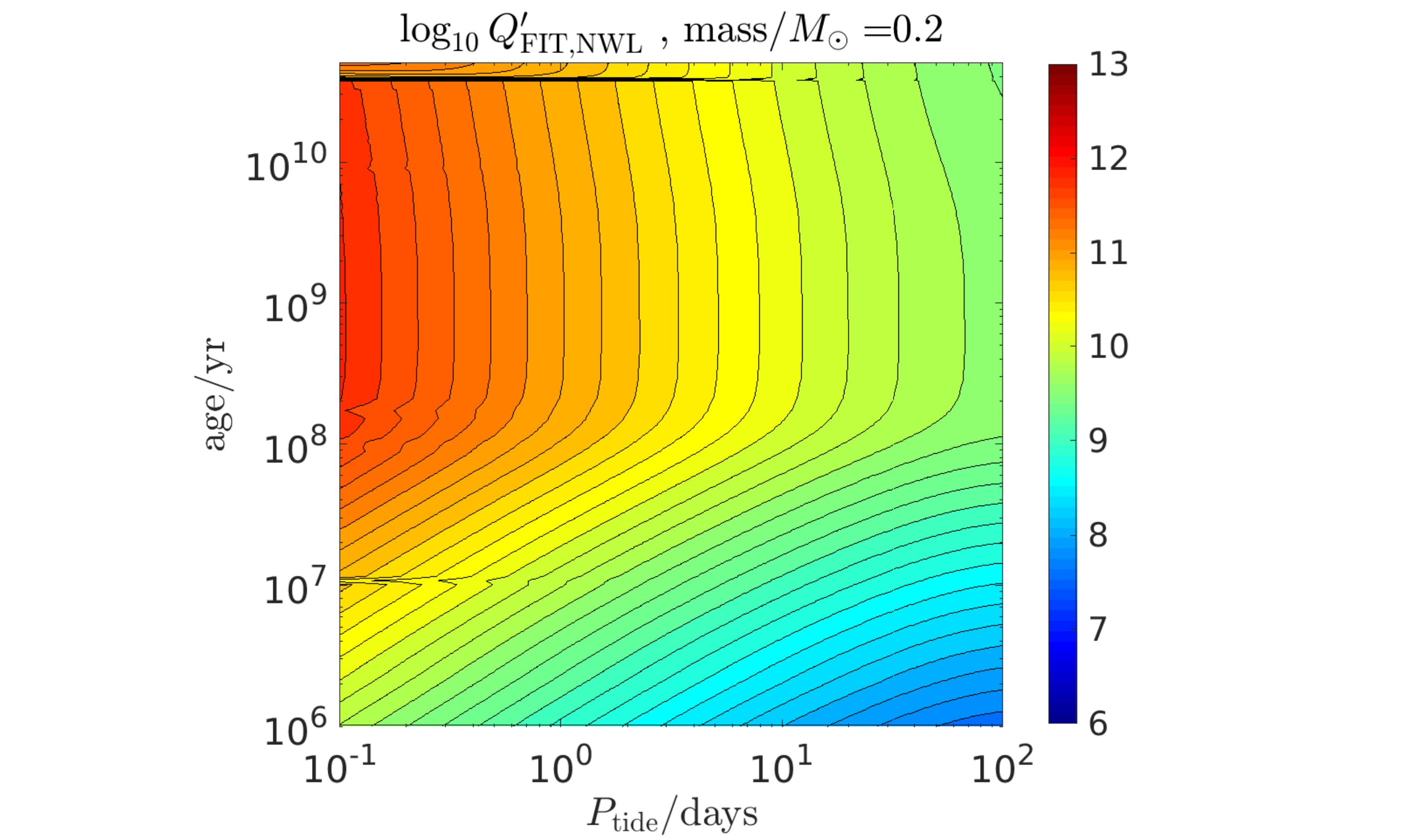}}
    \subfigure{\includegraphics[trim=3.5cm 0cm 5.5cm 0cm,clip=true,width=0.27\textwidth]{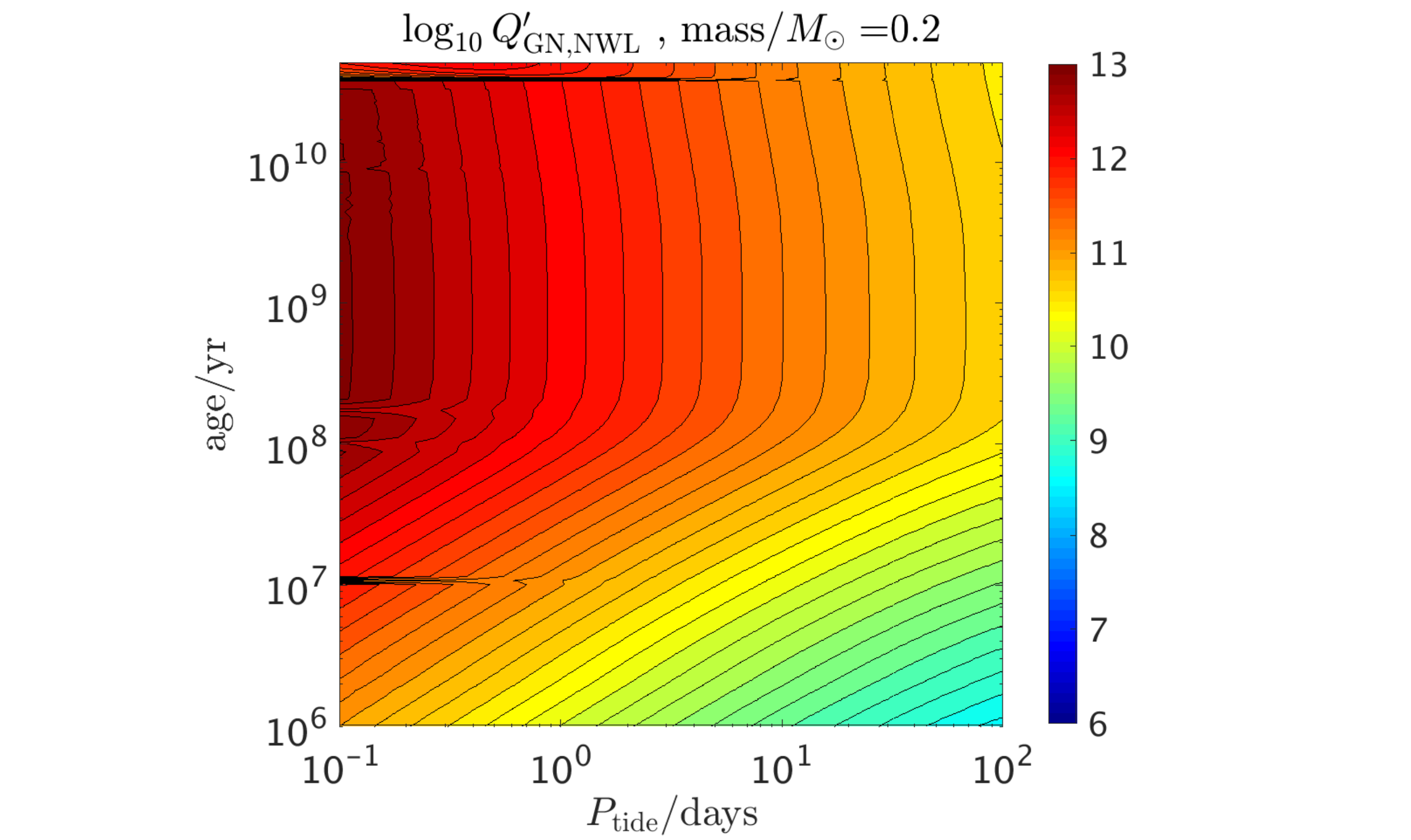}}
     \subfigure{\includegraphics[trim=3.5cm 0cm 5.5cm 0cm,clip=true,width=0.27\textwidth]{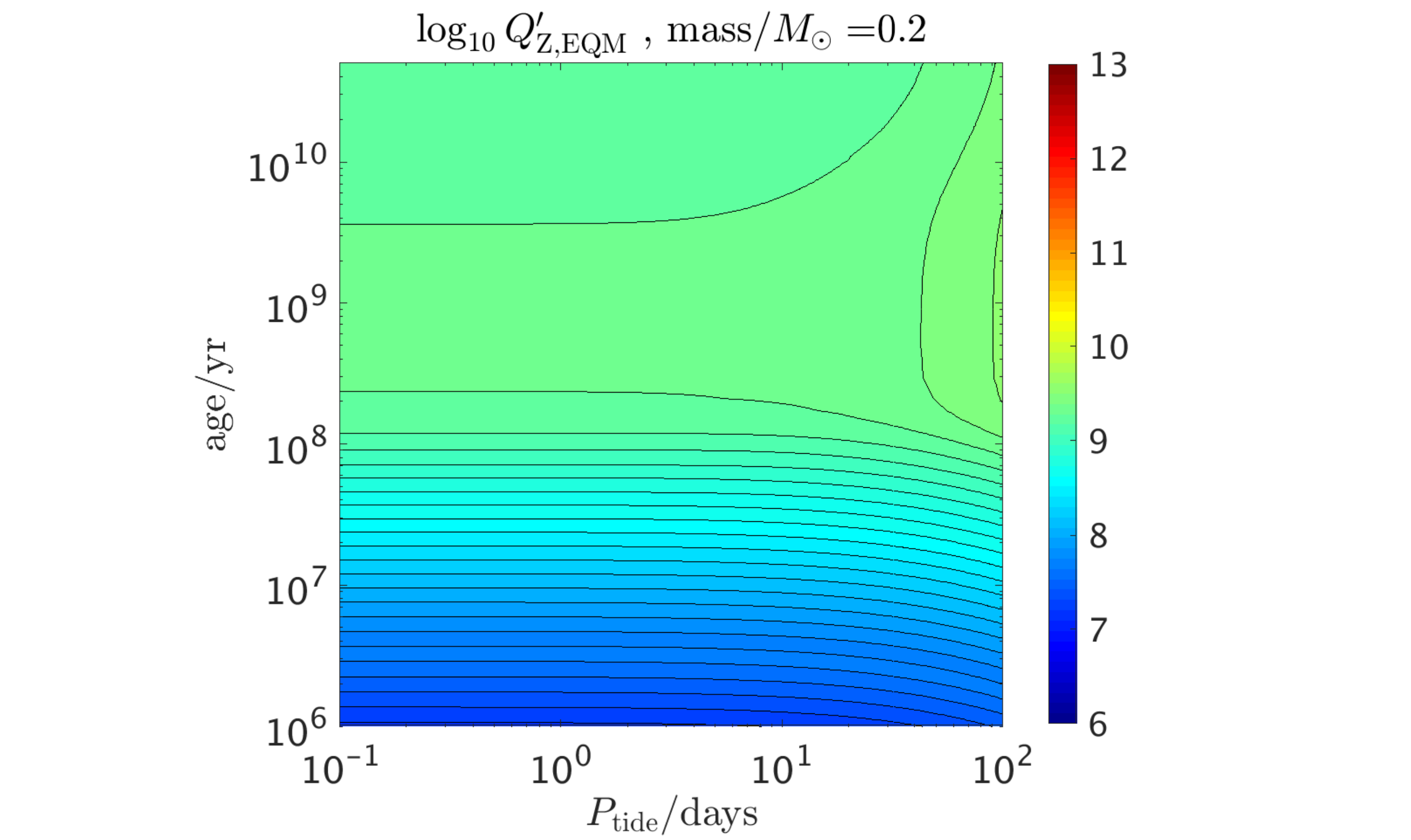}}
    \subfigure{\includegraphics[trim=3.5cm 0cm 5.5cm 0cm,clip=true,width=0.27\textwidth]{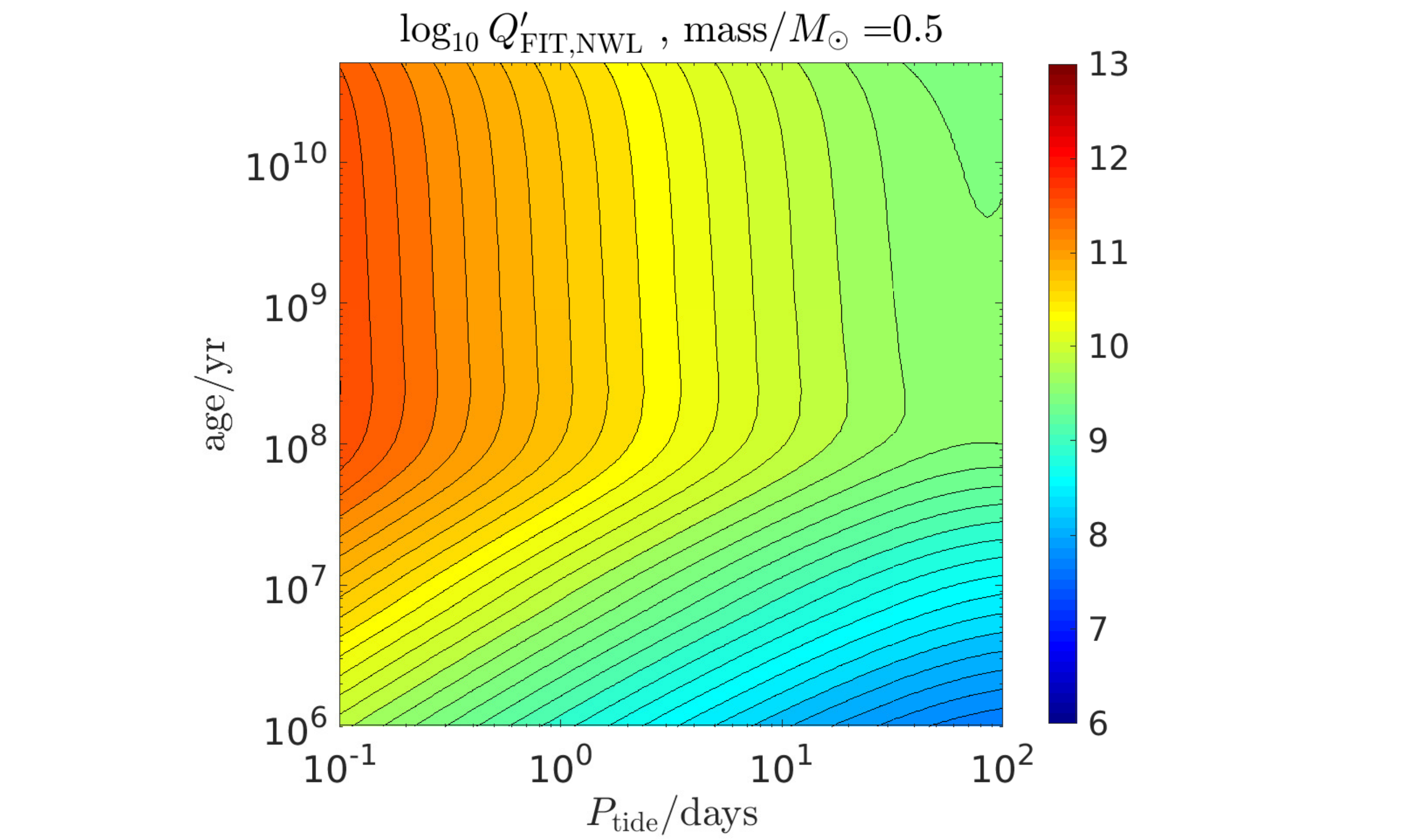}}
    \subfigure{\includegraphics[trim=3.5cm 0cm 5.5cm 0cm,clip=true,width=0.27\textwidth]{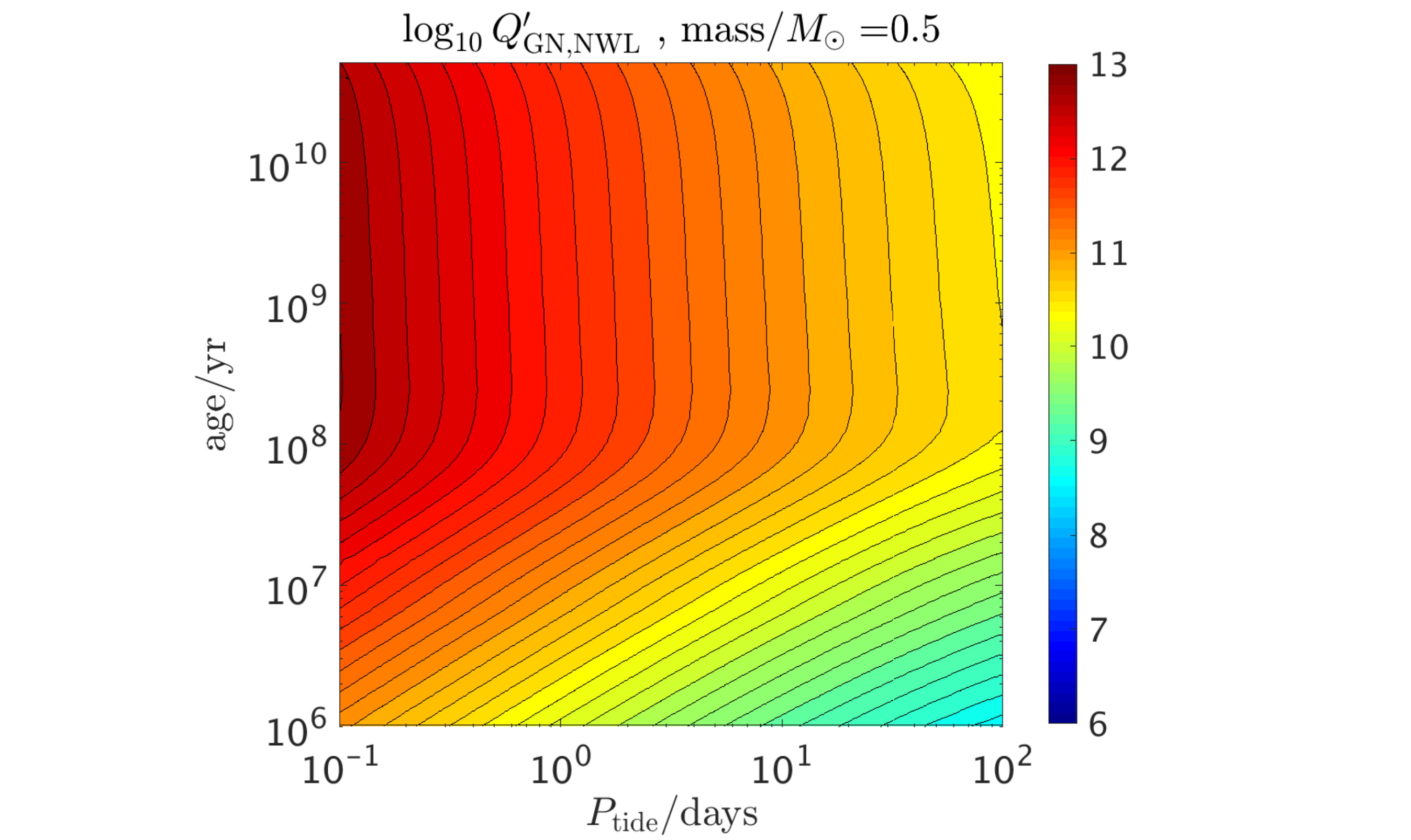}}
     \subfigure{\includegraphics[trim=3.5cm 0cm 5.5cm 0cm,clip=true,width=0.27\textwidth]{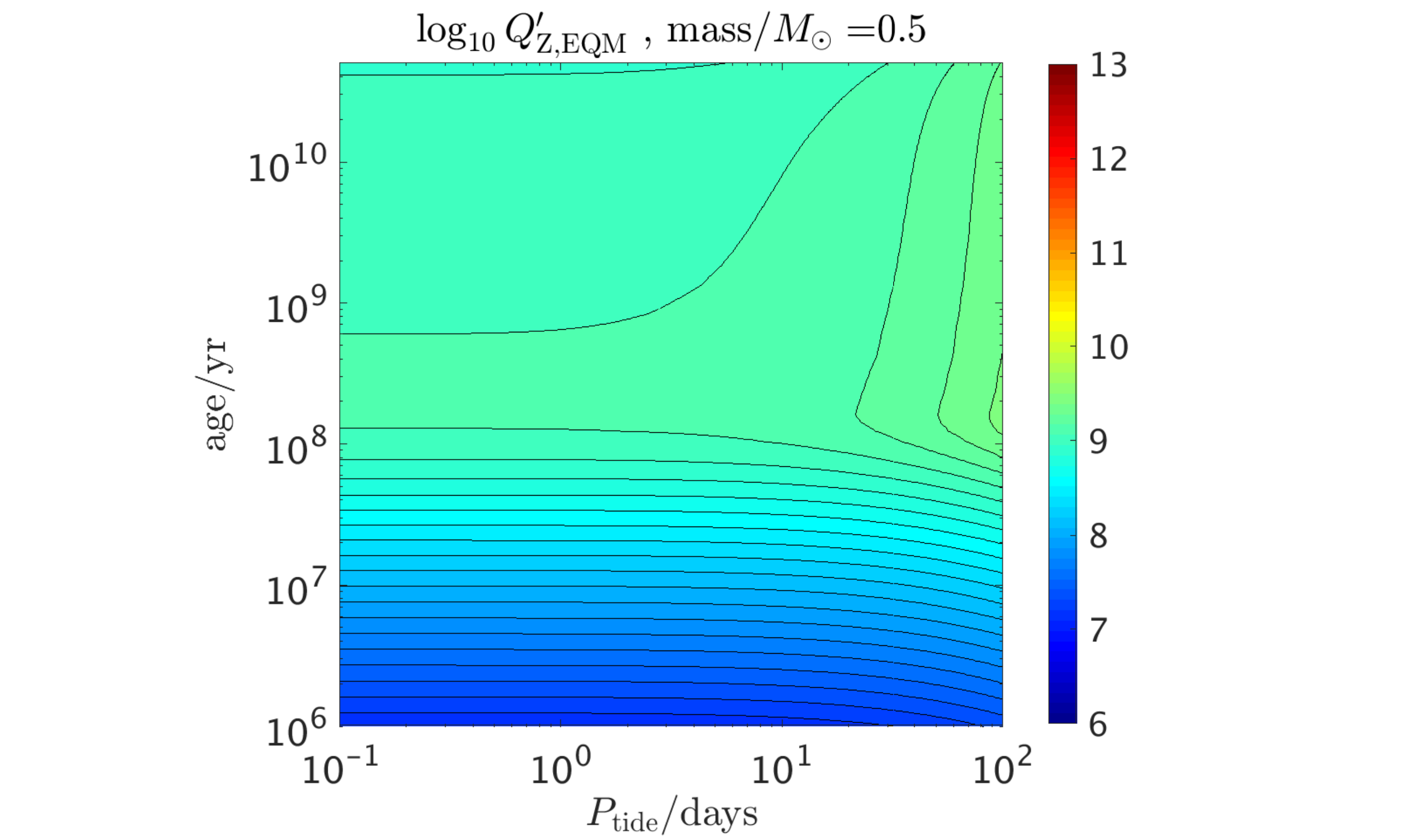}}
    \subfigure{\includegraphics[trim=3.5cm 0cm 5.5cm 0cm,clip=true,width=0.27\textwidth]{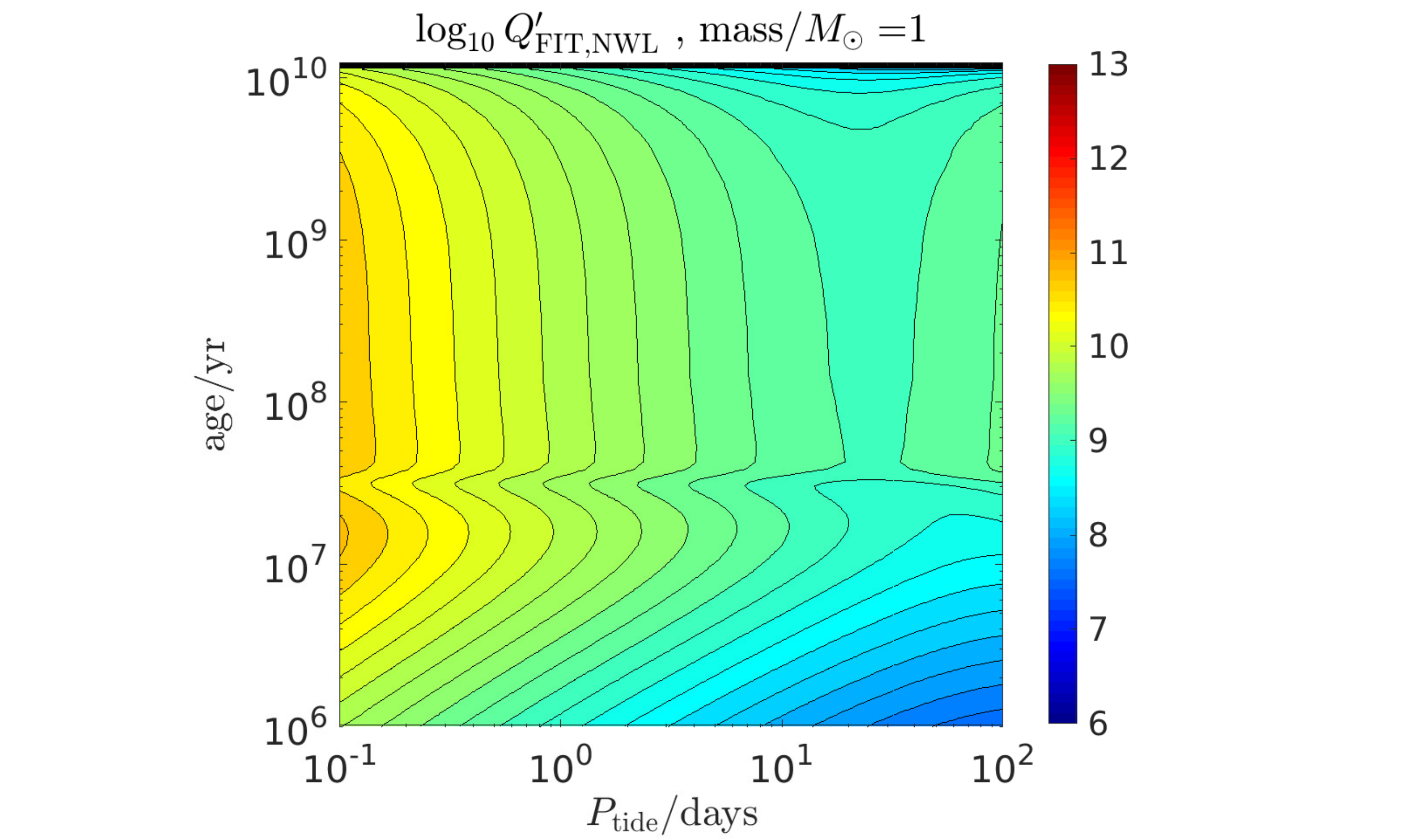}}
    \subfigure{\includegraphics[trim=3.5cm 0cm 5.5cm 0cm,clip=true,width=0.27\textwidth]{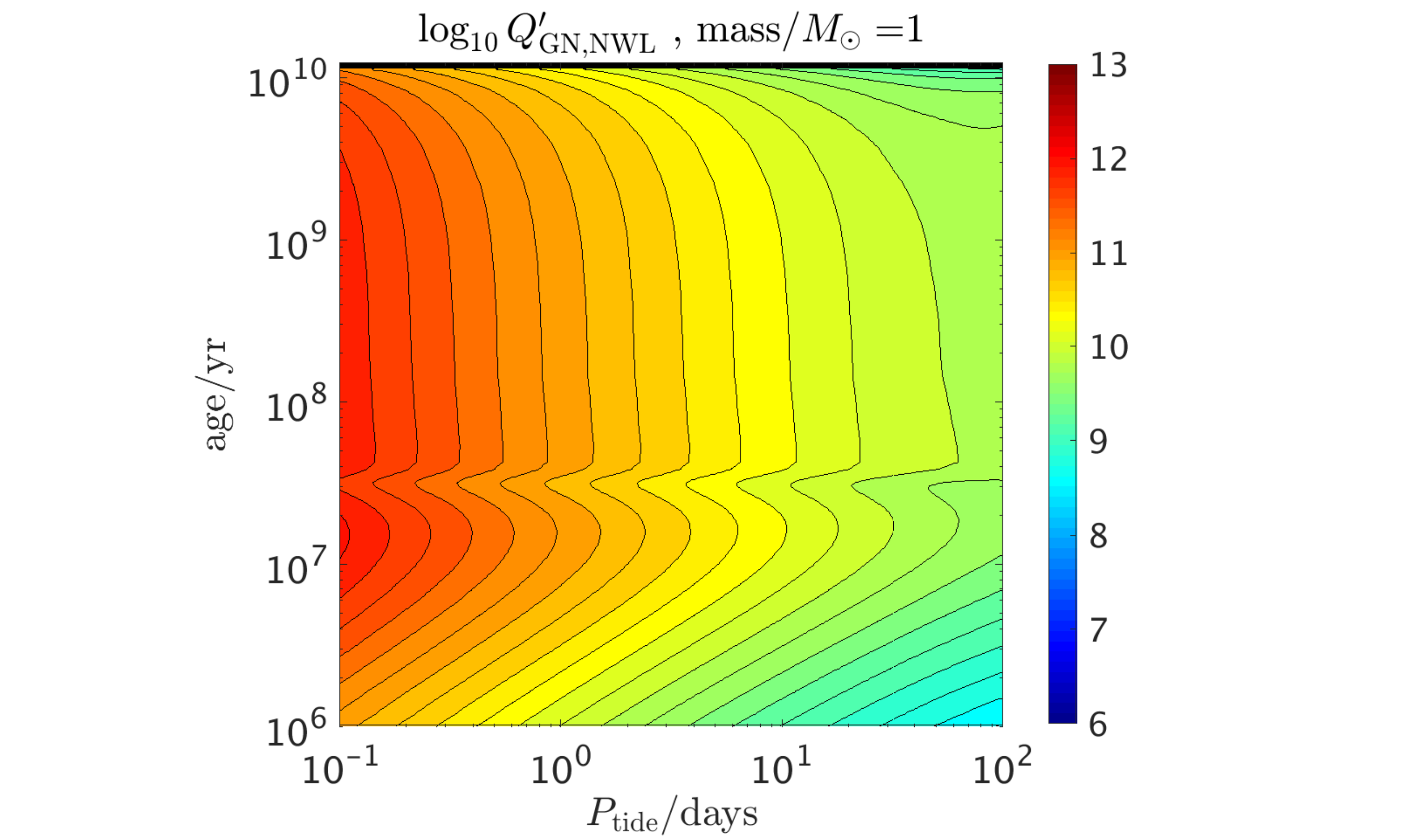}}
     \subfigure{\includegraphics[trim=3.5cm 0cm 5.5cm 0cm,clip=true,width=0.27\textwidth]{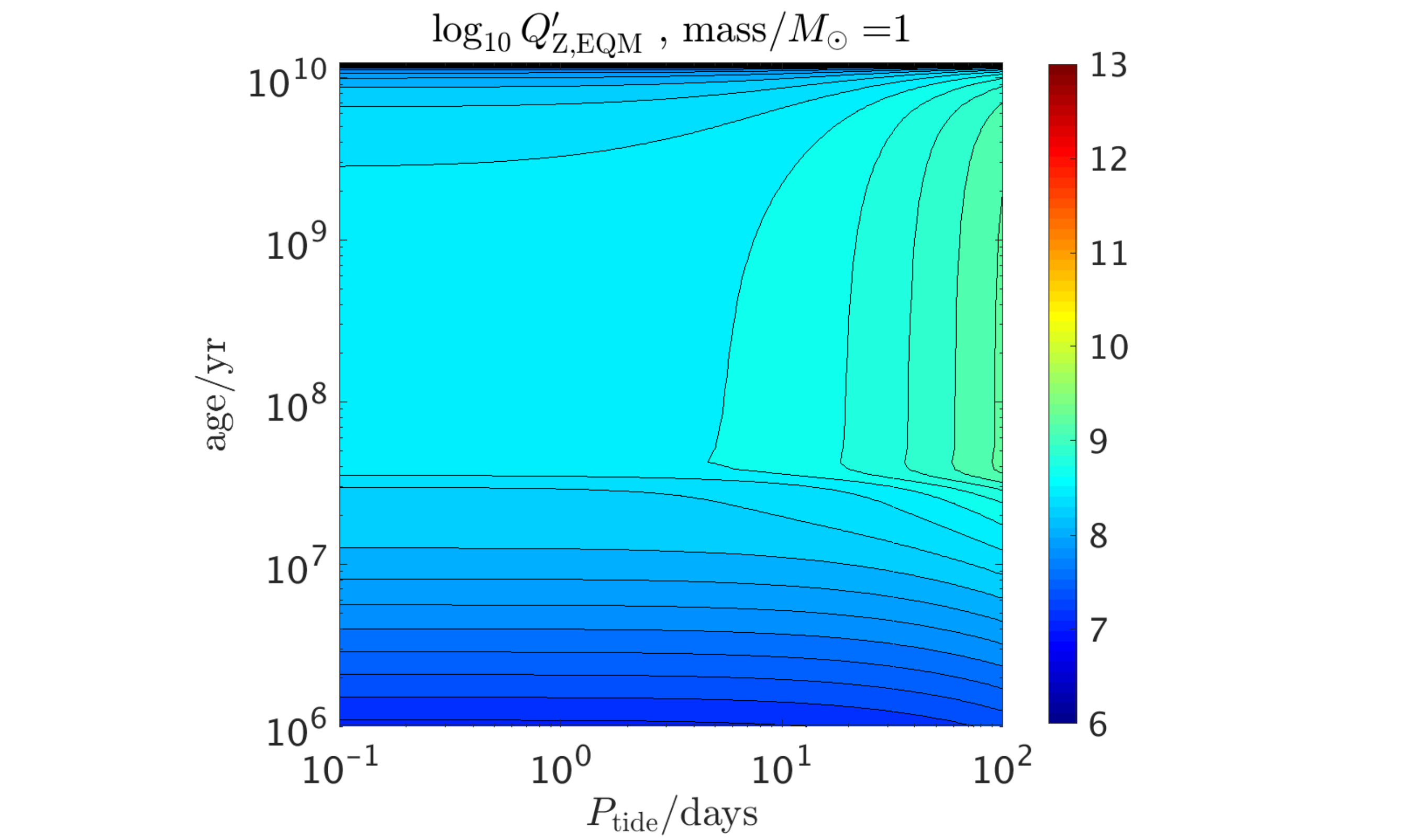}}
     \subfigure{\includegraphics[trim=3.5cm 0cm 5.5cm 0cm,clip=true,width=0.27\textwidth]{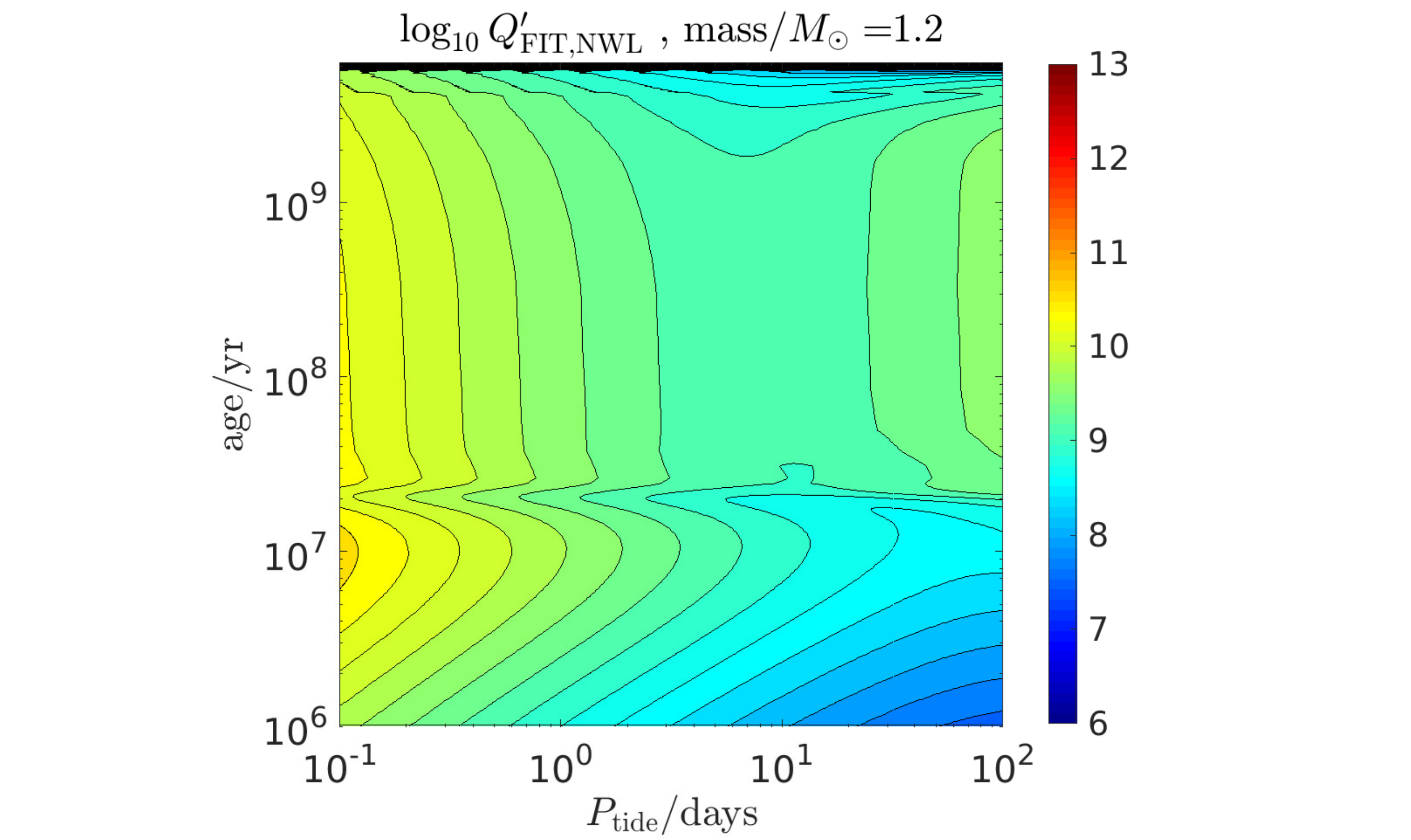}}
    \subfigure{\includegraphics[trim=3.5cm 0cm 5.5cm 0cm,clip=true,width=0.27\textwidth]{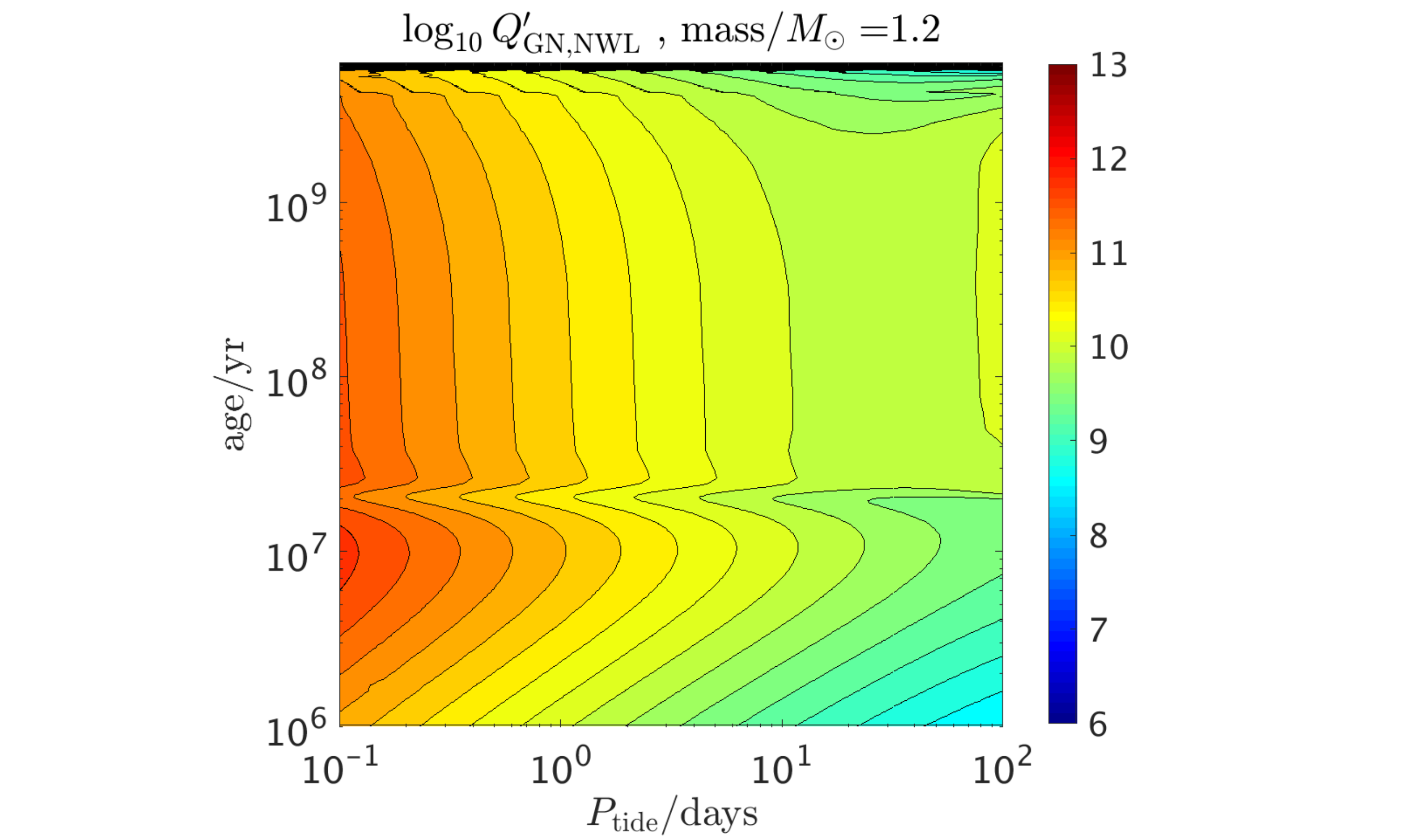}}
     \subfigure{\includegraphics[trim=3.5cm 0cm 5.5cm 0cm,clip=true,width=0.27\textwidth]{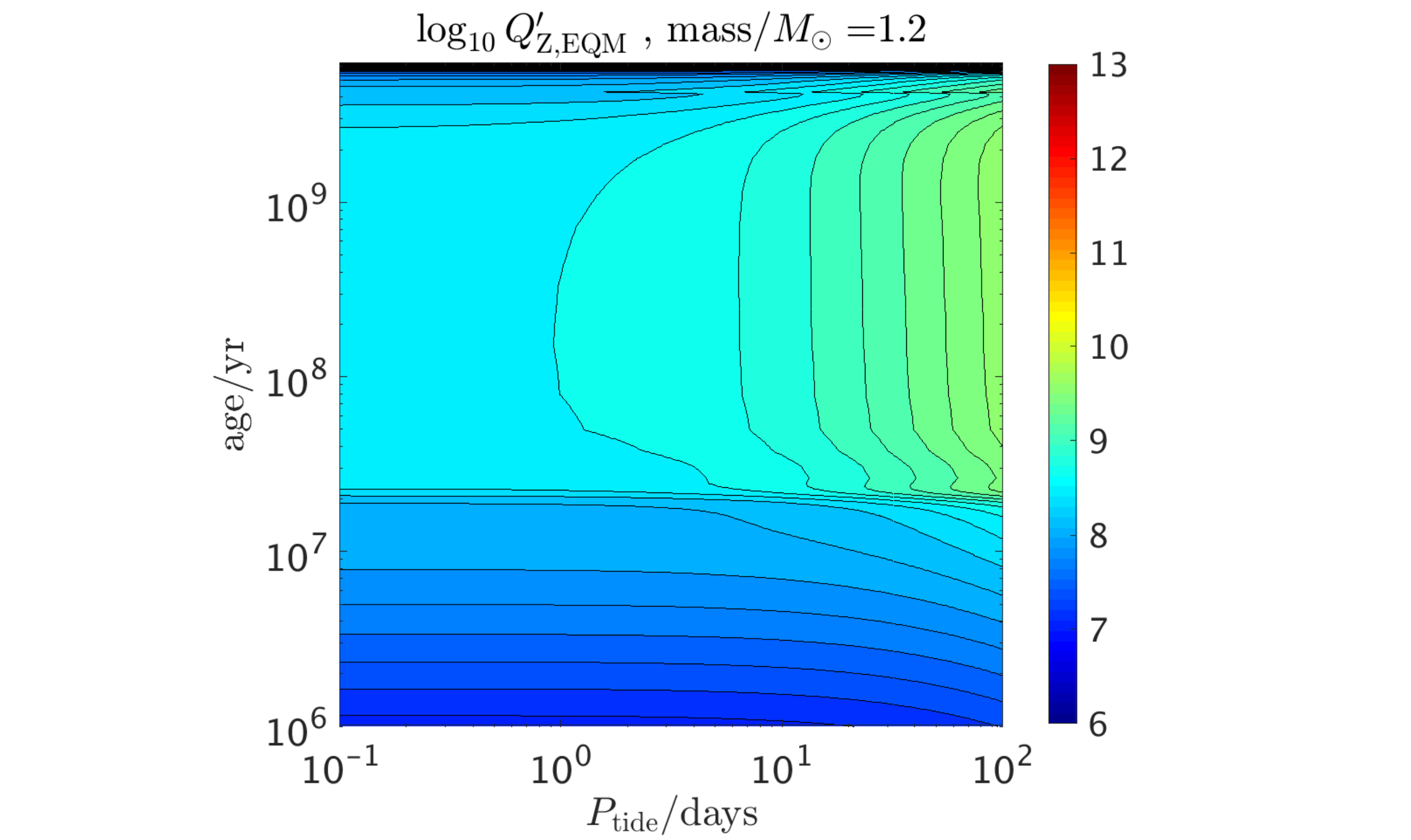}}
     \subfigure{\includegraphics[trim=3.5cm 0cm 5.5cm 0cm,clip=true,width=0.27\textwidth]{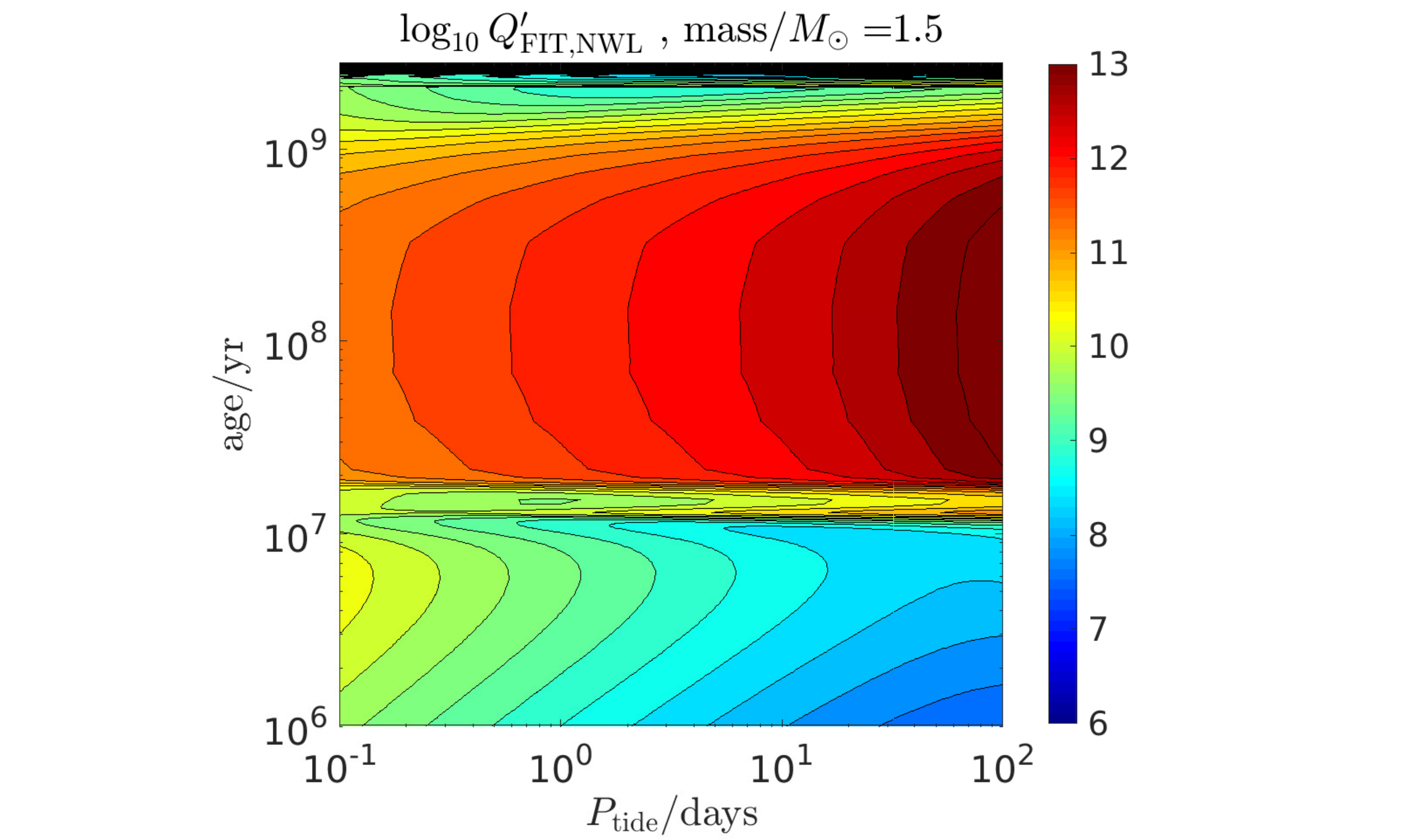}}
    \subfigure{\includegraphics[trim=3.5cm 0cm 5.5cm 0cm,clip=true,width=0.27\textwidth]{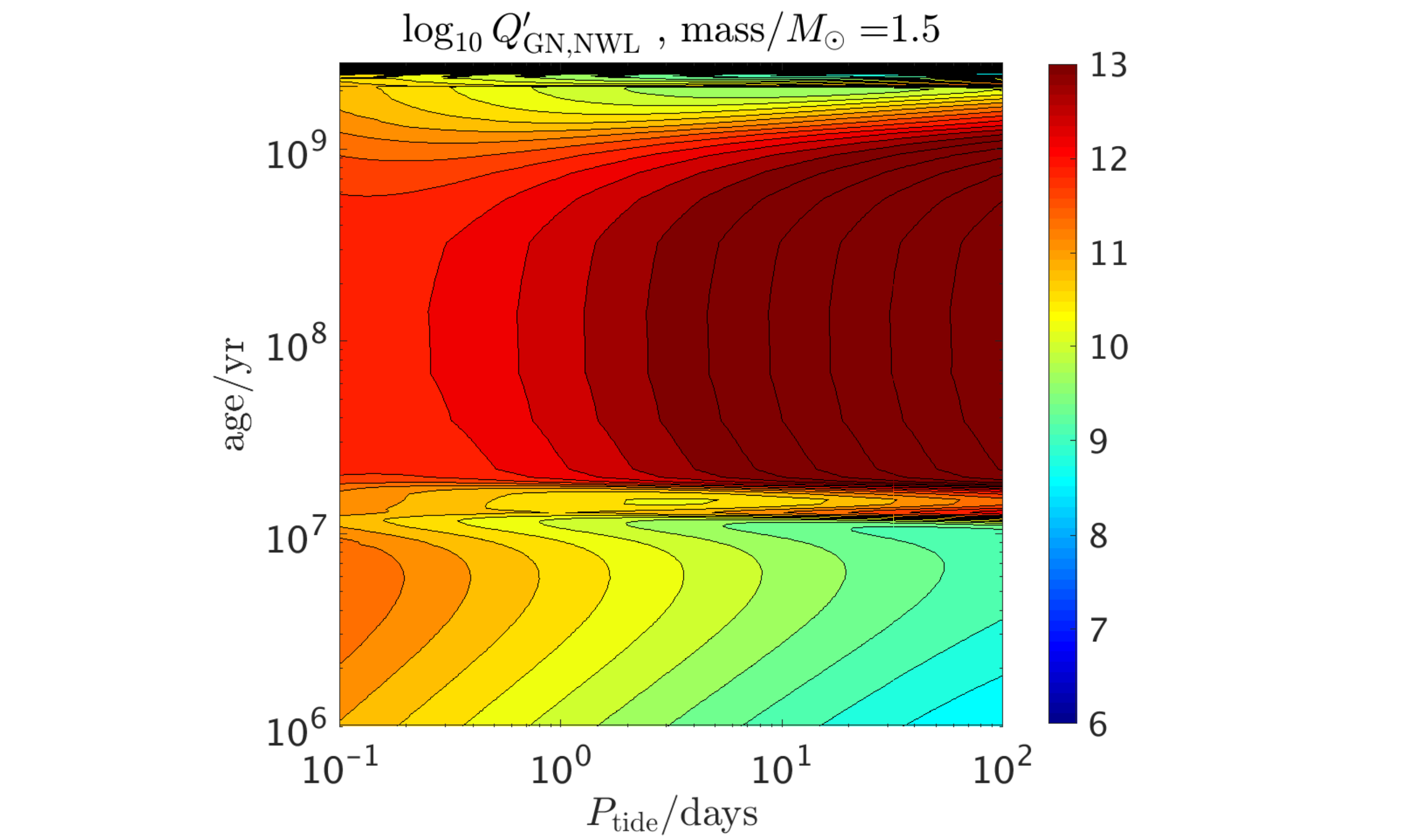}}
     \subfigure{\includegraphics[trim=3.5cm 0cm 5.5cm 0cm,clip=true,width=0.27\textwidth]{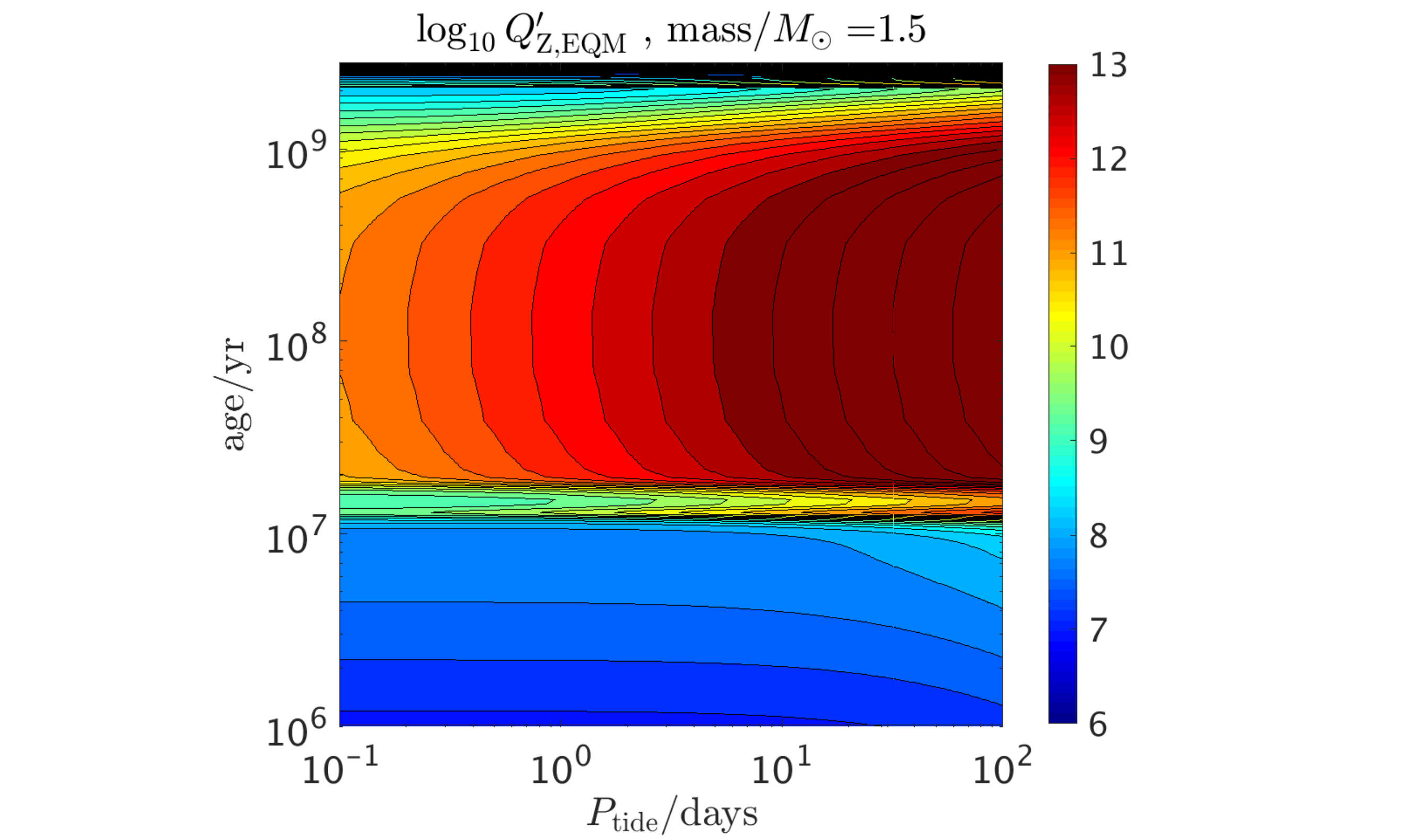}}
    \end{center}
  \caption{$Q'_\mathrm{eq}$ resulting from dissipation of the correct equilibrium tide (NWL) in convective envelopes as a function of age and tidal period. First column: $\nu_\mathrm{FIT}$ acting on NWL \citep{DBJ2020a}. Second: same for $\nu_\mathrm{GN}$. Third: EQM tide with $\nu_\mathrm{Z}$ for reference. Note that the latter can over-predict the dissipation by 1-3 orders of magnitude for fast tides and we would not advocate its use in inferring the rates of tidal evolution. Only the first 2 columns show results from the correct equilibrium tide in convection zones for two different viscosity prescriptions that are approximately consistent with the range of results from hydrodynamical simulations in the short tidal period regime.}
  \label{QpNWL}
\end{figure*}

We now turn to explore further the variation in $Q^\prime_\mathrm{eq}$ as a function of both stellar mass and age in Fig.~\ref{QpNWL}. In this figure, we plot contours of $\log_{10}Q^\prime_\mathrm{eq}$ as a function of tidal period (in days) and stellar age (in yrs) for various stellar models with masses $M/M_\odot=0.2, 0.5, 1, 1.2$ and 1.5 following their evolution, using the various viscosity prescriptions indicated. These models span a wide range from fully-convective to solar-type to F-type stellar models. For each stellar mass, this figure interpolates data from up to 400 snapshots at various times in the evolution of each star. The first column indicates results from $\nu_\mathrm{FIT}$ and the second instead uses $\nu_\mathrm{GN}$.

Fig.~\ref{QpNWL} shows the same major trends already shown in Fig.~\ref{EQMNWLdisscomp} for main-sequence ages (of order Gyrs). We note here though that the model with $1.2M_\odot$ (not previously shown in Fig.~\ref{EQMNWLdisscomp}) is somewhat more dissipative than the $1M_\odot$ model for this mechanism. The third column indicates results using the standard equilibrium tide along with $\nu_Z$. This over-predicts the dissipation (i.e.~leads to smaller $Q^\prime_\mathrm{eq}$) over NWL and other viscosity prescriptions by 1-3 orders of magnitude for short tidal periods, depending on stellar mass and age. As a result, we advocate against using this model for modelling tidal evolution, based on our current understanding.

This figure shows that the most efficient equilibrium tide dissipation occurs during the pre-main sequence phase when the star is mostly or fully convective. There is also a blip in the contours around $10^7-10^8$ yrs corresponding with zero age main sequence.
The subsequent $Q'_\mathrm{eq}$ does not show significant evolution on the main sequence, but can vary significantly when the star evolves off the main sequence. Efficient dissipation is also observed in this figure for the latest ages in the models with 1.2 and 1.5$M_\odot$. The F-type stellar model with $1.5M_\odot$ has very weak dissipation on the main sequence, and the values obtained for $Q'_\mathrm{eq}$ are quite similar for each viscosity prescription. This is a combined result of larger convective velocities (meaning weak differences between the different viscosity prescriptions, and only a weak frequency-reduction), and the very low densities of the thin envelopes of these stars.

\section{Dissipation due to inertial waves in convection zones}
\label{IWresults}

\begin{figure}
  \begin{center}
    \subfigure{\includegraphics[trim=3cm 0cm 4.5cm 1cm,clip=true,width=0.45\textwidth]{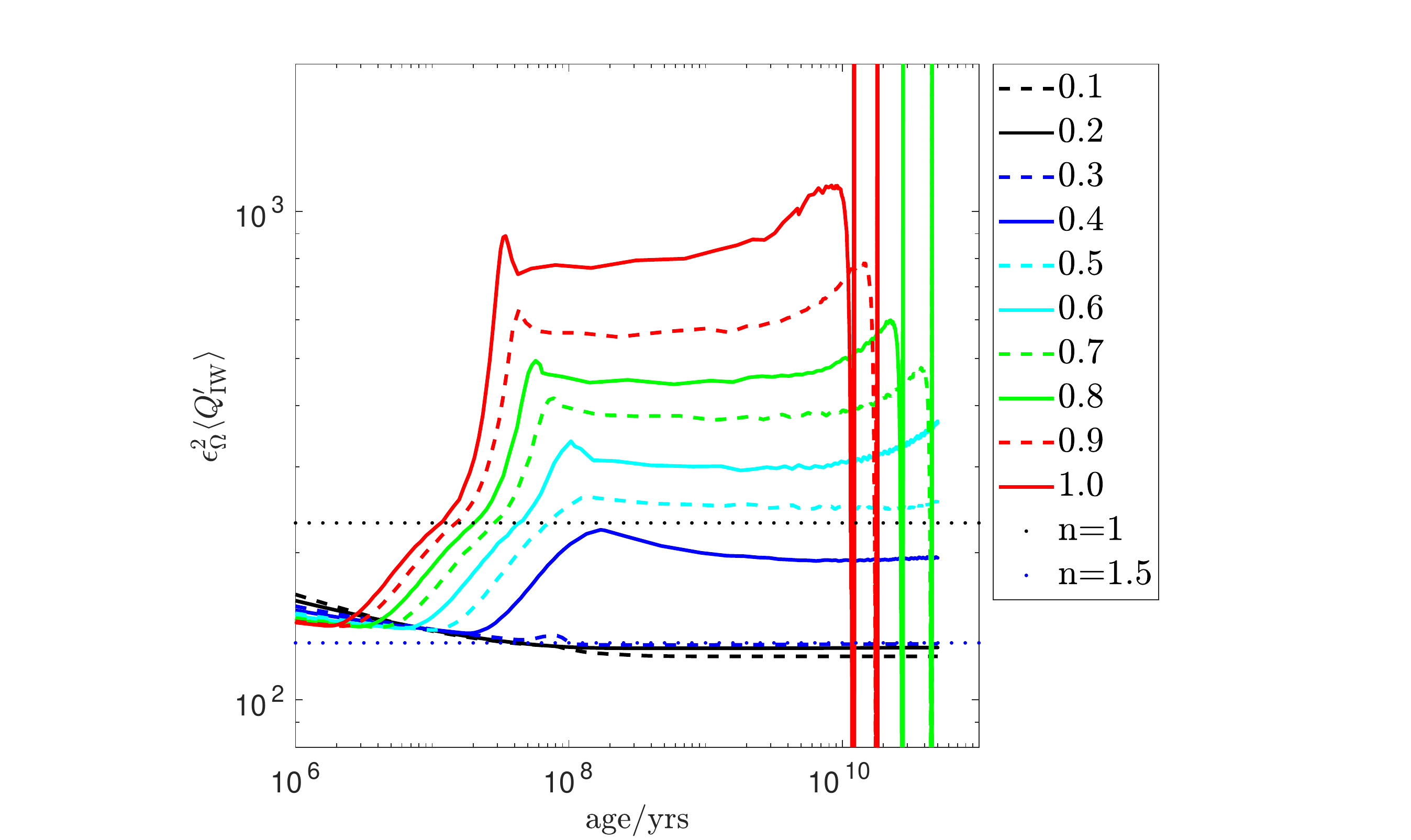}}
    \subfigure{\includegraphics[trim=3cm 0cm 5cm 1cm,clip=true,width=0.45\textwidth]{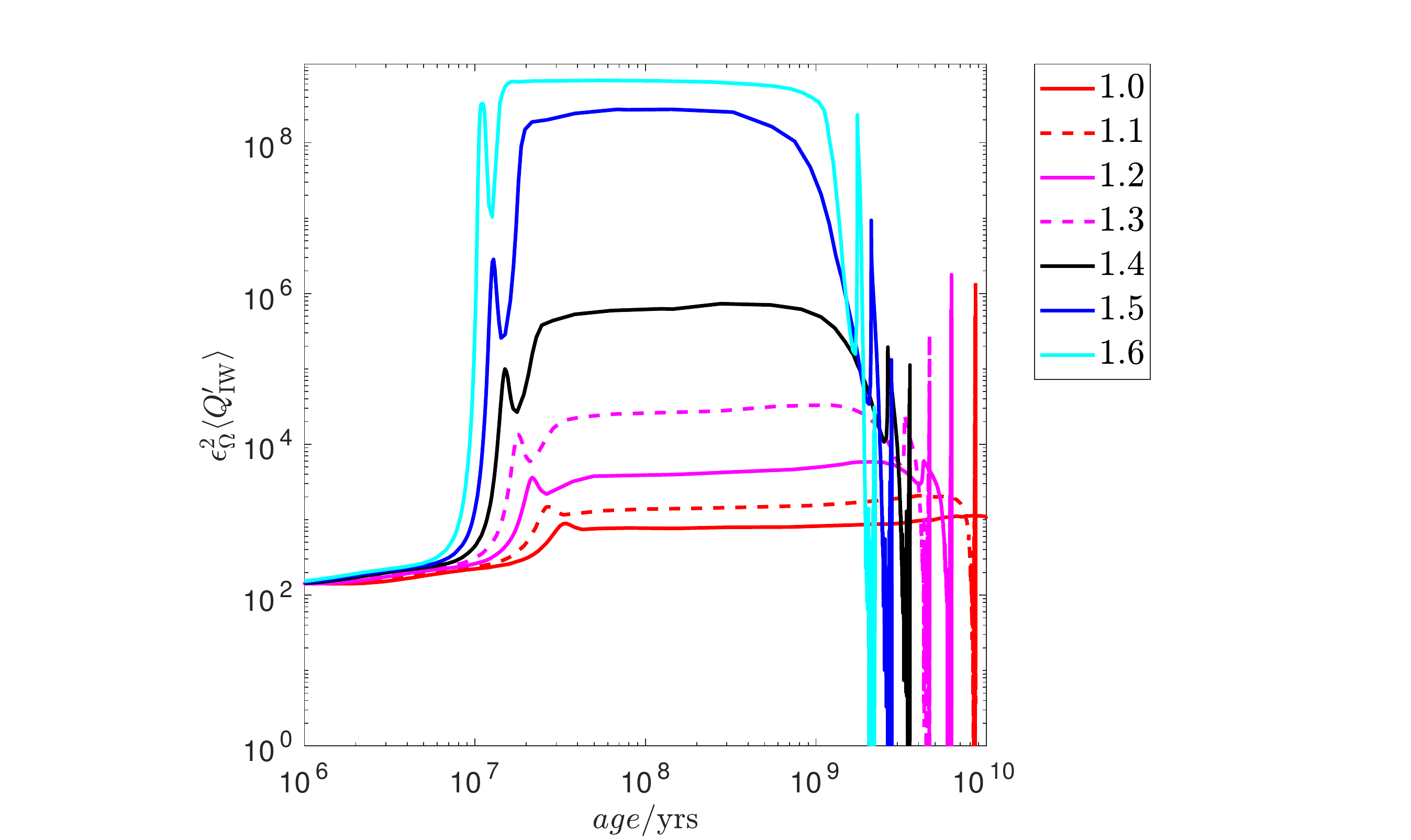}}
    \end{center}
  \caption{$\langle Q'_{\mathrm{IW}}\rangle$ multiplied by $\epsilon_\Omega^2$ due to (frequency-averaged) dissipation of inertial waves in convective envelopes as a function of their age. Top: masses below and including $1M_\odot$. Bottom: masses above and including $1M_\odot$. Results for an $n=1$ and $n=1.5$ polytrope are plotted in the top panel for reference. Tidal dissipation of inertial waves is typically most efficient during the brief phases as these stars evolve off the main-sequence, and also in the pre-main sequence phases.}
  \label{QpIWcomp}
\end{figure}

\begin{figure}
  \begin{center}
    \subfigure{\includegraphics[trim=4cm 0cm 5cm 1cm,clip=true,width=0.45\textwidth]{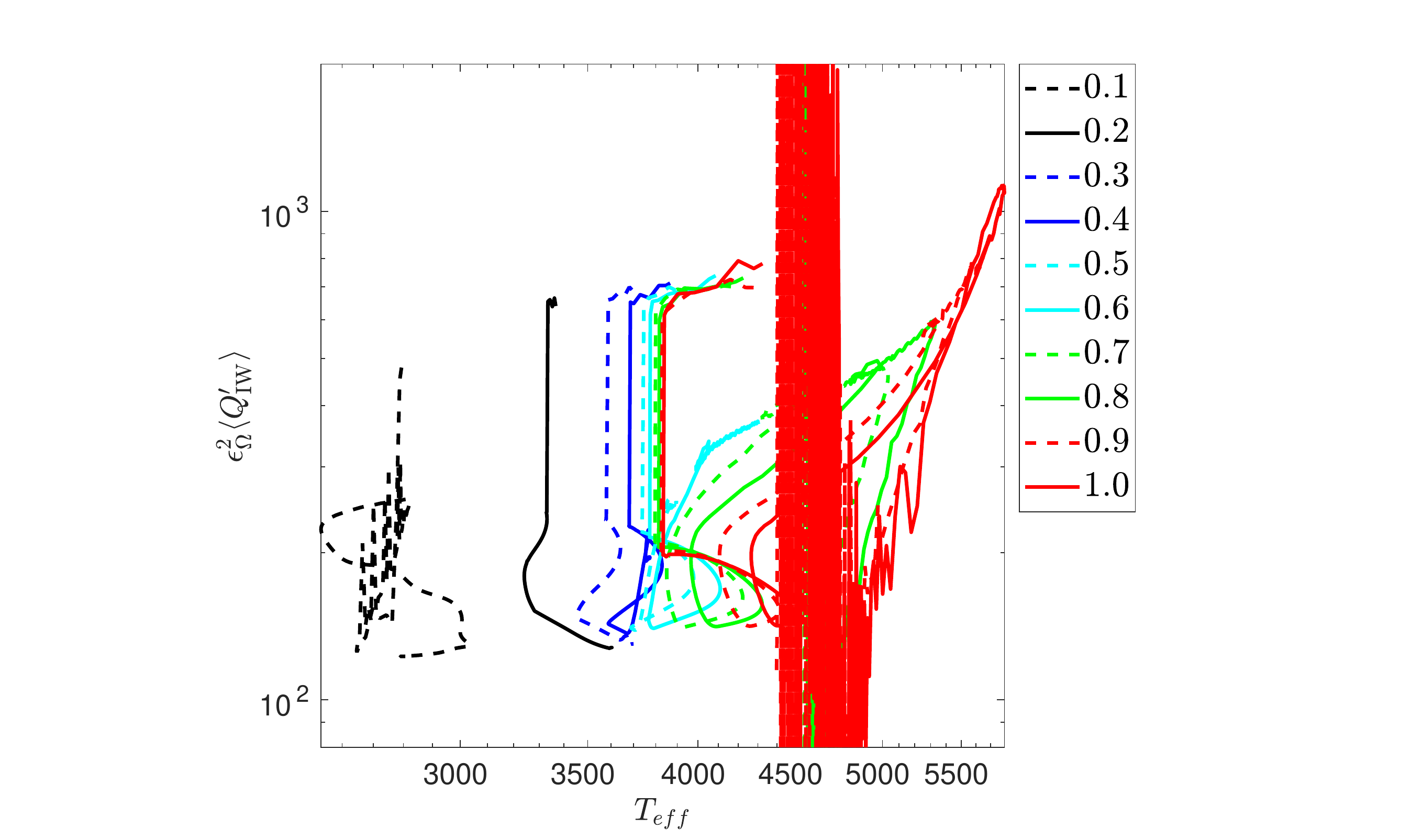}}
   \subfigure{\includegraphics[trim=4cm 0cm 5cm 1cm,clip=true,width=0.45\textwidth]{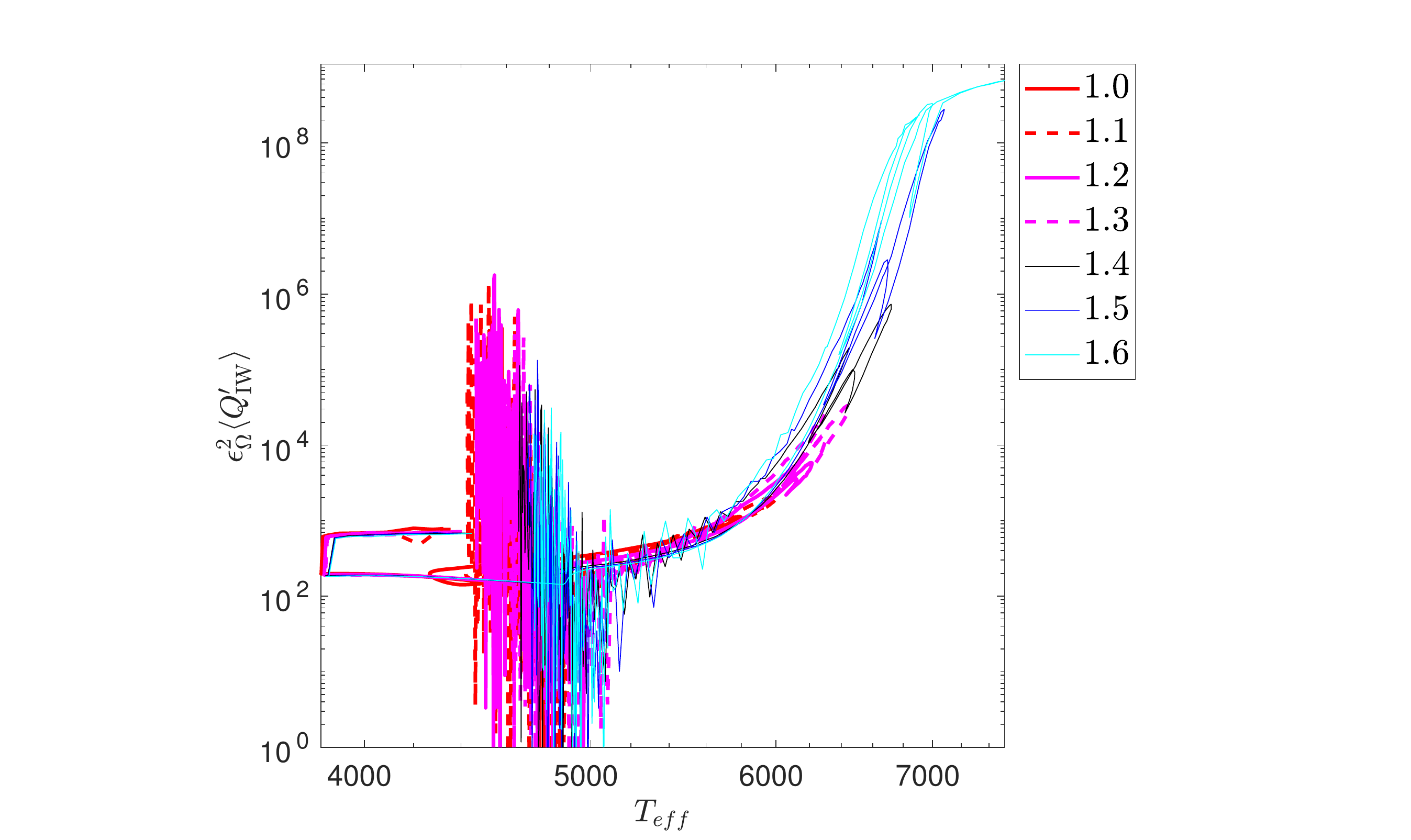}}
    \end{center}
  \caption{Same as Fig.~\ref{QpIWcomp} except results are shown as a function of stellar effective temperature $T_{eff}$, which is more directly observable.}
  \label{QpIWcompTeff}
\end{figure}

\begin{figure}
  \begin{center}
    \subfigure{\includegraphics[trim=3cm 0cm 5cm 1cm,clip=true,width=0.45\textwidth]{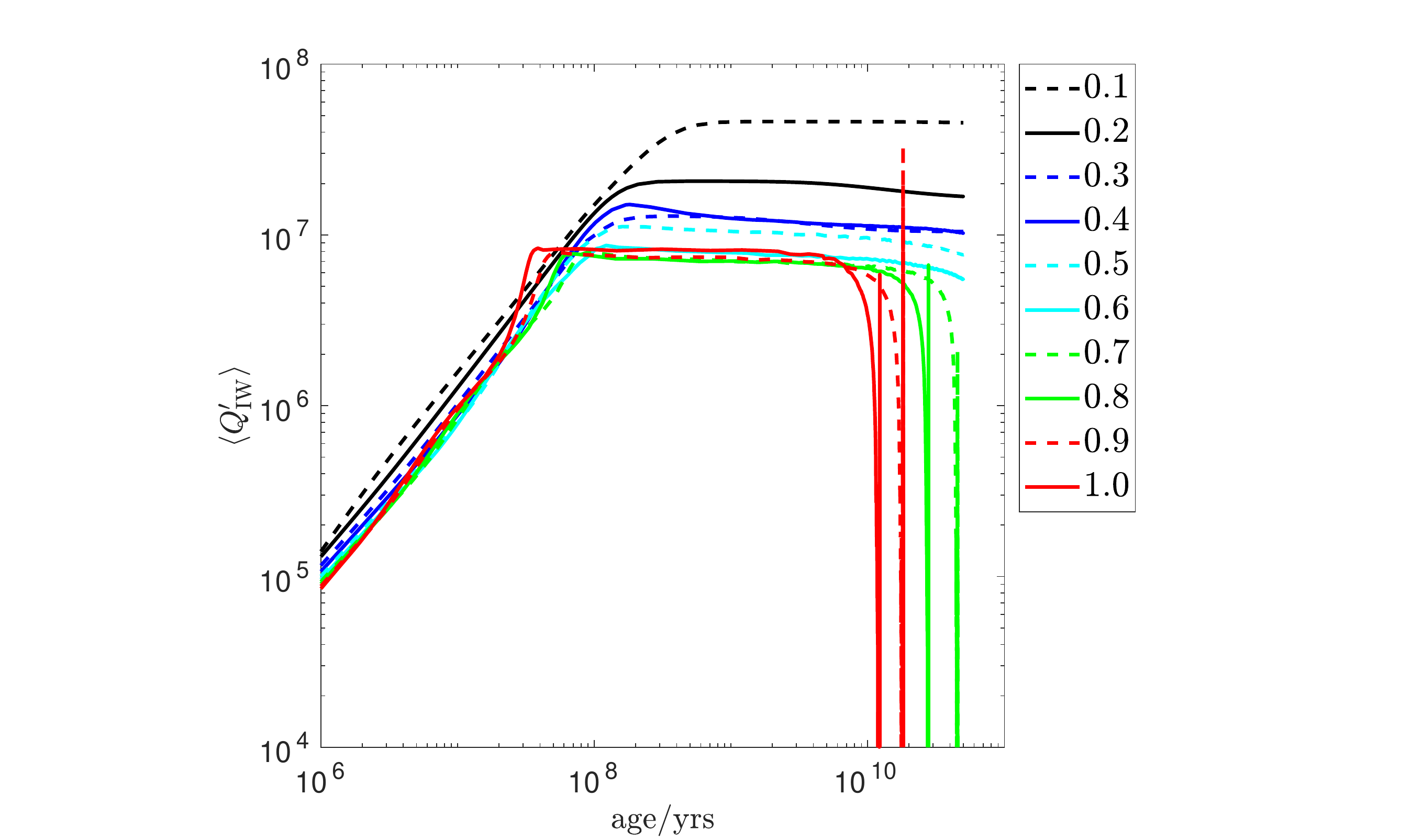}}
    \subfigure{\includegraphics[trim=3cm 0cm 5cm 1cm,clip=true,width=0.45\textwidth]{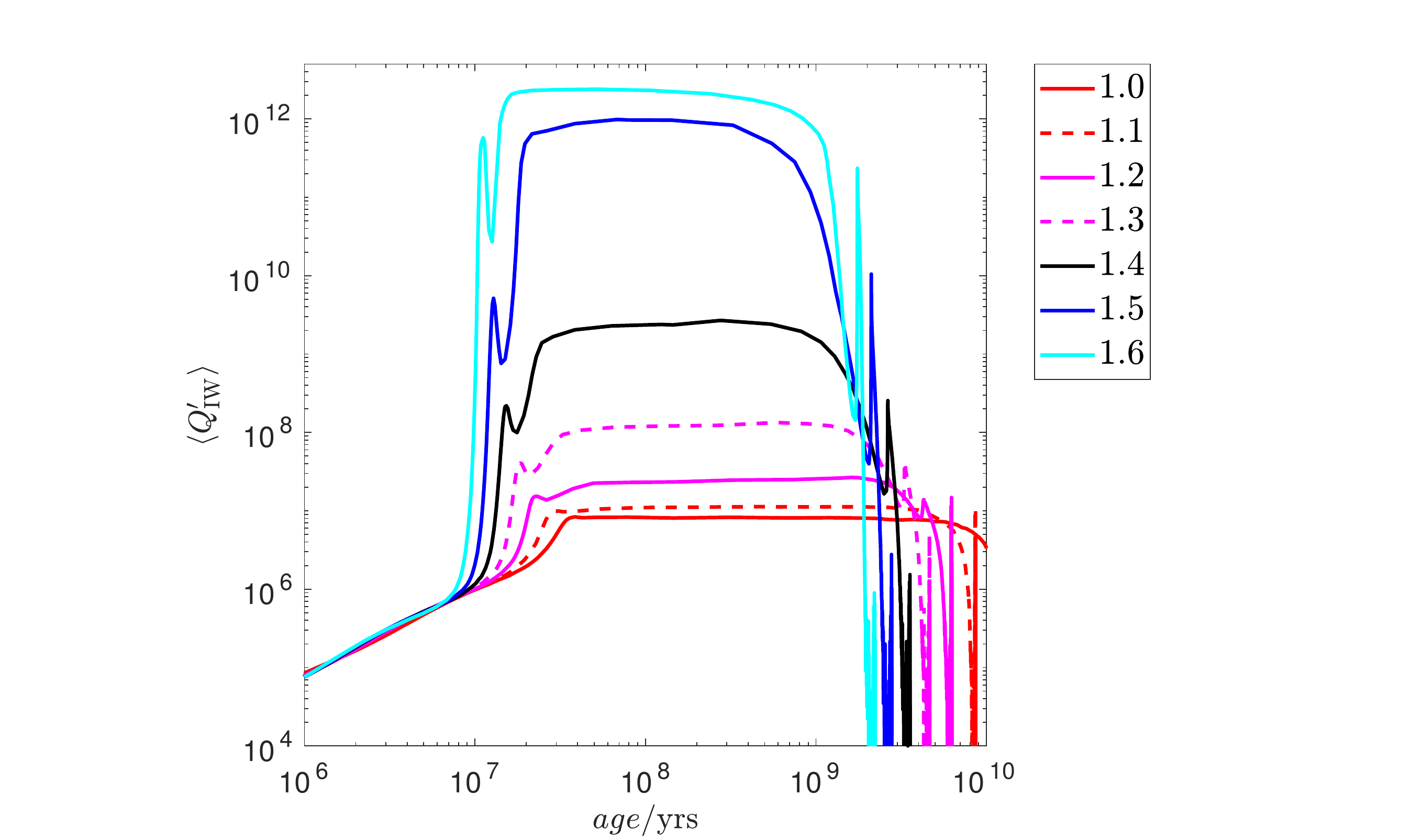}}
    \end{center}
  \caption{Same as Fig.~\ref{QpIWcomp} except that we have assumed a rotation period of 10 days in order to calculate $\epsilon_\Omega$ following the evolution of each star. We find $\langle Q_\mathrm{IW}'\rangle \approx 10^7 (P_\mathrm{rot}/10\mathrm{d})^{2}$ on the main sequence for most stars with $M<1.1 M_\odot$, with larger values in F-stars with thin convective envelopes.}
  \label{QpIWcomp2}
\end{figure}

Here we present results following the method outlined in \S~\ref{IW} for the frequency-averaged dissipation of inertial waves in convection zones. This represents a crude but useful measure of the tidal dissipation due to these waves. This mechanism only operates if the tidal period is longer than half the rotation period of the star. If this condition is not satisfied, then inertial waves cannot be excited in convection zones, and this mechanism will not apply.

In Fig.~\ref{QpIWcomp}, we show the evolution of the tidal quality factor $\langle Q'_{\mathrm{IW}}\rangle$ representing the frequency-averaged dissipation multiplied by 
$\epsilon_\Omega^2$ as a function of age. This indicates that $\langle Q'_{\mathrm{IW}}\rangle \propto \Omega^{-2}$, so that rapid rotators generally experience more efficient dissipation due to inertial waves when these waves are excited. In the top panel, we show this quantity computed following the evolution of various stars with masses smaller than, and including, 1 $M_\odot$. This includes fully convective stars and those with radiative cores and convective envelopes. Note that in fully convective stars regular inertial modes exist \citep{LF1999,Wu2005}, which can provide important contributions to tidal dissipation even in the absence of an inner radiation zone but only for a discrete set of frequencies \citep{Ogilvie2013}. Nevertheless, the frequency-averaged measure probably still quantifies the typical level of dissipation in this case. In the bottom panel, we show masses above, and including, 1 $M_\odot$, showing results for solar-type and F-type stars. When evaluated at the appropriate spin periods, the results of Fig.~\ref{QpIWcomp} compare quite well, though the numbers are slightly larger than, the typical $Q'$ obtained in the direct (i.e.~not frequency-averaged) linear calculations of \cite{OL2007} for a 1 $M_\odot$ and $0.5 M_\odot$ star, and those of \cite{BO2009} for F-stars with $M=1.2-1.5 M_\odot$. This supports the validity of the frequency-averaged formalism.

We observe that the most efficient inertial wave dissipation before 1 Gyr occurs during pre-main sequence phases for ages less than $10^7$ yrs for most stars, when $\epsilon_\Omega^2\langle Q'_{\mathrm{IW}}\rangle\approx 2\times 10^2$. This is found in both top and bottom panels. The fully convective stars with $M<0.4M_\odot$ possess similar values throughout their subsequent evolution, but other stars evolve to become less dissipative as they transition onto the main-sequence. Note that $\epsilon_\Omega^2\langle Q'_{\mathrm{IW}}\rangle$ generally increases with stellar mass on the main-sequence. We have also over-plotted results for an $n=1$ and $n=1.5$ polytrope on Fig.~\ref{QpIWcomp} for reference. The latter is a good model of the stars we have studied with $M=0.1-0.3M_\odot$, and its prediction for $\epsilon_\Omega^2\langle Q'_{\mathrm{IW}}\rangle$ closely agrees with those models. (The former may be a better model of even lower mass objects.) We have shown the same results as a function of stellar effective temperature $T_{eff}$ in Fig.~\ref{QpIWcompTeff}.

If the rotation period, mass and stellar radius of a star are known, Fig.~\ref{QpIWcomp} (or Fig.~\ref{QpIWcompTeff}) can be used to give a crude prediction for the resulting tidal quality factor due to inertial wave dissipation in convection zones. To complement Fig.~\ref{QpIWcomp}, we also plot $\langle Q'_{\mathrm{IW}}\rangle$ evaluated by assuming a rotation period of 10 days for each star in Fig.~\ref{QpIWcomp2} (i.e. we evaluate $\epsilon_\Omega$ for each model). These predictions can be straightforwardly scaled to another rotation period $P_\mathrm{rot}$ by multiplying $\langle Q'_{\mathrm{IW}}\rangle$ by $(P_\mathrm{rot}/10 \mathrm{d})^2$.

We wish to briefly return to Fig.~\ref{EQMNWLdisscomp} before moving on. This indicates that inertial wave dissipation, assuming a spin period of 10 d, is the most efficient tidal mechanism for tidal periods longer than about 5 d (depending on stellar mass) for which these waves can be excited. We believe this mechanism to be primarily responsible for binary circularization and synchronization, which we will explain further in \S~\ref{binaryimplications}. However, internal gravity wave dissipation is more efficient for short tidal periods when inertial waves are not excited, and is likely to be the dominant mechanism for planetary orbital decay (which we will explain further in \S~\ref{HJimplications}). We will discuss this mechanism in more detail in the next section.

We observe that $\langle Q'_{\mathrm{IW}} \rangle$ evolves along a common track for all stellar models during the pre-main sequence phases, and all models below $1.1 M_\odot$ have $\langle Q'_{\mathrm{IW}} \rangle\approx 10^7 (P_\mathrm{rot}/10 \mathrm{d})^2$ on the main-sequence. 
As these stars evolve off the main sequence, they become very dissipative according to this measure, with $\langle Q'_{\mathrm{IW}}\rangle$ even passing below 1 in some models. This is because both the depth of the convective envelope and the stellar radius increase dramatically. It indicates that tides become very efficient as the star evolves off the main sequence if inertial waves are excited. However, inertial waves may not always be excited in a given application because evolved stars typically rotate much slower than those early on the main sequence.

\begin{figure}
  \begin{center}
    \subfigure{\includegraphics[trim=2cm 0cm 3cm 1cm,clip=true,width=0.45\textwidth]{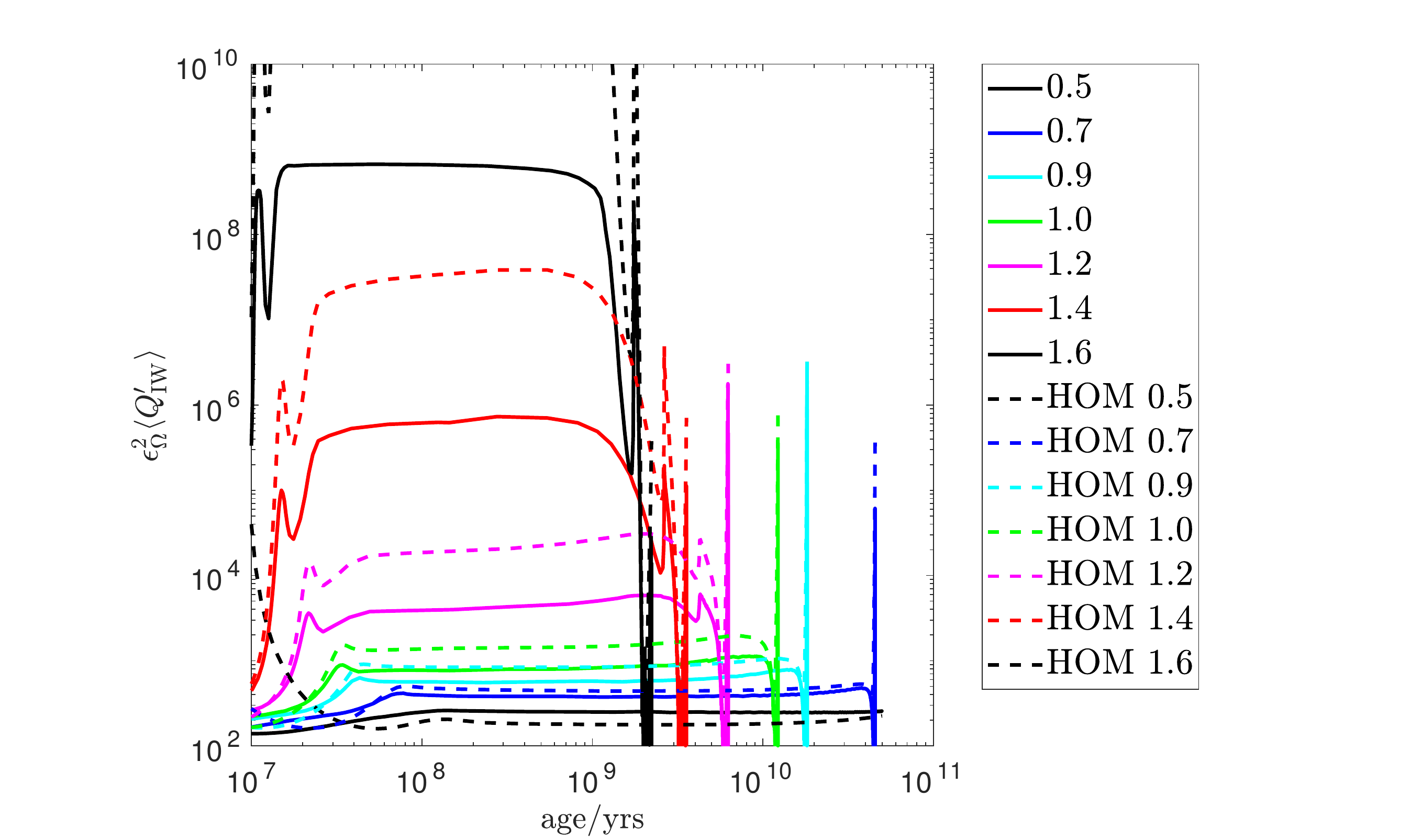}}
    \subfigure{\includegraphics[trim=3cm 0cm 3cm 1cm,clip=true,width=0.45\textwidth]{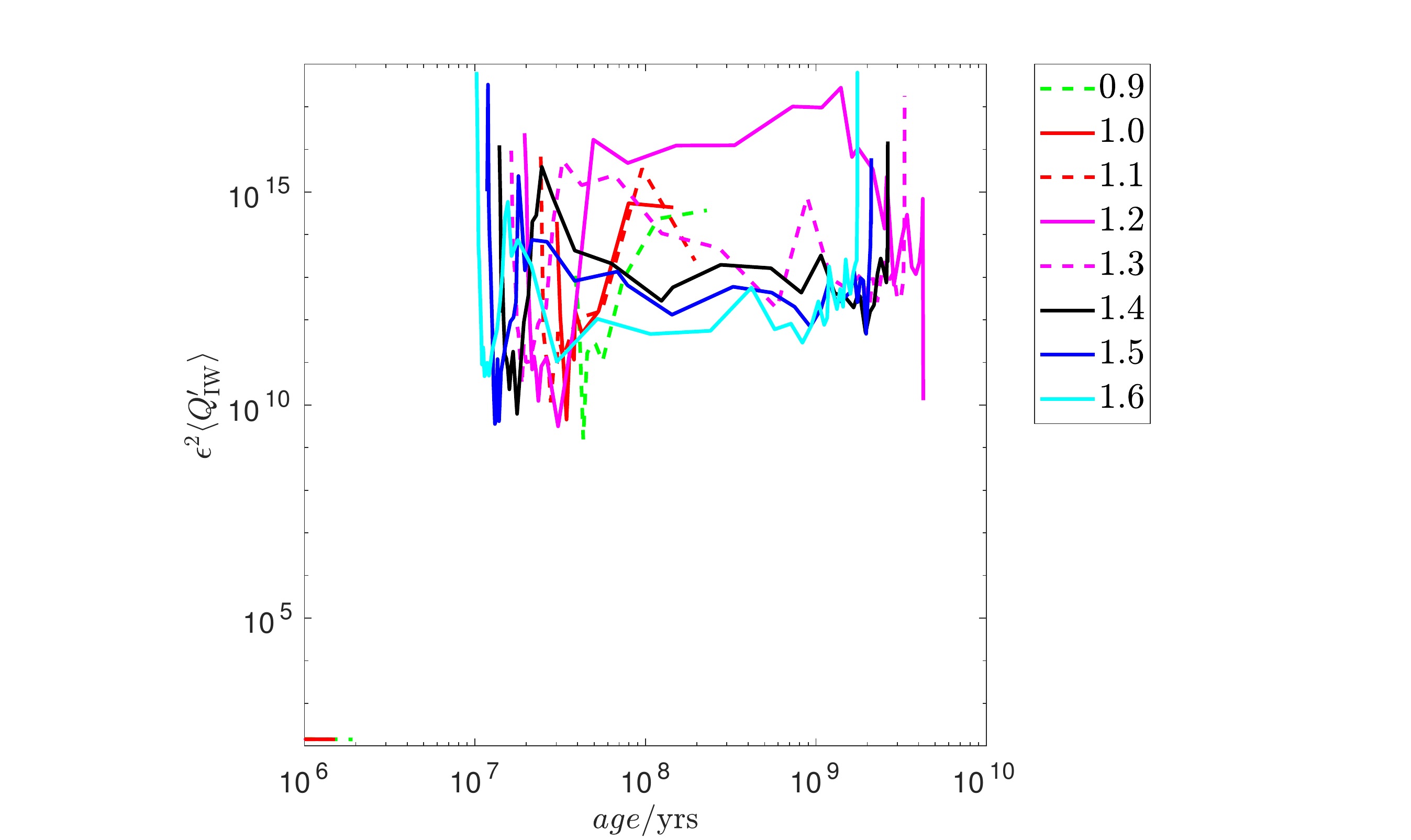}}
    \end{center}
  \caption{Top: $\epsilon_\Omega^2\langle Q'_{\mathrm{IW}} \rangle$ due to dissipation of inertial waves in convective envelopes based on the frequency-averaged formalism. Top: comparison with a simplified two-layer piece-wise homogeneous model (dashed lines) with the more realistic model considered in this work (solid lines). For most stars, the two-layer model works reasonably well on the main-sequence, generally under-predicting the resulting dissipation by a factor of 2. However, it should not be used for F-stars with masses above 1.2 $M_\odot$, where it can under-predict the dissipation by an order of magnitude or more. Bottom: Dissipation in convective cores using the full model, which (compared with Fig.~\ref{QpIWcomp}) is found to be much weaker than the dissipation in the envelope.}
  \label{QpIWcomp3}
\end{figure}

Similar calculations to those in this section have been presented by \cite{Mathis2015}, who adopted a simplified two-layer piece-wise homogeneous model of a star derived by \cite{Ogilvie2013}. The benefit of the two-layer model is that it is simple to apply and efficient to calculate in a stellar evolution code. However, its validity has not been assessed by comparing its predictions with the more realistic model considered in this work. In order to compare with \cite{Mathis2015}, we compute the dissipation using the same two-layer model as described in Eq.~\ref{IWTwoLayer}. The results are shown in the top panel of Fig.~\ref{QpIWcomp3} (using dashed lines), where they are compared with results using the realistic structure of the star (using solid lines) for a range of stellar masses from 0.5 to 1.6 solar masses. As a check on our implementation, we have checked that our two-layer predictions are consistent with Fig.~4 in \cite{Mathis2015}. Lower mass stars are not considered because they are generally fully convective, so a two-layer model does not properly represent their structure. In addition, in pre-main sequence phases when the star is fully convective, it does not make sense to apply the two-layer model.

Fig.~\ref{QpIWcomp3} shows that on the main sequence, for Gyr ages, the predictions of the two-layer model generally agree reasonably well with those of the more realistic model for stars with $0.5\leq M/M_\odot \leq 1.2$. Our results that account for the realistic structure of the star are generally more dissipative by a factor of less than approximately 2 in these models on the main sequence. However, the two-layer model performs poorly in F-stars with masses larger than $1.2M_\odot$, generally under-predicting the dissipation in these stars by an order of magnitude or more.

Since the model studied in this paper is not much more difficult to compute, even if it requires more computational work than the simplified two-layer model, we advocate its use over the two-layer model in future calculations modelling tidal interactions coupled with stellar evolution. 

Finally, we compute the corresponding inertial wave dissipation in the convective cores of stars with masses greater than $0.9M_\odot$ following their evolution in the bottom panel of Fig.~\ref{QpIWcomp3}. We observe that dissipation in convective cores is much weaker than in the envelope in these stars, with the exception of the most massive F-star with $M=1.6M_\odot$, where they are comparable. This (and the results for the equilibrium tide in the previous section) indicates that in most applications it is safe to ignore the contribution to tidal dissipation from convective cores for stars with masses in the range we have been studying.

\section{Dissipation due to internal gravity waves in radiation zones}
\label{IGWresults}

\begin{figure}
  \begin{center}
    \subfigure{\includegraphics[trim=3cm 0cm 5cm 0cm,clip=true,width=0.4\textwidth]{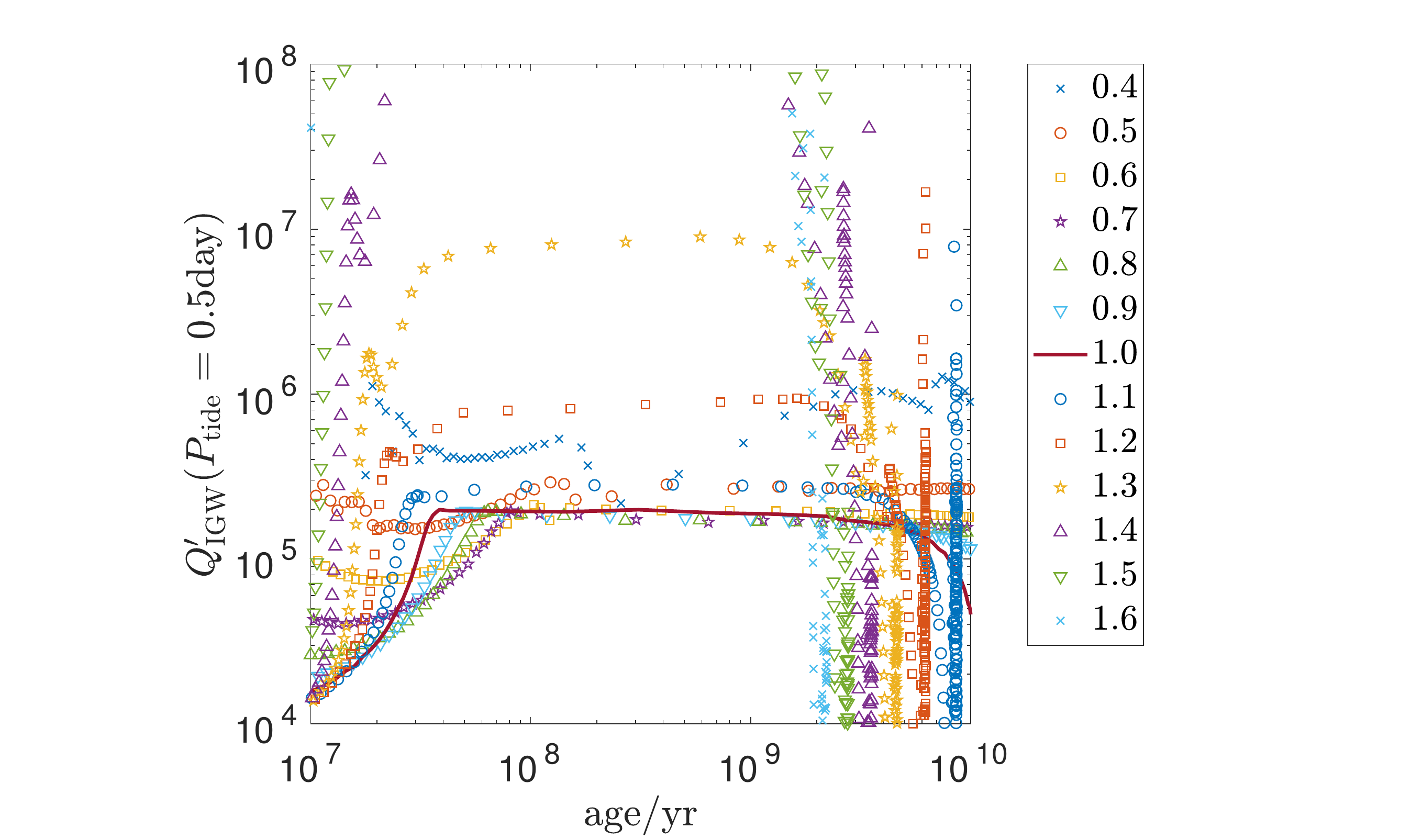}}
    \subfigure{\includegraphics[trim=3cm 0cm 5cm 0cm,clip=true,width=0.4\textwidth]{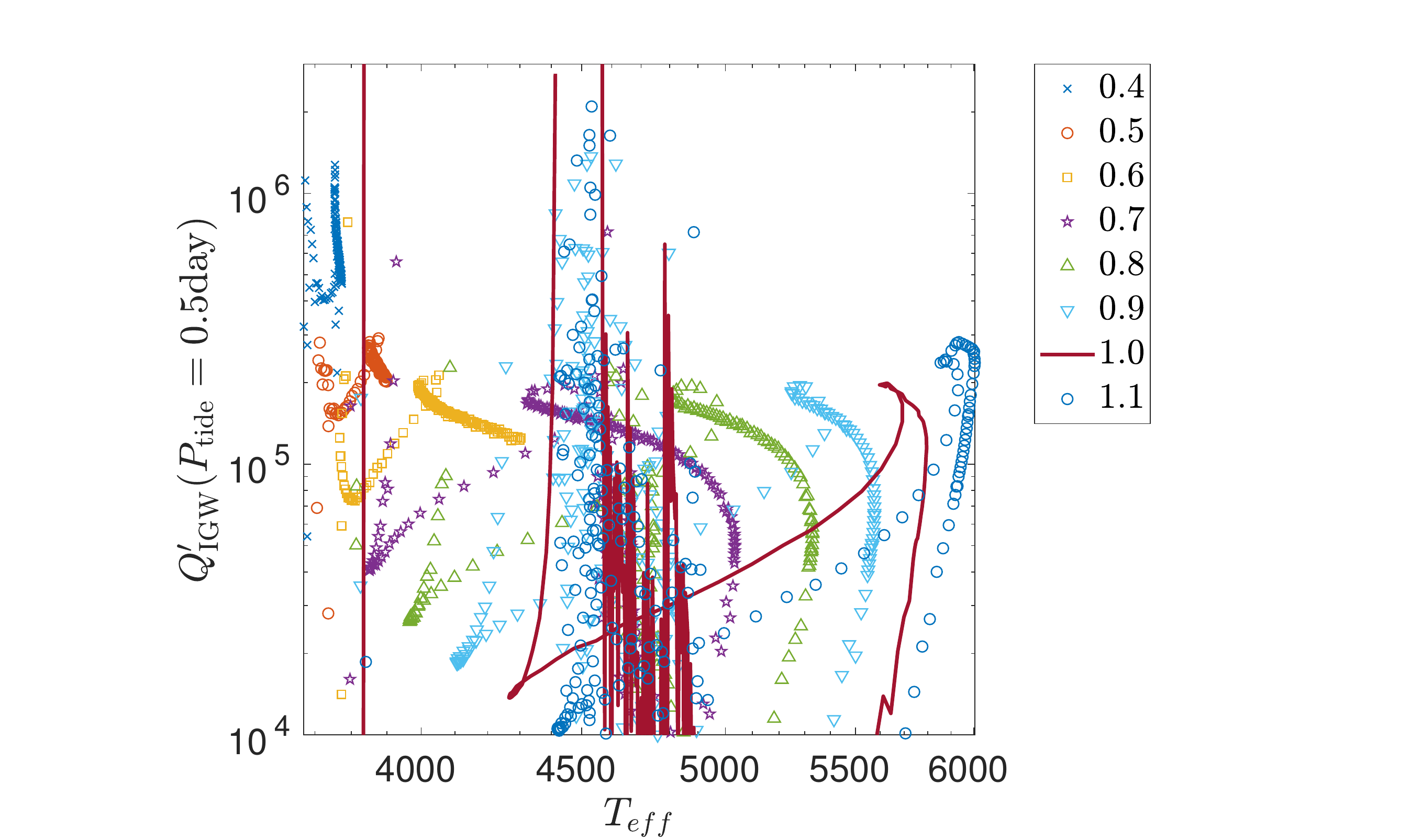}}
    \end{center}
  \caption{Top: $Q'_\mathrm{IGW}$ as a function of age (in yrs) for various stellar masses (indicated in the legend), resulting from internal gravity wave dissipation in radiation zones, under the assumption that these waves are launched from the radiative/convective interface and are subsequently fully damped. We have assumed a tidal period $P_\mathrm{tide}=0.5$ d, corresponding to a circularly orbiting hot Jupiter on a 1 d orbit. Note that the period-dependence is: $Q'_\mathrm{IGW}\propto \left(\frac{P_\mathrm{tide}}{0.5 \mathrm{d}}\right)^{\frac{8}{3}}$, so these results can be straightforwardly extrapolated to other $P_\mathrm{tide}$. Bottom: same but showing the variation as a function of the stellar effective temperature $T_{eff}$ in $K$, restricting the range to the masses indicated.}
  \label{QpIGWcomp}
\end{figure}

\begin{figure}
  \begin{center}
   \subfigure{\includegraphics[trim=3cm 0cm 5cm 0cm,clip=true,width=0.4\textwidth]{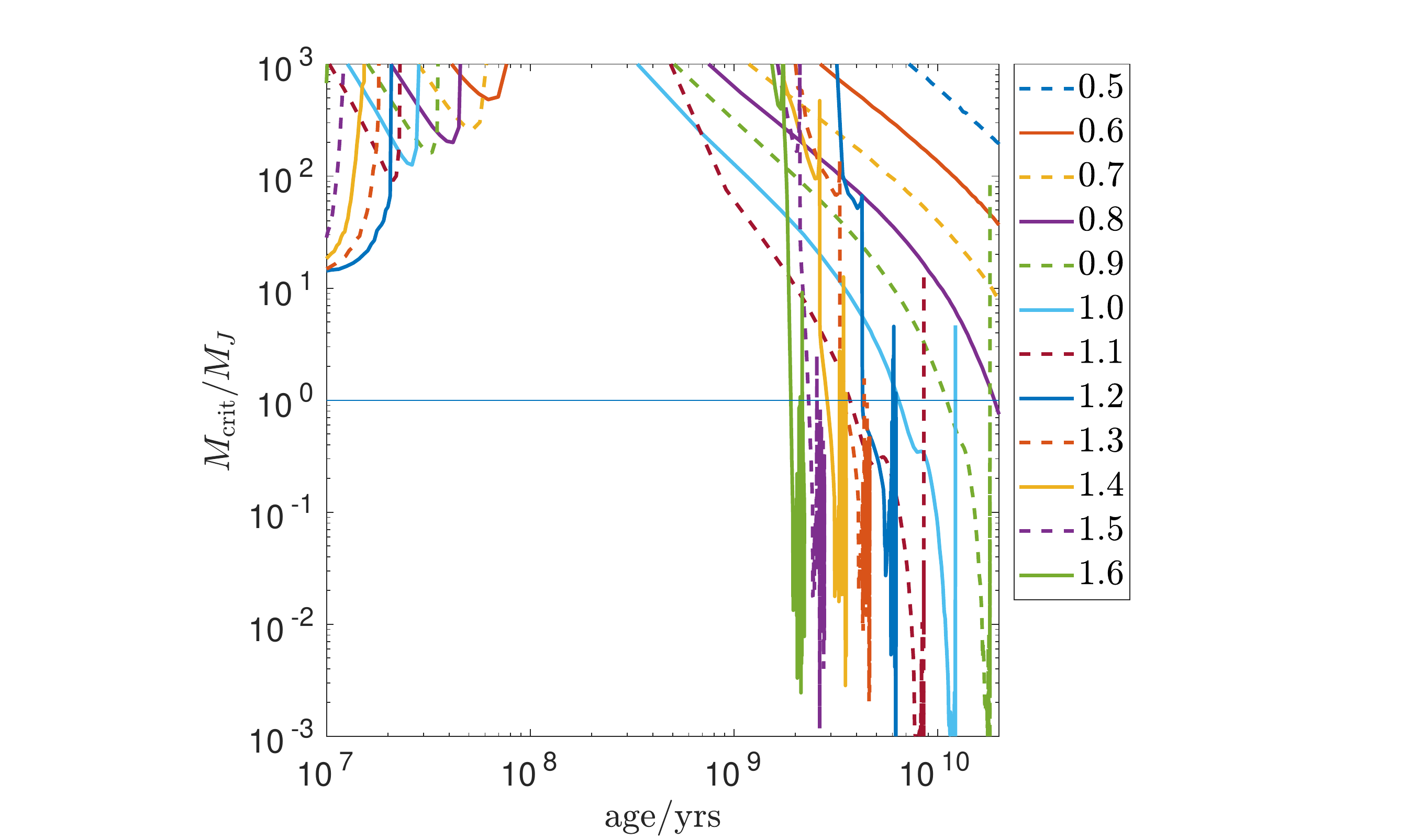}}
   \subfigure{\includegraphics[trim=3cm 0cm 5cm 0cm,clip=true,width=0.4\textwidth]{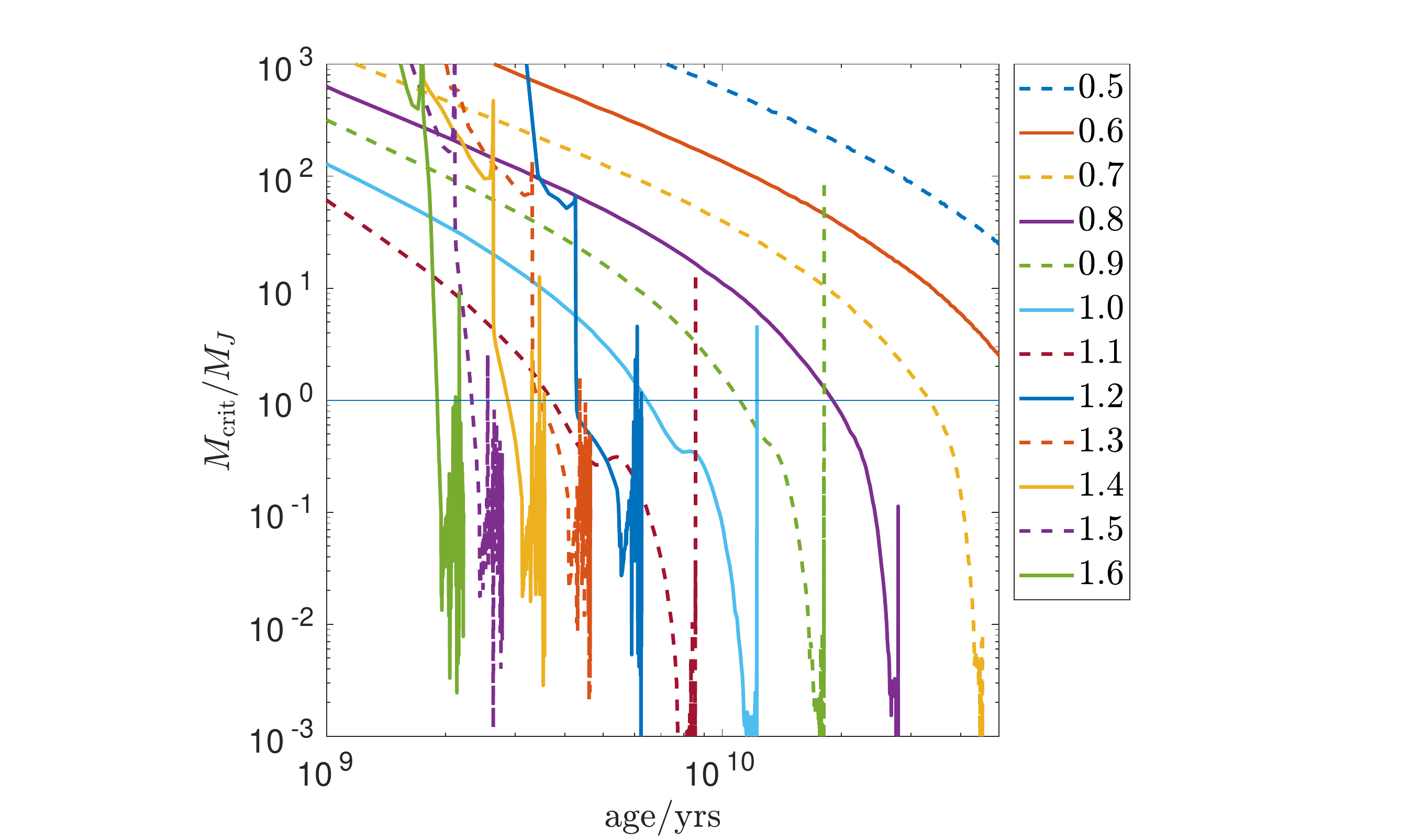}}
    \end{center}
  \caption{Critical planetary mass $M_\mathrm{crit}$ in Jupiter-masses as a function of age (in yrs) for a range of stellar masses (indicated in the legend). When $M_p>M_\mathrm{crit}$, the internal gravity waves are predicted to break in the radiation zone, assuming a circularly orbiting planet with a 1 d period (this criterion is weakly sensitive to orbital period). When $M_p>M_\mathrm{crit}$, we would then predict the tidal quality factor to behave according to Fig.~\ref{QpIGWcomp}.}
  \label{QpIGWcomp2}
\end{figure}

\begin{figure}
  \begin{center}
    \subfigure{\includegraphics[trim=3cm 0cm 5cm 0cm,clip=true,width=0.4\textwidth]{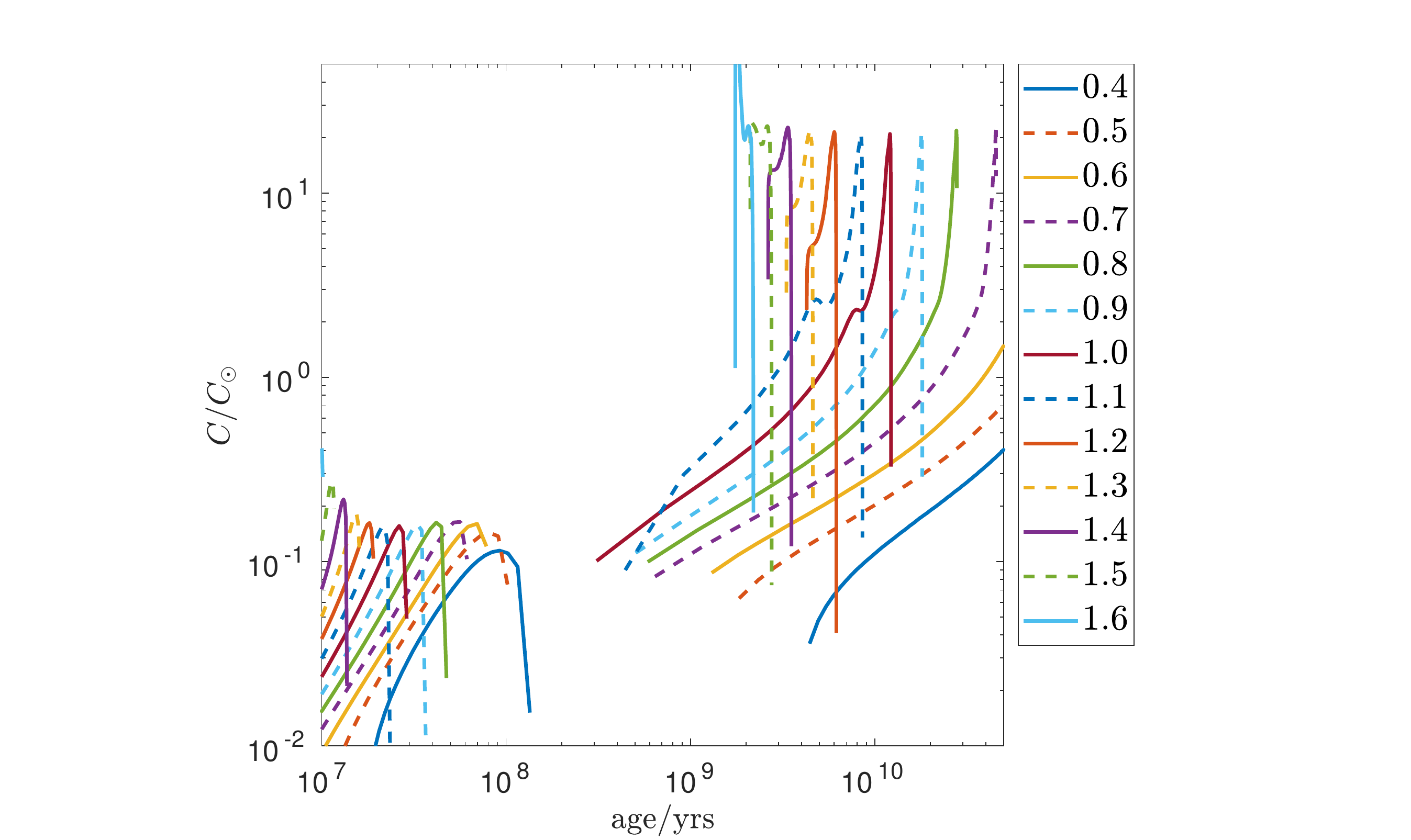}}
    \end{center}
  \caption{Strength of stable stratification at the centre of the star, where $N=Cr$ near $r\sim 0$, as a function of age (in yrs) for a range of stellar masses (indicated in the legend). When a line is shown, this indicates that the core is radiative. Note that $C$ increases with age due to nuclear burning, and this variation is the primary contribution to the trends observed in Fig.~\ref{QpIGWcomp2}.}
  \label{CoCSun}
\end{figure}

In this section we present results for the dissipation due to tidally-excited internal gravity waves in radiation zones. We assume that these waves are launched from the radiative/convective interface and are then subsequently fully damped, following the procedure described in \S~\ref{IGW}.

In the top panel of Fig.~\ref{QpIGWcomp} we show the predicted variation in $Q'_\mathrm{IGW}$ for a tidal period of $0.5$ d due to this mechanism as a function of age, showing results for various stellar masses spanning the range of stars that possess interior radiation zones (but they may also have convective cores). A tidal period of 0.5 d is chosen because this is relevant for computing the orbital decay of a 1 d hot Jupiter around a slowly rotating star. This mechanism predicts a robust period-dependence, with $Q'_\mathrm{IGW}\propto P_\mathrm{tide}^{\frac{8}{3}}$, under the assumption that the launching region buoyancy profile can be modelled as a linear function of the distance from the interface. Since this period-dependence is known, we can explore the variation in $Q'_\mathrm{IGW}$ as a function of stellar mass and age for a fixed tidal period according to Eq.~\ref{QIGW}. This is equivalent to showing the variation of $R\mathcal{G}/GM^2$ with stellar mass and age. It is then straightforward to extrapolate the results of Fig.~\ref{QpIGWcomp} to different tidal periods. In the bottom panel of this figure, we show the same prediction as a function of the stellar effective temperature, which is more directly observable.

Fig.~\ref{QpIGWcomp} shows that on the main sequence, stars with masses in the range $0.5-1.1M_\odot$ all possess similar values
\begin{eqnarray}
\label{QIGWpred}
Q'_\mathrm{IGW}\approx [1,3]\times 10^5 \left(\frac{P_\mathrm{tide}}{0.5 \mathrm{d}}\right)^{\frac{8}{3}}.
\end{eqnarray}
It may seem surprising that such a wide range of stars possess similar dissipative properties, though this has been noticed before \citep{BO2010}. The model with $M=0.4M_\odot$ shows greater variability but has $Q'_\mathrm{IGW}\sim 10^5-10^6$ for this tidal period. $Q'_\mathrm{IGW}$ increases with stellar mass for F-type stars on the main sequence, with $M=1.2M_\odot$ having $Q'_\mathrm{IGW}\approx 10^6$, $M=1.3M_\odot$ having $Q'_\mathrm{IGW}\approx 10^7$, and $M=1.4M_\odot$ having $Q'_\mathrm{IGW}\approx 10^8$, and so on, for this tidal period. This indicates that this mechanism is less efficient in F-stars, so that their radiation zones are generally less dissipative than those of K and G-stars. This may partly explain why the most massive short-period hot Jupiters are preferentially found around F-stars.

All stars are somewhat more dissipative in pre-main sequence phases than on the main sequence, but they attain their maximum dissipation as they evolve off the main sequence. Very small values of $Q'_\mathrm{IGW}\sim 10^0-10^1$ for this tidal period can be achieved as the stars evolve off the main sequence, once again demonstrating that this is the most efficient period for tidal dissipation in their evolution, for all tidal mechanisms that we have studied. Such efficient dissipation arises from a reduction in $\mathcal{G}$ by 2-3 orders of magnitude while $R$ increases by nearly 1 order of magnitude. Such efficient dissipation may be important to explain the destruction of short-period planets as stars evolve towards the end of the main sequence. (This enhancement of the tidal dissipation is qualitatively consistent with the behaviour reported in Fig.~3 of \cite{SunArrasWenberg2018}, though a direct comparison is not straightforward.)

In Fig.~\ref{QpIGWcomp2}, we show the predicted critical planetary mass required for wave breaking $M_\mathrm{crit}$ in Jupiter-masses as a function of age for the same range of stellar masses as Fig.~\ref{QpIGWcomp}. When the stars have radiative cores we use Eq.~\ref{Anl} to determine the critical planetary mass, and when stars have convective cores we instead use Eq.~\ref{xidrkr2}, assuming a 1 day circular orbit in both cases. The bottom panel shows a zoomed-in version that is restricted to ages beyond 1 Gyr.

This figure demonstrates clearly that the critical planetary mass required for wave breaking decreases sharply with the age of the star. This is because $M_\mathrm{crit}\propto C^{-\frac{5}{2}}$, so there is a strong dependence on $C$, the strength of the stratification at the centre, which increases as the star evolves. The evolution of $C$ normalised by $C_\odot$ (at the current age) is shown in Fig.~\ref{CoCSun}, indicating that this is primarily responsible for the evolution observed in Fig.~\ref{QpIGWcomp2}. 

We predict from Fig.~\ref{QpIGWcomp2} that $M_\mathrm{crit}$ will fall below $1 M_J$ in all stars with masses larger than $0.9 M_\odot$ before 10 Gyr. This indicates that those planets with short enough periods are likely to be engulfed during the main-sequence lifetime (or subgiant phase) of their stars. As these stars evolve off the main sequence, the minimum mass required for wave breaking is shown to decrease to $10^{-3}-10^{-2}M_J$. We therefore predict hot super-Earths and Neptunes to also initiate wave breaking in the radiation zones of their stars at later stages in their evolution, but lower mass planets with approximately Earth's mass may survive. The long-term survival of planets more massive than a few Earth masses on short-period orbits may therefore be prevented by this mechanism before the star leaves the main sequence.

There is a general trend that more massive stars require smaller planetary masses to initiate wave breaking, except for a jump around $1.1-1.2 M_\odot$ which is due to the onset of a convective core (which acts to reduce the geometrical focusing effect at the centre, thereby reducing the maximum wave amplitude attained).

A strong prediction of Fig.~\ref{QpIGWcomp2} is that wave breaking is expected for planetary-mass companions at some point during the evolution of all stars with $M\gtrsim 0.9 M_\odot$ as they evolve on the main sequence (or subgiant phase). However, this is not expected until the star has evolved sufficiently to acquire a strong enough stratification $N$ in the radiation zone due to nuclear burning. When wave breaking occurs, we predict the resulting tidal quality factor in Fig.~\ref{QpIGWcomp}. We will explore the consequences of this mechanism for the survival and orbital decay timescales of short-period hot Jupiters in the next section.

\section{Implications for the orbital decay of hot Jupiters}
\label{HJimplications}

\begin{figure*}
  \begin{center}
    \subfigure[$M/M_{\odot}=0.5,\,\text{age/yr}=3.34\times10^9$]{\includegraphics[trim=1cm 0cm 3.2cm 0cm,clip=true,width=0.4\textwidth]{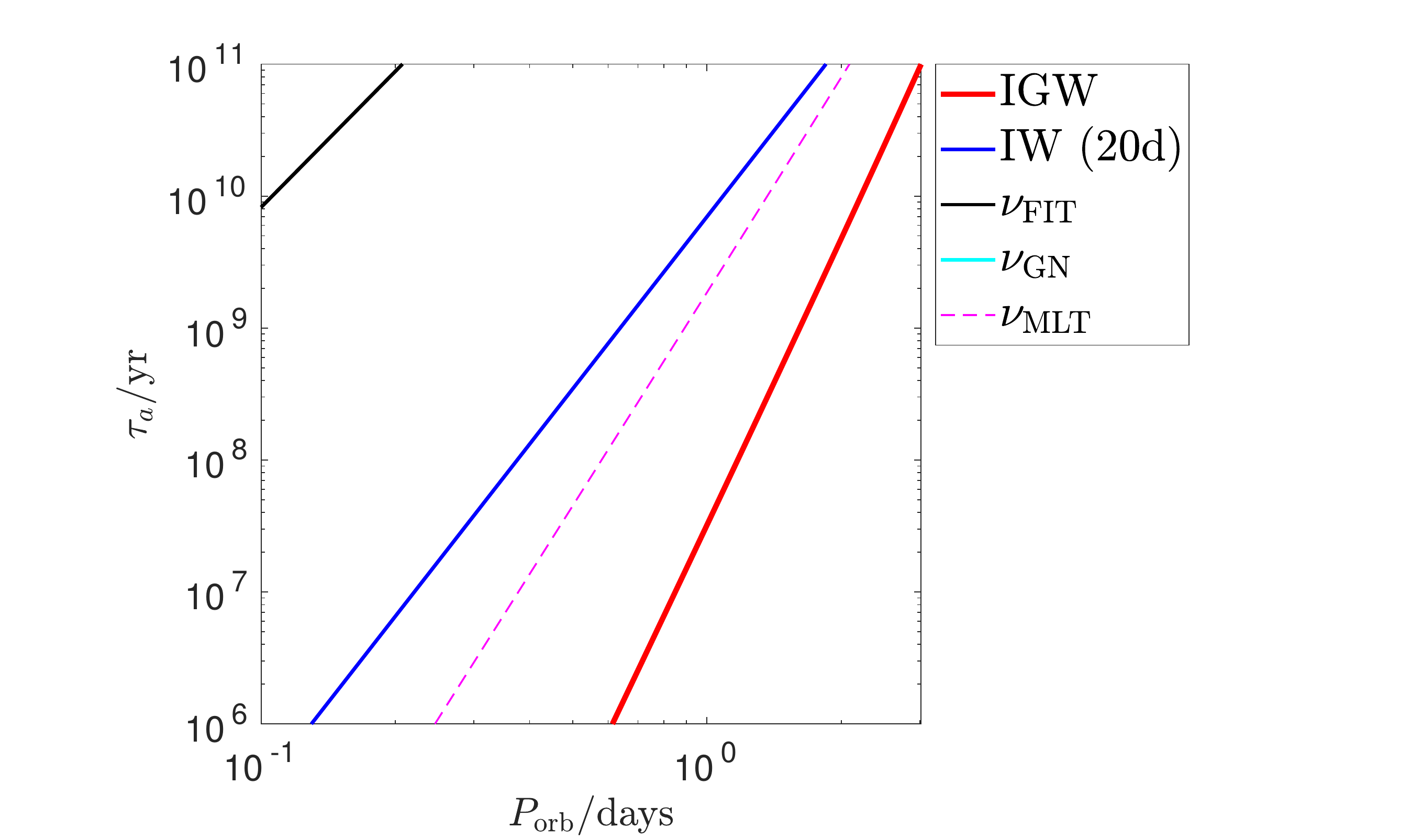}} 
    \subfigure[$M/M_{\odot}=0.5,\,\text{age/yr}=3.34\times10^9$]{\includegraphics[trim=1cm 0cm 3cm 0cm,clip=true,width=0.4\textwidth]{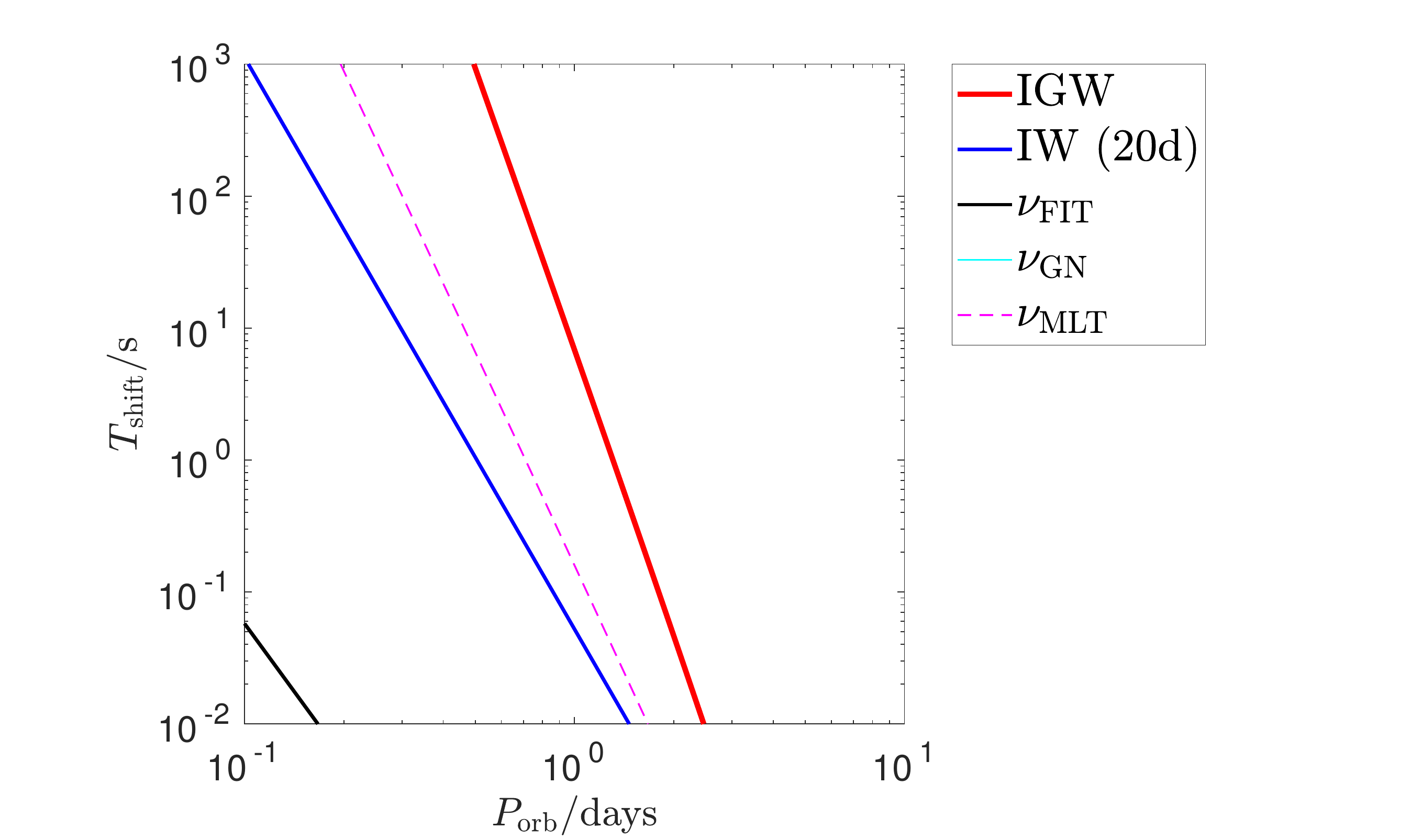}}
    \subfigure[$M/M_{\odot}=1,\,\text{age/yr}=4.70\times10^9$]{\includegraphics[trim=1cm 0cm 3.2cm 0cm,clip=true,width=0.4\textwidth]{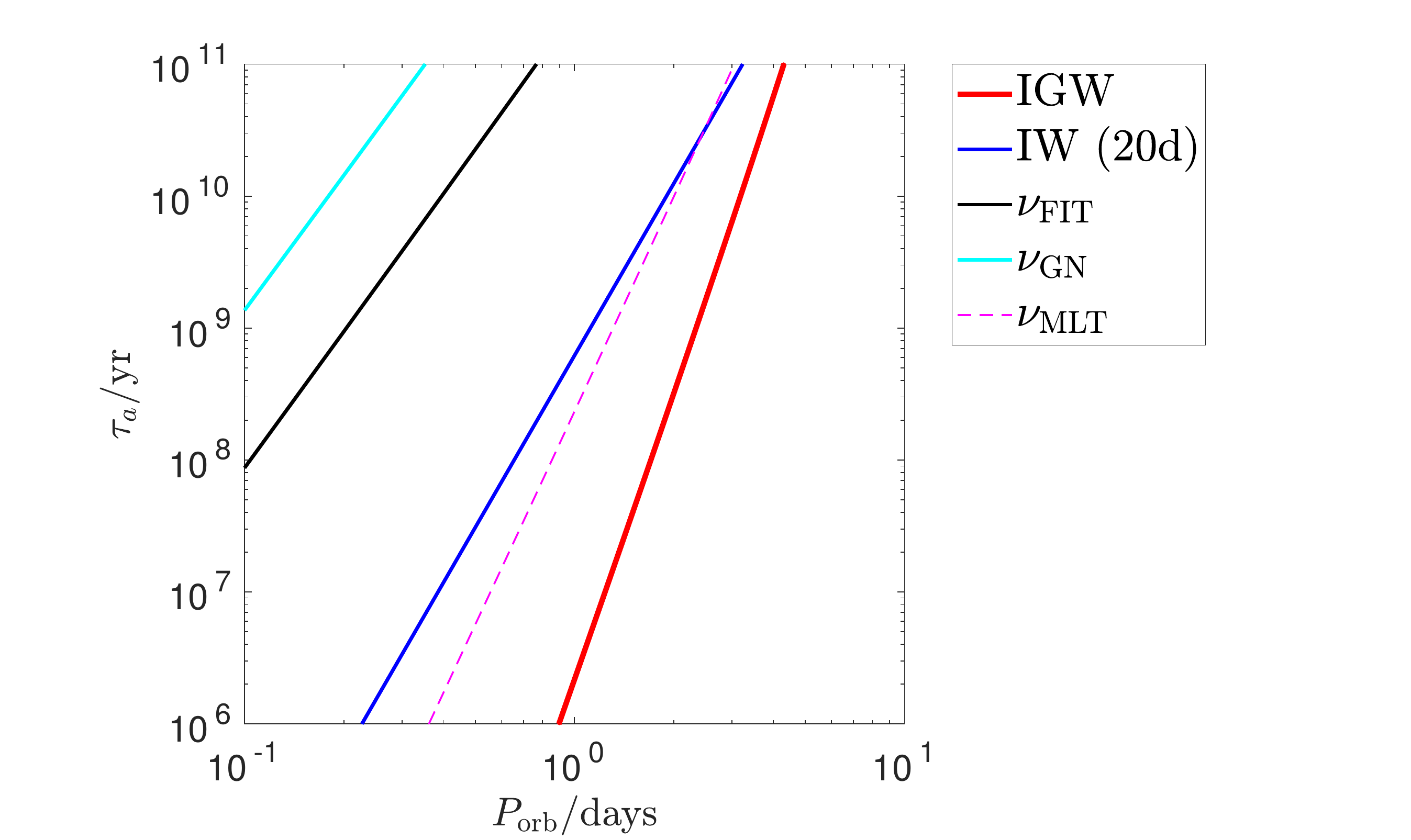}}
    \subfigure[$M/M_{\odot}=1,\,\text{age/yr}=4.70\times10^9$]{\includegraphics[trim=1cm 0cm 3cm
    0cm,clip=true,width=0.4\textwidth]{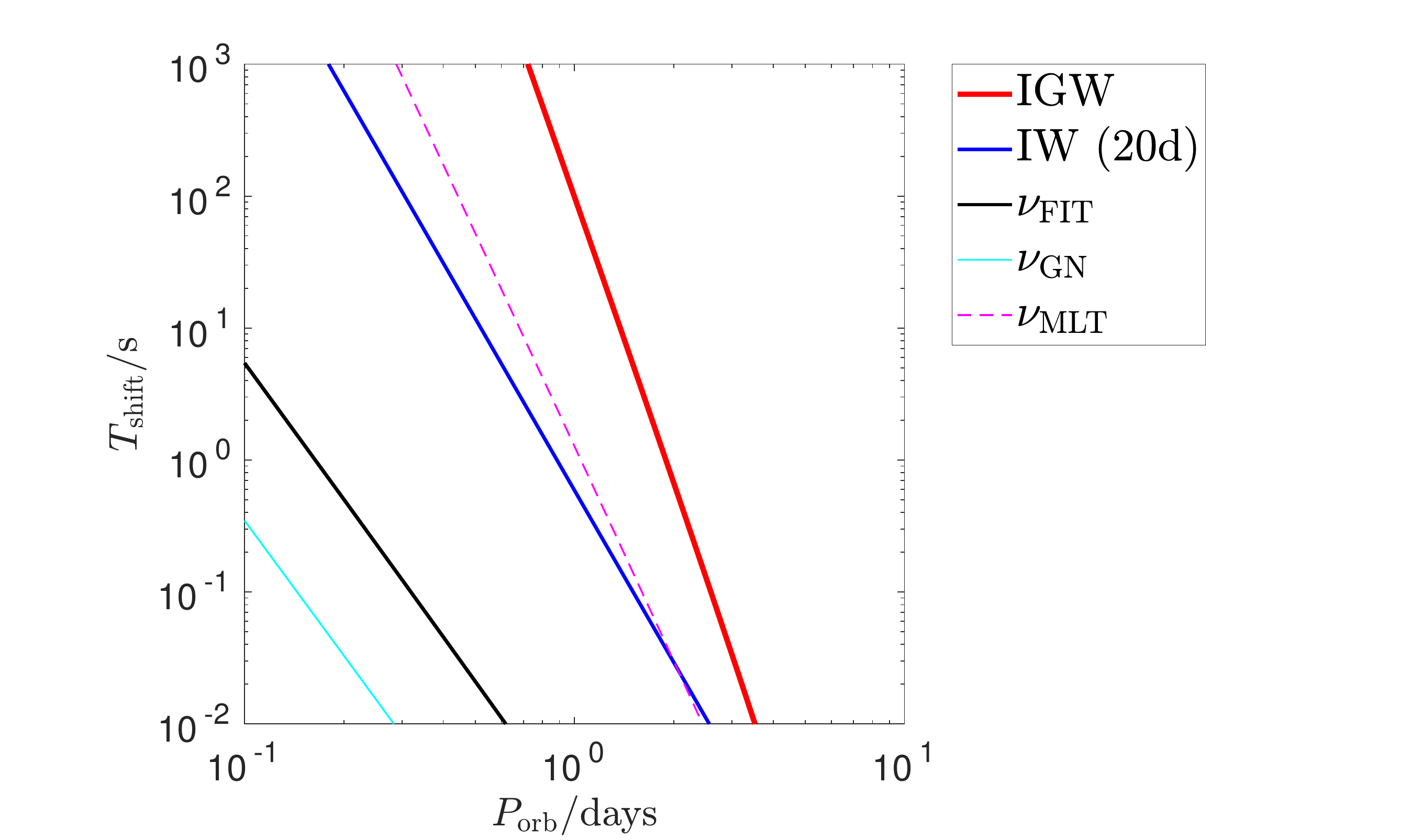}}
    \subfigure[$M/M_{\odot}=1.4,\,\text{age/yr}=1.29\times10^9$]{\includegraphics[trim=1cm 0cm 3.2cm 0cm,clip=true,width=0.4\textwidth]{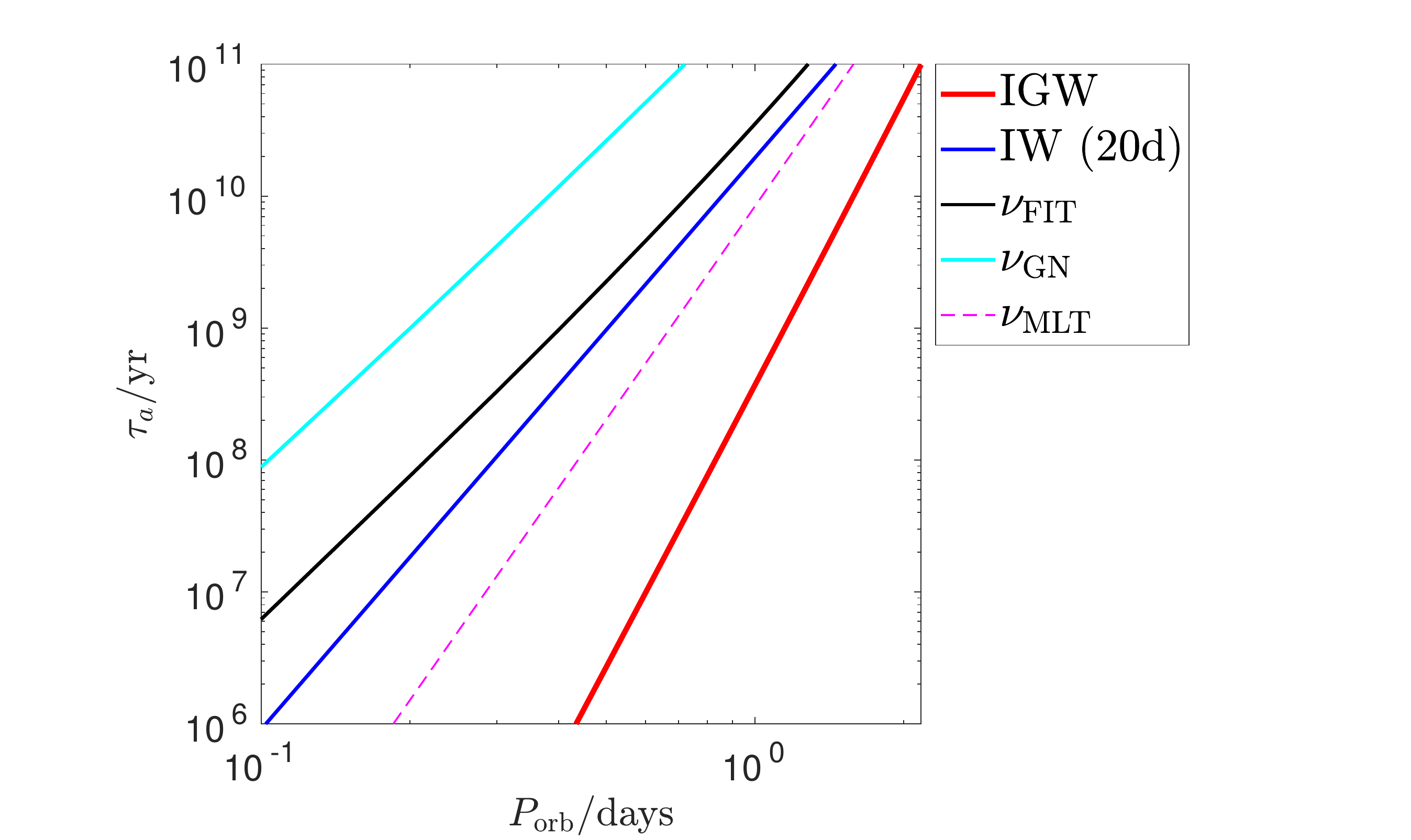}}
    \subfigure[$M/M_{\odot}=1.4,\,\text{age/yr}=1.29\times10^9$]{\includegraphics[trim=1cm 0cm 3cm 0cm,clip=true,width=0.4\textwidth]{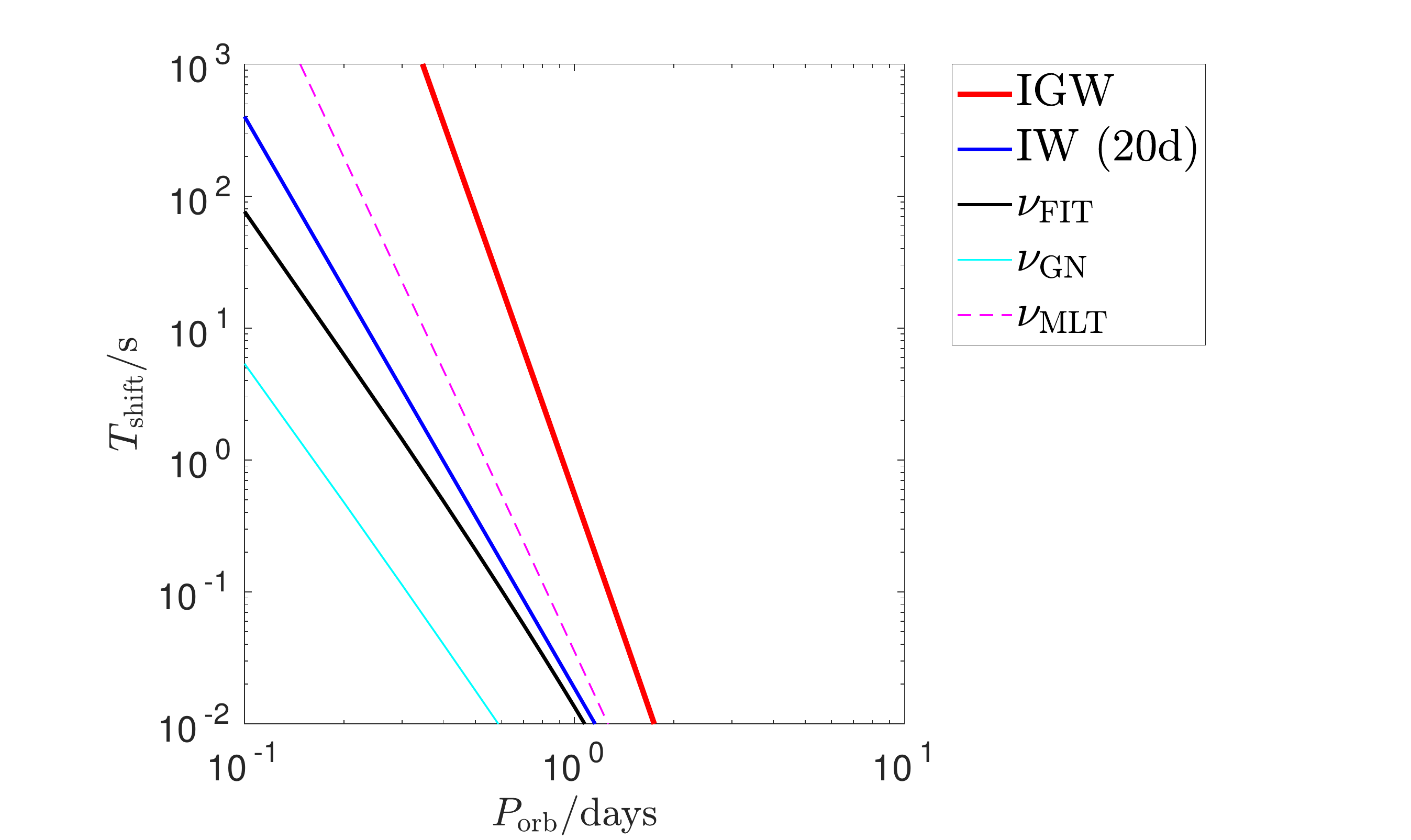}}
    \end{center}
  \caption{Left: predictions for orbital decay timescales of a $1M_J$ hot Jupiter on a circular orbit as a function of orbital period $P_\mathrm{orb}$ around the stars indicated in the caption. Right: corresponding predictions for the shift in transit arrival times $T_\mathrm{shift}$ for observations over a decade. The most efficient mechanism is dissipation of internal gravity waves (IGW) in radiation zones under the assumptions outlined in \S~\ref{IGW}. Inertial wave (IW) prediction is shown for reference assuming $P_\mathrm{rot}=20$ d; it should not be applied to predict planetary orbital evolution unless $P_\mathrm{orb}>10$ d so these waves can be excited. Based on our current understanding of convective damping of equilibrium tides, this mechanism provides a negligible contribution to planetary orbital decay on the main sequence.}
  \label{taua}
\end{figure*}

We now apply our results to the orbital decay of hot Jupiters, which is now starting to be detected or constrained from variations in transit arrival times \citep[e.g.][]{Maciejewski2016,Wilkins2017,Patra2017,Yee2020}. Based on our current understanding, the most important mechanism for the orbital decay of hot Jupiters orbiting slowly rotating main-sequence stars is gravity wave dissipation, which we discussed in \S~\ref{IGWresults}. Since the shortest-period hot Jupiters for which orbital decay could be detected orbit slowly rotating stars, and in any case, $P_\mathrm{orb}$ is typically so short that it is likely to be much smaller than $P_\mathrm{rot}$, inertial waves are not usually excited by tidal forcing (since $P_\mathrm{tide}>P_\mathrm{rot}/2$), so the mechanism we discussed in \S~\ref{IWresults} won't apply. We will confirm below that equilibrium tide dissipation on the main sequence is also negligible, based on the results of \S~\ref{NWLdissipation}, though it is likely to be important in evolved stars \citep[e.g.][]{Mustill2012}.

We apply our results from \S~\ref{IGWresults} by considering a hot Jupiter on a circular equatorial orbit around a slowly rotating star. For this problem, the relevant tidal frequency is $\omega=2(n-\Omega)$, and tidal evolution tends to cause the planet to spiral into the star\footnote{In principle, a massive planet with large enough orbital angular momentum can instead spin up the star to a tidal quasi-equilibrium state \citep{Hut1980}, though ongoing tidal evolution is still expected because of stellar magnetic braking \citep{BO2009,DamianiLanza2015}}. The orbital decay rate (assuming $n>\Omega$) is
\begin{eqnarray}
\frac{\mathrm{d}\ln a}{\mathrm{d} t} &=&  -\frac{9}{2} n \left(\frac{M_p}{M}\right)\left(\frac{R}{a}\right)^5\frac{1}{Q'} \\
 &=& -\frac{9\pi}{Q'}\left(\frac{M_p}{M}\right)\left(\frac{M}{M+M_p}\right)^{\frac{5}{3}}\frac{P_\mathrm{dyn}^{\frac{10}{3}}}{P_\mathrm{orb}^{\frac{13}{3}}}.
\end{eqnarray}
For the gravity wave mechanism such that $Q'=Q'_\mathrm{IGW}$, the right hand side is proportional to $a^{-21/2}$, which indicates an accelerating orbital decay. The resulting orbital decay time is
\begin{eqnarray}
\tau_a &=& -\frac{2}{21}\left(\frac{\mathrm{d}\ln a}{\mathrm{d} t}\right)^{-1} \\
 &\approx&2.3 \,\mathrm{Myr} \left(\frac{M_J}{M_P}\right)\left(\frac{M_\odot}{M}\right)\left(\frac{P_\mathrm{orb}}{1\mathrm{d}}\right)^7, 
\end{eqnarray}
for a $1M_J$ hot Jupiter orbiting the current Sun, based on applying Eq.~\ref{QIGW1}, and using the results of Fig.~\ref{QpIGWcomp}.

We can also predict the associated shift in transit arrival time, i.e. transits should occur earlier by \citep{Birkby2014,Wilkins2017}
\begin{eqnarray}
T_\mathrm{shift} &=& \frac{27}{8}n\left(\frac{M_p}{M}\right)\left(\frac{R}{a}\right)^5\left(Q'\right)^{-1} T_\mathrm{dur}^2, \\
&\approx& 98.4 s \left(\frac{P_\mathrm{orb}}{1 \mathrm{d}}\right)^{-\frac{21}{3}}\left(\frac{M_p}{M_J}\right)\left(\frac{T_\mathrm{dur}}{10 \,\mathrm{yr}}\right)^2,
\end{eqnarray}
where $T_\mathrm{dur}$ is the duration of the observations (e.g. decadal timescales). In the second line we have evaluated this for the current Sun, assuming $Q'=Q'_\mathrm{IGW}$ based on Eq.~\ref{QIGW1}, for a 1 $M_J$ hot Jupiter on a 1 d orbit. Note that this quantity is strongly period dependent, partly because this mechanism predicts $Q'_\mathrm{IGW}\propto P_\mathrm{orb}^{8/3}$. For stars with masses in the range $0.5-1.1 M_\odot$ on the main-sequence, we obtain similar $Q'_\mathrm{IGW}$ (see Fig.~\ref{QpIGWcomp}), approximately independently of stellar mass and age. This means that the above prediction can be applied to other stars by simply changing $M$ and $R$.

We show predictions for orbital decay timescales (left) and shifts in transit times (right) in Fig.~\ref{taua} for a $1M_J$ planet (circularly) orbiting three illustrative main sequence stellar models in our chosen mass range by applying our previous results for $Q'$ for each tidal mechanism. This shows predictions from gravity wave damping (IGW), inertial wave damping (IW), as well a convective damping of equilibrium tides using three different viscosity prescriptions, for illustration. Note that inertial wave damping is shown to indicate the typical level of dissipation due to this mechanism (assuming $P_\mathrm{rot}=20$ d), but this mechanism does not operate in slowly rotating stars for these tidal periods. Fig.~\ref{taua} shows that convective damping of equilibrium tides for the viscosity prescription that best matches the latest simulations ($\nu_\mathrm{FIT}$) plays a negligible role in the orbital decay of hot Jupiters. In all cases, internal gravity wave dissipation, assuming the waves to be fully damped, is the dominant mechanism in stars with interior radiation zones. This mechanism is less efficient in hot ($M=1.4 M_\odot$) and very cool ($M=0.5 M_\odot$) stars and is most efficient for $M\sim M_\odot$.

\cite{HamerSchlaufman2019} and \cite{HamerSchlaufman2020} demonstrated with a population-wide analysis that hot Jupiters are eventually destroyed by tides during the main sequence lifetimes of their host stars, but that ultra-short period planets with much lower masses (i.e.~comparable with Earth's mass) may survive. Our results in \S~\ref{IGWresults} that wave breaking is predicted to occur for planets with masses larger than $10^{-2}-10^{-3}M_J$ at some point on the main sequence, and that this results in efficient $Q'_\mathrm{IGW}$, may partly explain this result for short orbital periods. It would be worthwhile to apply our results to explore the implications of planetary survival in more detail. \cite{SchlaufmanWinn2013} previously found evidence for tidal destruction of hot Jupiters around subgiant stars (out to $P_\mathrm{orb}<200$ d), which requires very efficient dissipation $Q'\sim 10^2-10^3$. Our results in \S~\ref{IWresults} and \ref{IGWresults} and Fig.~\ref{QpIGWcomp} indicate that inertial waves may be able to provide such efficient dissipation briefly as the stars evolve off the main sequence, depending on the rotation rate of the star, so it would be worthwhile to revisit this mechanism in more detail with a population-wide study.

In the following we apply our results to various observed short-period hot Jupiters. Most of these are systems in which shifts in transit arrival times have been either detected or constrained, following \cite{Patra2020}, or those in which future detection appears favourable. We report our predictions in Table~\ref{TableHJ}, as well as the parameters and outputs in our representative stellar models used to fit each star. In some cases our models have slightly different radii or effective temperatures than the observed values. Further work is required to fit each star as accurately as possible, accounting for the necessary tweaking of input parameters and stellar physics (e.g.~mixing processes), which may lead to quantitatively different predictions to this table (probably more significant for $M_\mathrm{crit}$ than $Q'_\mathrm{IGW}$). We have made no serious attempt to do this here, and each system warrants a dedicated study. However, the overall picture we present below for each system is likely to be robust.

\underline{\textbf{WASP-4b}}: is a 1.2 $M_J$ orbiting a $0.93M_\odot$ star in $1.34$ d. The planet was observed to appear early in the TESS observations of \cite{Bouma2019} (supported by \citealt{Southworth2019}), indicating possible orbital decay consistent with $Q'\sim 4-8\times 10^4$. However, \cite{Bouma2020} later found WASP-4 to be accelerating towards Earth due to interaction with a wide-orbiting massive companion, and the resulting Doppler effect may instead cause the shift in transit times. Our best fit stellar models do not clearly predict wave breaking (since $M_{\mathrm{crit}}\approx 2-7 M_J$), but there is uncertainty in the age of the star ($5-7$ Gyr), and if the star is older wave breaking may be expected. In these models we predict $Q'_\mathrm{IGW}\sim 3-4\times 10^5$ if gravity waves are fully damped, so it appears very difficult to explain the observed $Q'$ purely from tidally-driven orbital decay for this system, which indirectly supports the findings of \cite{Bouma2020}.

\underline{\textbf{WASP-12b}}: is a $1.4M_J$ planet orbiting a metal-rich F-star with $1.35-1.43 M_\odot$ in $1.09$ d \citep{Hebb2009}. This planet was first detected to be undergoing possible tidally-driven orbital decay by \cite{Maciejewski2016}, and this has since been confirmed by various subsequent studies \citep{Patra2017,Maciejewski2018,Maciejewski2020a,Yee2020,Patra2020}, which are consistent with $Q'\approx 2\times 10^5$. Previous theoretical work on this system by \cite{Chernov2017} and \cite{Weinberg2017} indicated that the observed decay rate can be explained using gravity wave dissipation depending on the specific stellar model. We have been unable to find any model that matches all of the observational constraints for WASP-12, in agreement with \cite{2019Bailey}, who performed a more extensive parameter survey for this system. However, we do find that wave breaking is not expected in the main sequence models that we have considered, and is only expected if the star is a subgiant with a lower mass $M\sim 1.2M_\odot$ \citep{Weinberg2017,2019Bailey}. We have listed three different MESA models for this star in the table. We confirm that wave breaking is not expected in the two main sequence models, so the preferred explanation is for WASP-12 to be a subgiant with a radiative core, in which wave breaking can operate. In addition, our main-sequence model with $M=1.43M_\odot$, as quoted by \cite{Patra2020}, predicts $Q'_\mathrm{IGW}>2\times 10^6$, which is in conflict with observations. Our lower mass stellar models with $M=1.35M_\odot$ (or subgiant with $M=1.22M_\odot$) are however able to predict $Q'_\mathrm{IGW}$ that closely match the observations on the main sequence (even if they have too low $T_\mathrm{eff}$). This system is an excellent test of both stellar evolution and tidal theory. A more detailed comparison of theory and observations is therefore warranted for this system, and further observations would be very helpful to constrain the stellar properties.

\underline{\textbf{WASP-18b}}: is a massive $11.4 M_J$ planet orbiting an  $1.46 M_\odot$ F-star in only $0.94$ d \citep{Hellier2009}. Previous observations have indicated that $Q'>1.3\times 10^6$ \citep{Wilkins2017,Maciejewski2020,Patra2020}. The stellar mass appears to be uncertain in this system (both $M=1.24$ and $1.46$ have been reported), and our predictions for $M_\mathrm{crit}$ (and to a lesser extent $Q'$) are sensitive to stellar mass. We find $M=1.46 M_\odot$ quoted in \cite{Patra2020} to be too high to produce the observed stellar radius for the quoted ages, so we instead consider $M=1.24 M_\odot$ \citep[as in][]{Wilkins2017}. For this model we predict the planet to be insufficiently massive to cause wave breaking in the star, which is consistent with e.g.~\cite{BO2010} and \cite{Wilkins2017}. If the planetary orbit is really decaying due to this mechanism, then we predict $Q'_\mathrm{IGW}\approx 3\times 10^6$ (with the exact value sensitive to stellar mass and age), which would not be clearly detected using the existing observations. We strongly advocate future observations of this system to determine whether the planetary orbit really is decaying at such a rate. This would provide a good test of both stellar modelling and tidal theory.

\underline{\textbf{WASP-19b}}: is an ultra-short period $1.14 M_J$ planet orbiting an $0.94 M_\odot$ star in only $0.79$ d \citep{Hebb2010}. The age of the star has been quoted as 7-9 Gyr, and our predictions for $M_\mathrm{crit}$ are very sensitive to the exact age. Our best fit stellar model has $9.3$ Gyr, and predicts $M_\mathrm{crit}=0.8 M_J$, such that wave breaking is expected, but wave breaking would not be predicted for the younger ages. We predict $Q'_\mathrm{IGW}\sim 0.6-0.8\times 10^5$, which is much smaller than the observational constraint \citep{Patra2020} (see also \citealt{Petrucci2020}). This would be consistent with theory only if the star is somewhat younger, in which case wave breaking would not be expected (so that such small $Q'_\mathrm{IGW}$ would not apply). Further observations of this system would be very useful to further constrain tidal theory.

\underline{\textbf{WASP-43b}}: is a $2.03 M_J$ planet orbiting a $0.72 M_\odot$ star in only $0.81$ d \citep{Hellier2011}. Based on our best fitting stellar model, wave breaking is not expected at 5 Gyr and so we do not predict orbital decay to be detectable. If the gravity waves are otherwise fully damped, in agreement with \cite{Chernov2017}, we predict $Q'_\mathrm{IGW}$ that is similar to the observational constraint \citep{Patra2020}. Further observations of this system would be useful to constrain tidal theory. If the orbit is observed to decay, since wave breaking is not expected, weakly nonlinear mechanisms may instead be responsible \citep[e.g][or perhaps other mechanisms that efficiently damp gravity waves]{BO2011,EssickWeinberg2016}.

\underline{\textbf{WASP-72b}}: is a $1.55 M_J$ planet orbiting a $1.39 M_\odot$ star in $2.22$ d \citep{Gillon2013}. While wave breaking is likely to occur in this star for the inferred ages, the currently predicted $Q'_\mathrm{IGW}>10^{12}$ would result in negligible tidal evolution, and so we predict orbital decay will not be observed in this system. The existing observations do not yet provide good enough constraints to verify this prediction \citep{Patra2020}.

\underline{\textbf{WASP-103b}}: is a $1.51 M_J$ planet orbiting a $1.21 M_\odot$ star in $0.93$ d \citep{Gillon2014}. The star likely has a convective core, and wave breaking is not expected in the reported age range, so based on this we do not expect planetary orbital decay to be detected. If gravity waves are however fully damped, we predict $Q'_\mathrm{IGW}$ that is a bit larger than the minimum value constrained by observations \citep{Maciejewski2018,Patra2020}. Further observations of this system would be useful to constrain tidal theory.

\underline{\textbf{WASP-114b}}: is a $1.77 M_J$ planet orbiting a $1.29 M_\odot$ star in $1.55$ d \citep{Barros2016}. This star probably has a convective core and wave breaking is not expected, so we do not predict orbital decay to be detected. If the gravity waves are however fully damped, we predict $Q'_\mathrm{IGW}\sim 3\times 10^6$.

\underline{\textbf{WASP-121b}}: is a $1.18 M_J$ planet orbiting a $1.35 M_\odot$ star in $1.27$ d \citep{Delrez2016}. Again, this star probably has a convective core and wave breaking is not expected. If the gravity waves are however fully damped, we predict $Q'_\mathrm{IGW}\sim 2\times 10^7$, which may be difficult to test observationally.

\underline{\textbf{WASP-122b}}: is a $1.28 M_J$ planet orbiting a $1.24 M_\odot$ star in $1.71$ d \citep{Turner2016}. Again, this star probably has a convective core and wave breaking is not expected. If the gravity waves are however fully damped, we predict $Q'_\mathrm{IGW}\sim 3.5\times 10^5$, which could be detected observationally in the future. Further observations of this system would be useful to constrain tidal theory.

\underline{\textbf{WASP-128b}}: is a very massive $37 M_J$ object orbiting a $1.16 M_\odot$ star in $2.21$ d \citep{Hodzic2018}. Given the large planetary mass we predict wave breaking even though the star likely has a convective core. However, our prediction for the resulting $Q'_\mathrm{IGW}$ is strongly dependent on the stellar rotation period. If the star rotates slowly, $Q'_\mathrm{IGW}\sim 4\times 10^6$ is possible, but if the star is close to being synchronized with $P_\mathrm{rot}\sim 3$ d we instead predict $Q'_\mathrm{IGW}\sim 2\times 10^8$ or larger. If the star rotates sufficiently rapidly that $P_\mathrm{rot}\sim 3$ d, inertial waves would be excited in the star, and we predict a typical $\langle Q'_\mathrm{IW}\rangle\sim 1.4\times 10^6$ (though the actual $Q'$ may differ from this frequency-averaged value). In this system, it is possible that the planet could spin up the star to synchronism and that the planetary orbit would then decay on the magnetic braking timescale \citep[e.g][]{BO2009,DamianiLanza2015}. Further observations to constrain $P_\mathrm{rot}$, and to seek possible evidence of orbital evolution would be useful to constrain theoretical predictions.

\underline{\textbf{NGTS-6b}}: is a $1.33 M_J$ planet orbiting an old $0.79 M_\odot$ star in $0.88$ d \citep{Vines2019}. Wave breaking is not predicted in this star for its current age, so we do not clearly predict rapid orbital decay. However, if gravity waves are fully damped, we predict $Q'_\mathrm{IGW}\sim 10^5$, which could be detectable with future observations. We note that we found it difficult to reproduce the observed stellar radius in our models. Further observations of this system would be useful to constrain tidal theory.

\underline{\textbf{NGTS-7Ab}}: is a very massive $62 M_J$ object orbiting a very young $0.48 M_\odot$ star in only $0.68$ d \citep{Jackman2019}. It is likely that the star rotates rapidly because of its young age, and perhaps because of tidal synchronization. In this system we expect inertial wave dissipation to be dominant. Gravity wave breaking is not expected, but efficient $Q'_\mathrm{IGW}$ is obtained if the gravity waves are fully damped. If the star has been synchronized, we expect the orbit to evolve on the magnetic braking timescale \citep{BO2009,DamianiLanza2015}. This would be another excellent target for future observations to test tidal theory.

\underline{\textbf{NGTS-10b}}: is an ultra-short period $2.16 M_J$ planet orbiting an old $0.7 M_\odot$ star in only $0.77$ d \citep{McCormac2020}. Wave breaking and rapid orbital decay is not predicted for the current age, but if gravity waves are fully damped, we predict $Q'_\mathrm{IGW}\sim 10^5$. This would be another excellent target for future observations to test tidal theory.

\underline{\textbf{HAT-P-23b}}: is a $2.09 M_J$ planet orbiting a $1.13 M_\odot$ star in $1.21$ d \citep{Bakos2011}. The star probably has a convective core and wave breaking is not expected, so we do not predict detectable orbital decay. This is consistent with current observations. However, if gravity waves are fully damped we predict $Q'_\mathrm{IGW}\sim 3.5\times 10^5$, which is close to the observational constraint \citep{Maciejewski2018,Patra2020}, and would predict large $T_\mathrm{shift}$. Further observations of this system would be useful to constrain tidal theory.

\underline{\textbf{HATS-18b}}: is a $1.98 M_J$ planet orbiting a $1.04 M_\odot$ star in $0.84$ d \citep{Penev2016}. Wave breaking is expected in this star for the later ages (our best fit model has 4.6 Gyr), and we predict the planetary orbit to decay with $Q'_\mathrm{IGW}\sim 10^5$. This is a very promising candidate to search for orbital decay, and we strongly advocate further observations of this system to test tidal theory. Tidal spin-up of the star has also been suggested, which is hard to reconcile with gravity wave damping (other tidal mechanisms are probably negligible) unless the mechanisms of angular momentum transport within the star are particularly efficient.

\underline{\textbf{KELT-16b}}: is a $2.75 M_J$ planet orbiting a $1.21 M_\odot$ star in $0.97$ d \citep{Oberst2017}. The star likely has a convective core, and wave breaking is not expected, so we do not clearly predict observable orbital decay. If the waves are however somehow fully damped we predict $Q'_\mathrm{IGW}\sim 7\times 10^5$, which is an order of magnitude larger than the current observational constraint \citep{Maciejewski2018,Patra2020}. This may be detectable in this system with further observations, so we strongly advocate these to constrain tidal theory.

\underline{\textbf{TRES-3b}}: is a $1.91 M_J$ planet orbiting a $0.92 M_\odot$ star in $1.31$ d \citep{Odonovan2007}. Wave breaking is not predicted at the current age, though it is predicted as the star evolves. If the waves are however somehow fully damped we predict $Q'_\mathrm{IGW}\sim 6.5\times 10^5$, which is larger than the current observational constraint \citep{Mannaday2020}. This may be detectable with future observations, which would be useful to constrain tidal theory.

\underline{\textbf{OGLE-TR-56b}}: is a $1.39 M_J$ planet orbiting a $1.29 M_\odot$ star in $1.21$ d \citep{Udalski2002,Sasselov2003}. The star likely has a convective core, and wave breaking is not expected, so we do not predict observable orbital decay. This may explain the survival of the planet. However, if the waves are fully damped, we predict $Q'_\mathrm{IGW}\sim 2\times 10^6$, which might be difficult to observe but is compatible with current observational constraints \citep{Patra2020}.

\underline{\textbf{WTS-2b}}: is a $1.12 M_J$ planet orbiting a $0.82 M_\odot$ star in $1.02$ d \citep{Birkby2014}. Wave breaking is not predicted at the young inferred age of this star. However, we predict that if the waves are however somehow fully damped $Q'_\mathrm{IGW}\sim 2\times 10^5$. This could be detectable with future observations, which would be useful to constrain tidal theory.

\smallskip
Our results suggest that it is not a coincidence that the most massive hot Jupiters tend to be found around F-stars with convective cores, and that there are very few systems in which wave breaking is predicted and a very short-period massive planet is observed (WASP-19b and HATS-18b are possible exceptions to this rule). Gravity wave damping is also typically much less efficient in F-stars with $1.2-1.6 M_\odot$ relative to solar mass stars, which may together explain why such stars preferentially harbour close-in massive hot Jupiters. Those around lower mass stars may have been destroyed by wave breaking on the main sequence. Other mechanisms that operate for smaller planetary masses may also be important, such as passage through a resonance (that could initiate wave breaking), or weakly nonlinear interactions \citep[e.g][]{BO2011,Weinberg2012,EssickWeinberg2016}, which should motivate further work.

\section{Implications for binary circularization and synchronization}
\label{binaryimplications}

\begin{figure*}
  \begin{center}
    \subfigure[$M/M_{\odot}=0.2,\,\text{age/yr}=2.93\times10^9$]{\includegraphics[trim=1cm 0cm 3cm 0cm,clip=true,width=0.4\textwidth]{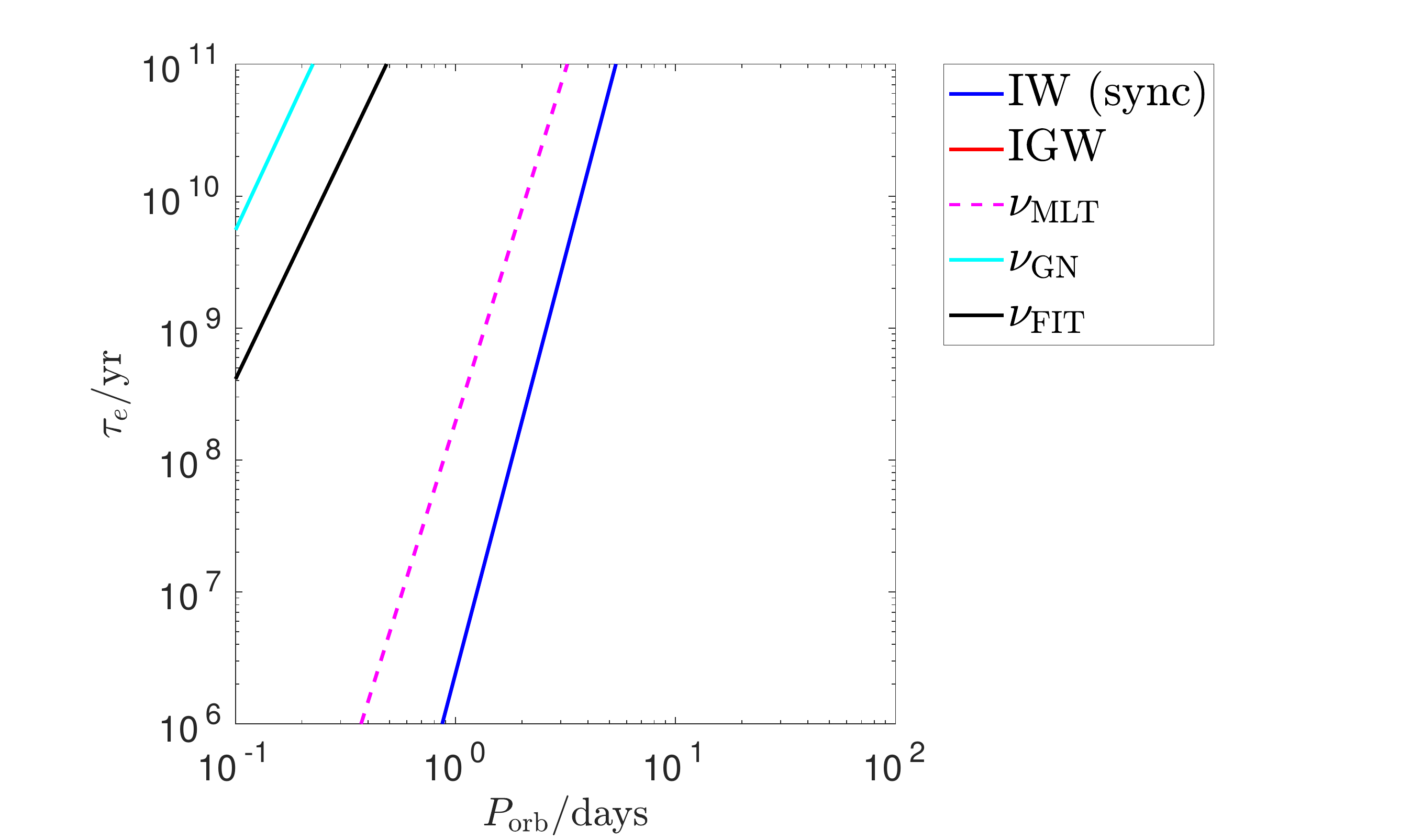}}
    \subfigure[$M/M_{\odot}=0.2,\,\text{age/yr}=2.93\times10^9$]{\includegraphics[trim=1cm 0cm 3cm 0cm,clip=true,width=0.4\textwidth]{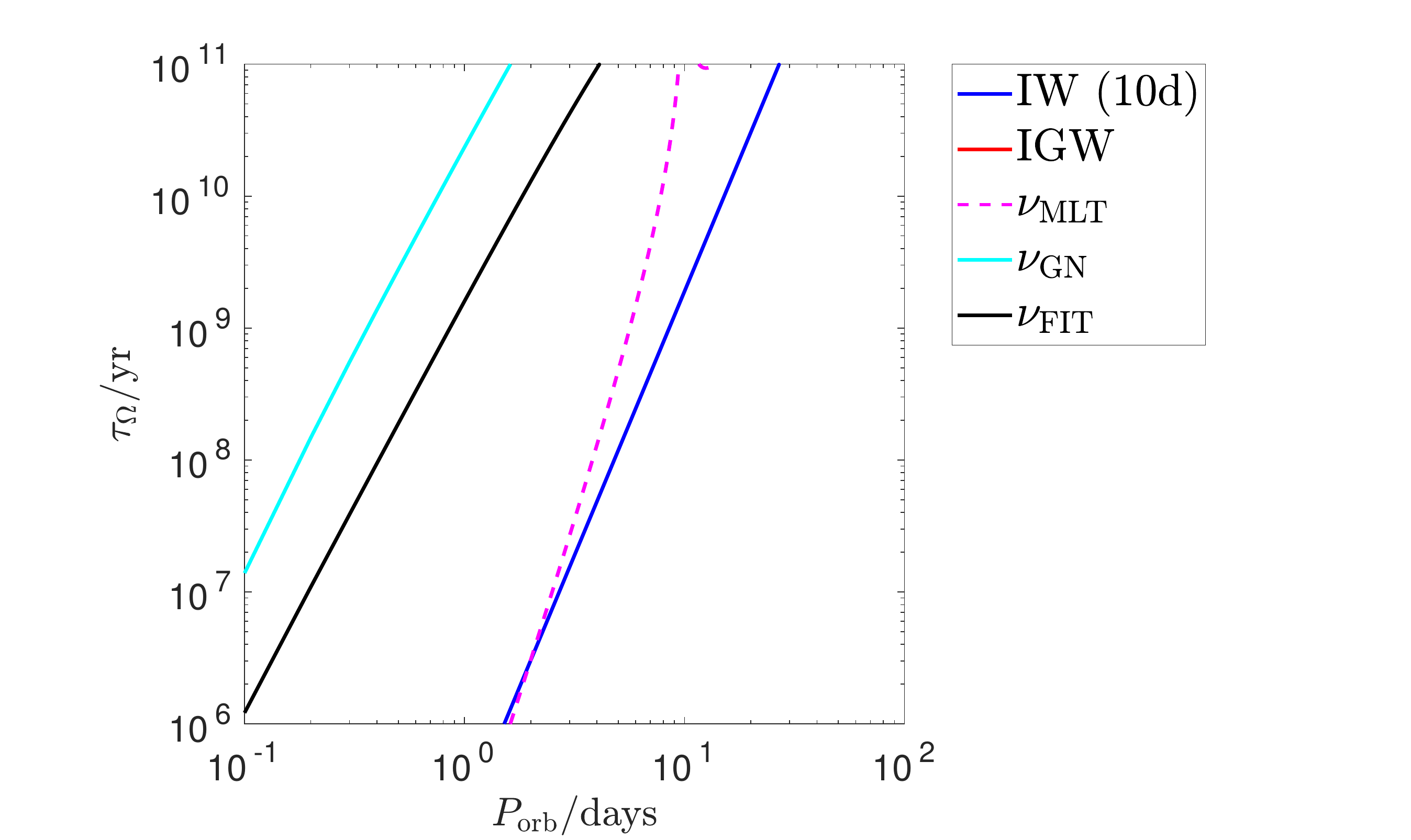}}
    \subfigure[$M/M_{\odot}=1,\,\text{age/yr}=4.70\times10^9$]{\includegraphics[trim=1cm 0cm 3cm 0cm,clip=true,width=0.4\textwidth]{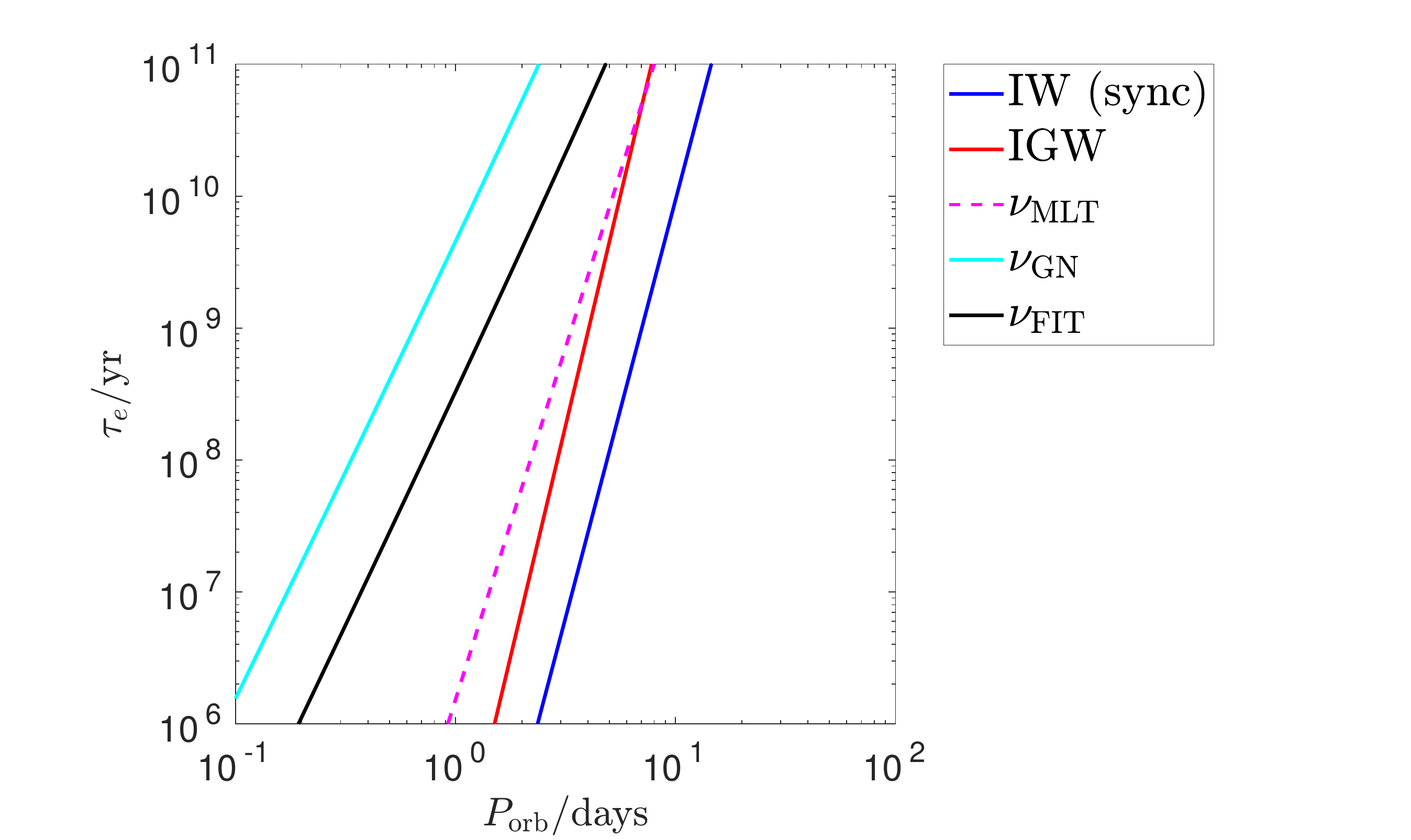}}
    \subfigure[$M/M_{\odot}=1,\,\text{age/yr}=4.70\times10^9$]{\includegraphics[trim=1cm 0cm 3cm
    0cm,clip=true,width=0.4\textwidth]{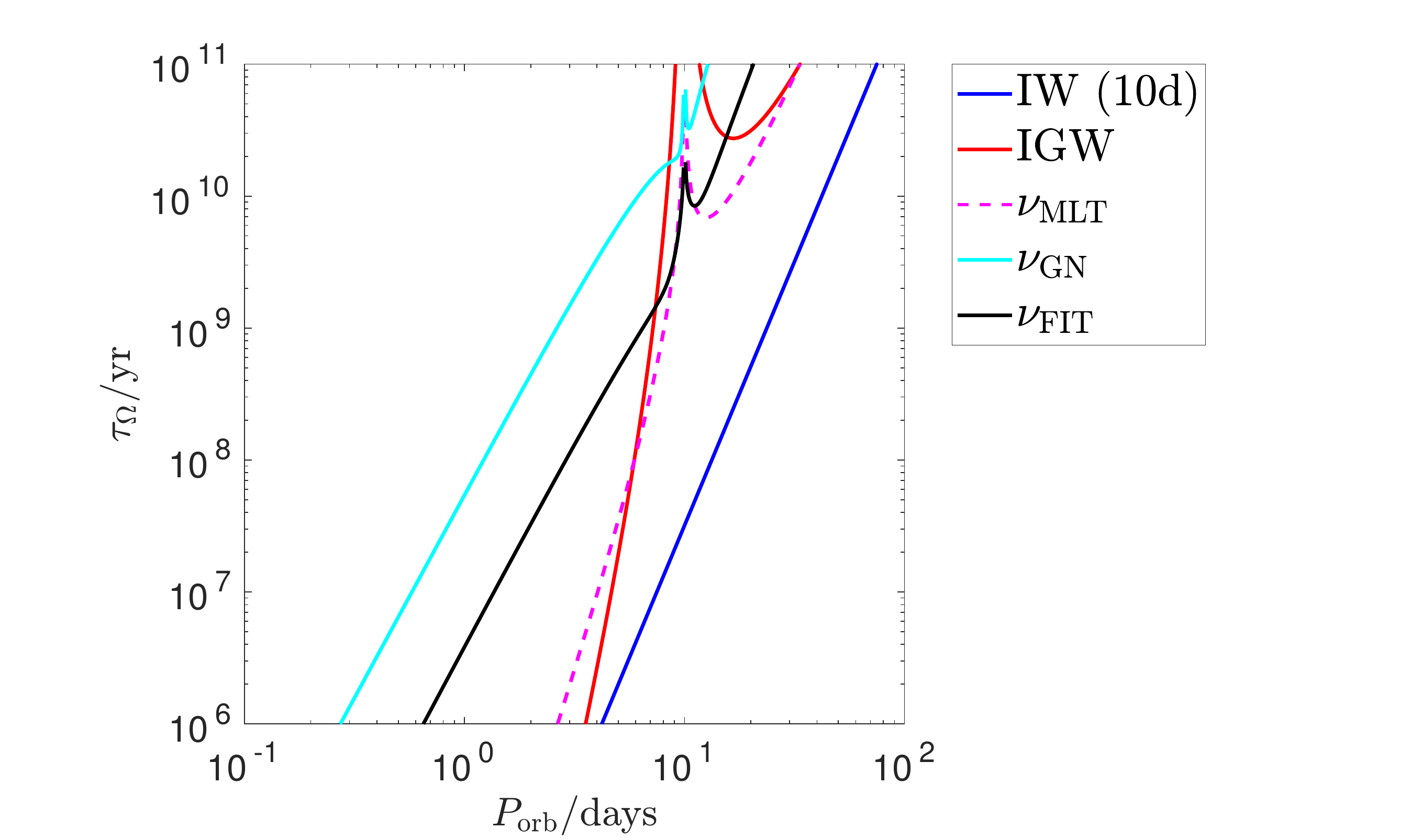}}
    \subfigure[$M/M_{\odot}=1.6,\,\text{age/yr}=1.03\times10^9$]{\includegraphics[trim=1cm 0cm 3cm 0cm,clip=true,width=0.4\textwidth]{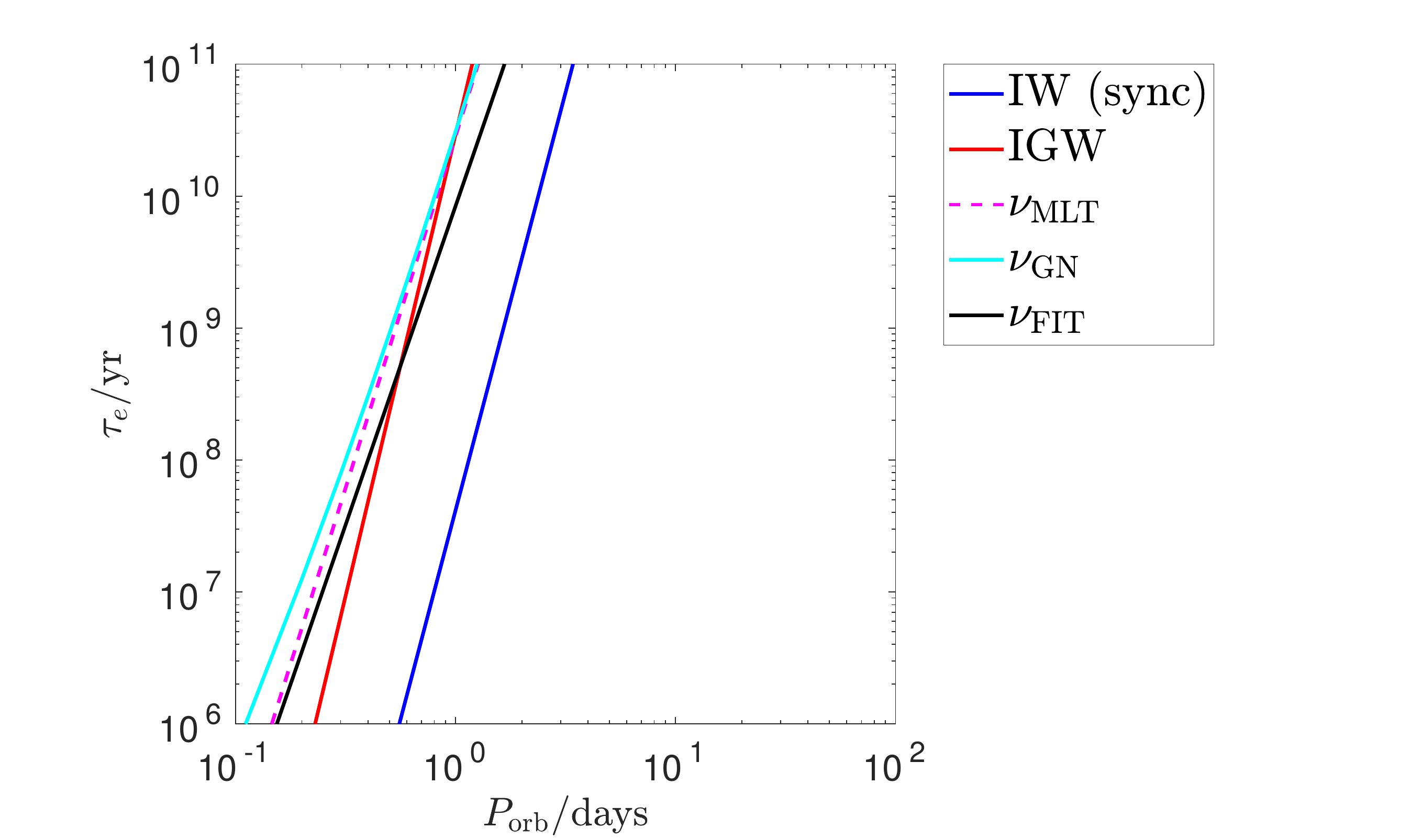}}
    \subfigure[$M/M_{\odot}=1.6,\,\text{age/yr}=1.03\times10^9$]{\includegraphics[trim=1cm 0cm 3cm 0cm,clip=true,width=0.4\textwidth]{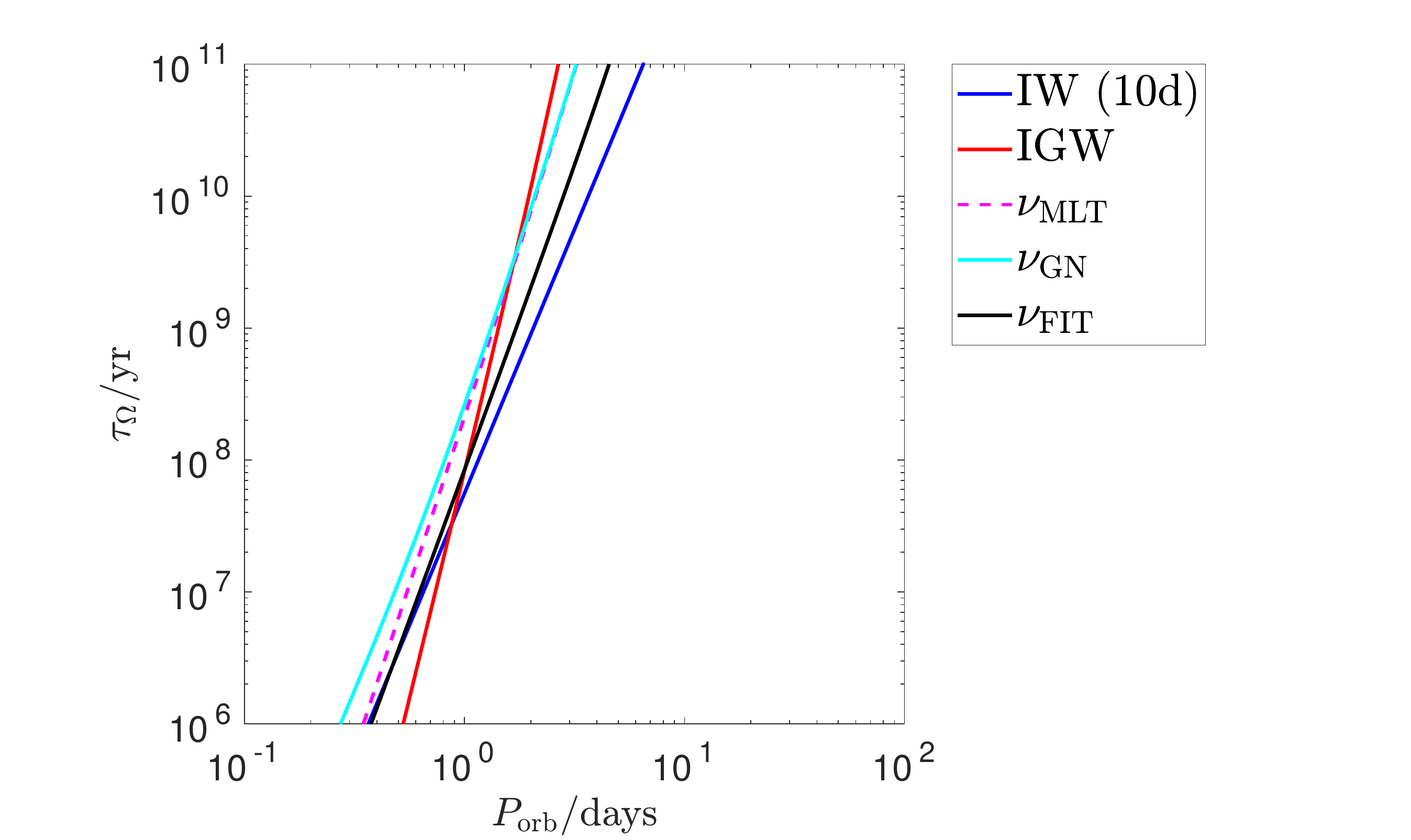}}
    \end{center}
  \caption{Predictions for tidal circularization timescales $\tau_e$ (left; assuming $P_\mathrm{orb}=P_\mathrm{rot}$) and spin synchronization timescales $\tau_\Omega$ (right; assuming $P_\mathrm{rot}=10$ d) in various stellar models indicated, assuming a companion of mass $M_2=M_\odot$. Inertial wave dissipation is by far the most efficient mechanism in both problems for these stars. Convective damping of equilibrium tides is shown to play only a minor role in binary circularization.}
  \label{taueOm}
\end{figure*}

In this section, we apply our results to compute timescales for circularization and synchronization of close binary stars. Once we have obtained $Q'$ due to each tidal mechanism, we may estimate the corresponding timescale for tidal synchronization of the stellar spin due to interaction with a companion star of mass $M_2$ by
\begin{eqnarray}
    \label{tauOm}
    \tau_\Omega &=& \frac{2 Q'}{9\pi  r_g^2}\left(\frac{M+M_2}{M_2}\right)^2\frac{P_\mathrm{orb}^4}{P_\mathrm{dyn}^2P_\mathrm{rot}} \\
    &\approx& 20 \,\mathrm{Myr} \left(\frac{P_\mathrm{orb}}{10\mathrm{d}}\right)^4\left(\frac{P_\mathrm{rot}}{5\mathrm{d}}\right),
\end{eqnarray}
where $r_g^2$ is the dimensionless squared radius of gyration\footnote{Defined by
\begin{eqnarray}
r_g^2 = \frac{8\pi}{GM R^2}\int_0^R\rho r^4\mathrm{d}r,
\end{eqnarray}
and this is reported for a selection of models in Table~\ref{Table}.} 
and we have evaluated this using inertial wave dissipation based on the results in \S~\ref{IWresults} which indicate that $\langle Q'_\mathrm{IW}\rangle \approx 10^7 (P_\mathrm{rot}/10 \,d)^2$ on the main sequence. We have assumed $M=M_2=M_\odot$ for this estimate. 

The corresponding circularization timescale, assuming small eccentricities and tidal dissipation in only the star of mass $M$, is estimated by
\begin{eqnarray}
\label{taue}
\tau_e&=& \frac{2}{63\pi} Q' \left(\frac{M}{M_2}\right)\left(\frac{M+M_2}{M}\right)^{\frac{5}{3}}\frac{P_\mathrm{orb}^{\frac{13}{3}}}{P_\mathrm{dyn}^{\frac{10}{3}}} \\
&\approx& 9.38\, \mathrm{Gyr} \left(\frac{P_\mathrm{orb}}{10\mathrm{d}}\right)^{\frac{19}{3}},
\end{eqnarray}
where we have evaluated this using inertial wave dissipation based on the results in \S~\ref{IWresults} using $M=M_2=M_\odot$. Similar dissipation is expected in stars in the range 0.8-1.2 $M_\odot$. Since $\tau_\Omega\ll \tau_e$, we have assumed spin-orbit synchronization for the purposes of this estimate, i.e.~$P_\mathrm{orb}=P_\mathrm{rot}$, but we note that assuming shorter $P_\mathrm{rot}$ would give shorter timescales. By itself, this estimate suggests that orbital circularization purely on the main-sequence due to inertial waves may not quite be sufficient to explain the observed circularization periods of binary stars, in agreement with \cite{OL2007}.

In Fig.~\ref{taueOm}, we show the predicted $\tau_\Omega$ and $\tau_e$ (assuming synchronization) for all tidal mechanisms studied in this paper in several stellar models spanning fully-convective to F-type stars. In each case we assume $M_2=M_\odot$. When inertial waves are excited, this figure demonstrates clearly that they are the dominant mechanism for binary synchronization and circularization for the adopted spin periods. (Note that because we have used the frequency-averaged measure, our predictions for inertial waves do not show the expected behaviour of vanishing dissipation for spin-orbit synchronization.) Convective damping of the equilibrium tide can play a role though when these waves are not excited \citep[e.g.][]{VB2020}. Gravity wave damping is typically the second most effective mechanism in stars with interior radiative zones. However, this mechanism will first spin up the radiative cores, and to spin up the whole star requires efficient mechanisms of angular momentum transport and coupling between radiation and convection zones.

We again observe in Fig.~\ref{taueOm} that tidal dissipation is generally most efficient (for the same rotation period) in ``solar-type" stars with $M\sim M_\odot$. This is because they have deep convective envelopes, so inertial waves can be efficiently excited in them. Hotter or much cooler stars in the mass range that we have studied here have longer evolutionary timescales (for the same $P_\mathrm{rot}$), and so we would predict them to undergo less rapid tidal evolution. This may explain why \cite{VanEylen2016} found binaries of solar-type (which they referred to as cool-cool binaries) tended to have smaller eccentricities than hot-cool or hot-hot binaries. It may also explain why some of the binaries in \cite{Triaud2017} with lower mass stars may not have undergone as rapid tidal circularization. However, we note that the mass-dependent predictions here may differ if stellar rotational evolution is properly accounted for rather than assuming the same $P_\mathrm{rot}$.

Previous work has mostly assumed that the mechanism responsible for binary synchronization and circularization is convective damping of the equilibrium tide \citep[e.g.][]{ZahnBouchet1989,Zahn2008,Nine2020}. In fact, while significant uncertainties remain, based on our best current understanding of tidal dissipation, we have shown that inertial wave dissipation is likely to be much more important prior to evolved stellar ages. Convective damping of the equilibrium tide could play a role in binary synchronization in cases when inertial wave damping does not operate \citep{VB2020} but it is likely to play a negligible role in binary circularization except for evolved stars \citep[e.g.][]{VerbuntPhinney1995}.

According to Eq.~\ref{tauOm} and Fig.~\ref{taueOm}, we would predict spin-orbit synchronization within 1 Gyr out to 20-30 days for stars with $0.5<M/M_\odot<1.2$ due to inertial waves. We similarly predict binary circularization on the main sequence out to 7 days within 1 Gyr for $M\sim M_\odot$ based on Eq.~\ref{taue} and Fig.~\ref{taueOm}. It would be worthwhile to apply this mechanism in more detail to see whether accounting for stellar evolution would enable the observed circularization periods of solar-type binaries to be explained \citep[e.g.][]{Meibom2005,Nine2020}.

We briefly comment that \cite{VerbuntPhinney1995} found that damping of the (incorrect) conventional equilibrium tide of \cite{Zahn1989} in the red giant phase nicely explains the observed circularization of evolved binaries containing white dwarfs. In these systems, the frequency reduction of the effective viscosity $\nu_E$ is usually thought to be less important because the orbital periods are relatively long relative to the convective timescales in the star, and inertial waves may not be important because of the slow stellar rotation. The use of the correct equilibrium tide here would probably reduce the dissipation by a factor of 2-3 for the same $\nu_E$. On the other hand, the latest simulations of \cite{DBJ2020a} find an enhancement in $\nu_E$ for low frequency tidal forcing over the classical MLT formulation that \cite{VerbuntPhinney1995} adopt, as modelled in our $\nu_\mathrm{FIT}$ (see Eq.~\ref{nuEFIT1} and Fig.~\ref{nuEplot}), by an O(1) factor (between 1.5 and 5 for $0.1 \lesssim |\omega|/\omega_c\lesssim 1$). These two factors probably cancel out so that their final predictions may be similar to those using our updated models. This can potentially provide evidence in support of $\nu_\mathrm{FIT}$ over previous prescriptions for $\nu_E$. It would be worthwhile to revisit this problem.

\section{Conclusions} \label{Conclusions}

We have studied tidal dissipation in low mass and solar-type stars with masses in the range $0.1$ to $1.6 M_\odot$ following their evolution, using stellar models computed with MESA. In convection zones, we compute convective damping of equilibrium tides, accounting for the frequency-dependence of the effective turbulent viscosity \citep[e.g.][]{DBJ2020a,VB2020a}. We compute dissipation of inertial waves (dynamical tides) in convection zones using a frequency-averaged formalism \citep{Ogilvie2013} that accounts for the realistic structure of the star for the first time. We compute dynamical tide dissipation in radiation zones by assuming that internal gravity waves are launched from the convective/radiative interface and are subsequently fully damped \citep{GD1998,BO2010}. This is likely to be justified when wave breaking occurs, and we calculate the critical planetary (or companion) mass required for wave breaking as a function of stellar mass and age.

The main contribution of this work is to provide theoretical predictions for tidal quality factors $Q'$ due to each tidal mechanism throughout the evolution of these stars as a function of stellar mass, age (or $T_{eff}$), rotation rate and tidal period. Our main results are shown in Figs.~\ref{QpNWL}, \ref{QpIWcomp} and \ref{QpIGWcomp}, respectively. We also provide predictions for tidal evolutionary timescales for the orbital decay of hot Jupiters (in \S~\ref{HJimplications}), and for the orbital circularization and spin synchronization of main-sequence binary stars (in \S~\ref{binaryimplications}). Furthermore, we have provided predictions for $Q'$ due to gravity wave dissipation, and the resulting transit timing variations due to tidally-driven orbital decay for a number of observed hot Jupiters which may tested observationally with e.g.~NGTS, TESS and PLATO. 

We find that tidal dissipation in stars, quantified here by the tidal quality factor $Q'$, varies by orders of magnitude as a function of stellar mass, age, rotation, tidal frequency and amplitude. However, based on our best current understanding of tidal flows, this variation is a clear prediction of the models which can be tested by observations. It is inappropriate to consider the tidal quality factor to be the same for all stars with different masses and ages (in general). The significant variation in $Q'$ is also likely to mean that it will be challenging to meaningfully infer tidal quality factors from statistical inferences based on entire populations \citep[e.g.][]{Hansen2010,Hansen2012,CollierCameronJardine2018,Penev2018}. We advocate a system-by-system comparison of theoretical predictions with observations as the cleanest way to test tidal theory.

Tidal dissipation due to internal gravity waves in radiation zones is the dominant mechanism for planetary orbital decay around slowly rotating main-sequence stars. This mechanism is expected to be efficient (as predicted by Fig.~\ref{QpIGWcomp}) when the planetary mass exceeds a critical value required for the waves to break. We find that the critical planetary mass required for tidally-excited gravity waves to break in the radiation zones of their stars is a strong function of stellar mass and age (as shown in Fig.~\ref{QpIGWcomp2}). For the current Sun, the critical mass for wave breaking is approximately 3 $M_J$ \citep{OL2007,BO2010}. But the critical mass decreases significantly with age, to attain a minimum value of approximately $10^{-3}-10^{-2} M_J$ as the star evolves towards the end of the main sequence, such that close-in giant planets are predicted to undergo tidally-driven orbital decay at some point in their lifetimes. This result may partly explain the observational trend of \cite{HamerSchlaufman2019}, that stars with hot Jupiters are on average younger than stars without them. In addition, since wave breaking is not predicted for approximately Earth mass planets, we predict the survival of many of the observed ultra-short period (approximately Earth-mass) planets. This is also consistent with the latest observational results \citep{HamerSchlaufman2020}. In addition, gravity wave damping may partly explain why the statistical analysis of hot Jupiter systems by \citet{Penev2018} inferred $Q'$ to increase towards the star: since planets with lower $Q'$ on the shortest-period orbits may have been rapidly destroyed by gravity wave damping, this process would leave only those systems in which gravity wave damping is inefficient. Further work is required to compare theory and observations to explore these possibilities.

Gravity wave damping can also explain the magnitude of the observed orbital decay rate for WASP-12 b \citep{Maciejewski2016,Patra2017,Yee2020} if the star is a subgiant \citep{Weinberg2017}. It remains difficult to reconcile the predicted absence of wave breaking in this system with the efficient dissipation that is inferred unless the star is a subgiant \citep{Chernov2017,Weinberg2017,2019Bailey}. We predict that this mechanism should not operate in WASP-18 b \citep{Wilkins2017,Maciejewski2020}, confirming prior theoretical expectations \citep{BO2010}, which may explain the lack of detected transit timing variations in this system (though if the gravity waves are damped we predict a $Q'$ similar to the observational constraint). Indeed, tidal dissipation is generally less efficient in F-type stars which have thin (and low density) convective envelopes and convective cores. Our results also suggest that it would be difficult for WASP-4 b to be due to tidally-driven orbital decay. We provide some new predictions for other hot Jupiter systems in \S~\ref{HJimplications} and Table~\ref{TableHJ}.

Tidal dissipation of equilibrium tides is typically strongly reduced by the frequency-dependent attenuation of the turbulent viscosity for short tidal periods. As a result, and contrary to popular belief,
this mechanism is unlikely to be important for planetary orbital decay on the main sequence \citep[e.g.][]{Rasio1996,DBJ2020a}. Our models are based on the latest hydrodynamical simulations of the interaction between tidal flows and convection \citep{DBJ2020,DBJ2020a,VB2020,VB2020a}, and our application of these results indicates that this mechanism is also probably much less effective for circularization of solar-type binaries than was previously believed \citep[e.g][]{Zahn1977,Zahn1989,Zahn2008}. Equilibrium tide damping could however play a role in binary synchronization in cases when inertial waves are not excited \citep[e.g][]{VB2020}, and for planetary destruction and orbital circularization involving evolved stars \citep[e.g.][]{VerbuntPhinney1995,Mustill2012,SunArrasWenberg2018}.

It has been realised that the conventional equilibrium tide of \cite{Zahn1966,Zahn1989} only applies in radiation zones, and does not correctly describe the tidal response in convection zones \citep{Terquem1998,GD1998,Ogilvie2014}. However, the conventional equilibrium tide is still commonly adopted to compute tidal dissipation in convection zones. In this paper, we have compared in detail the correct equilibrium (non-wavelike) tide with the conventional equilibrium tidal flow (in Appendix \ref{EQMCompsection}), as well as the resulting tidal dissipation due to turbulent convective viscosity (in Fig.~\ref{NWLdissipation}), in a range of stellar models. We confirm that the conventional equilibrium tide can differ non-negligibly from the correct equilibrium tide in convective regions. It also over-predicts tidal dissipation by a factor of 2-3. As a result, we advocate that the corrected version of the equilibrium tide used here should be employed instead of the conventional equilibrium tide.

Tidal dissipation of inertial waves in convective envelopes is typically the most efficient mechanism of tidal dissipation for main-sequence binary circularization and synchronization. However, it can only operate in rotating stars in circumstances when twice the tidal period exceeds the rotation period of the star. This condition is not often satisfied in hot Jupiter host stars, but it is satisfied in many binary systems. The resulting tidal quality factors vary over several orders of magnitude as a function of the stellar mass, age, and rotation rate, but with a typical value $\langle Q'_\mathrm{IW}\rangle\approx 10^7 (P_\mathrm{rot}/10 \mathrm{d})^2$ on the main sequence. We note that the statistical analysis of \cite{CollierCameronJardine2018} obtains $Q'\sim 2\times 10^7$ in the dynamical tide regime when inertial waves are excited, which is consistent with this result. The most efficient dissipation is obtained briefly as the star undergoes its evolution off the main sequence, when values of $\langle Q'_\mathrm{IW}\rangle$ that are smaller than 1 are possible (depending on the rotation rate of the star). This mechanism is then much less efficient in subsequent phases. In earlier stages, the most efficient dissipation is obtained for pre-main sequence (PMS) stars, where $\epsilon^2_\Omega \langle Q'_\mathrm{IW}\rangle\approx 10^2$ for all stars that we considered. Since PMS stars generally rotate more rapidly (i.e.~they have larger $\epsilon_\Omega^2$) than those on the main-sequence, this result will be strengthened significantly when our models are coupled with those governing stellar rotational evolution.

Several of our results on inertial wave dissipation agree with those of \cite{Mathis2015}, who adopted a simplified two-layer model to study the frequency-averaged tidal dissipation of inertial waves. We have compared the predictions of the more realistic model used here with those of the two-layer model. For solar-type stars on the main sequence, the two-layer model generally under-predicts the dissipation only by approximately a factor of 2, and so is a reasonably good model to adopt in that case, particularly in light of its simplicity. The two-layer model also correctly predicts PMS stars to be more dissipative than main sequence stars. However, its predictions can significantly differ from those of our model for F-stars, and for lower mass stars, indicating that the realistic structure of the star should be accounted for when computing tidal dissipation due to inertial waves when possible.

Further work is required to study the dissipation of tidally-excited gravity waves for planetary masses below the wave breaking threshold, where weaker nonlinear wave-wave interactions may instead be important \citep[e.g.][]{BO2011,Weinberg2012,EssickWeinberg2016}. Wave breaking should also be revisited with higher resolution simulations to determine whether the waves are efficiently absorbed by critical layers at high Reynolds numbers \citep{Alvan2013,Su2020}. The possibility of wave breaking caused by passage through a resonance, and resonant locking \citep[e.g.][]{Witte2002,Savonije2002,Fuller2017}, are also worth considering. Finally, it would be worthwhile to conduct a similar study as this one in models that account for the stellar rotational evolution, and in models that couple the orbital evolution with these models of tidal dissipation. Significant uncertainties remain regarding the mechanisms by which inertial waves in convective zones are dissipated in nonlinear and magnetised models \citep{FBBO2014,LO2018,Wei2018,Astoul2019}, and by the effect of rotation and density stratification on the convective turbulent viscosity \citep[e.g.][]{Stevenson1979,BDL2014,Mathis2016,Currie2020}.

To conclude, while significant theoretical uncertainties remain, current tidal theory is able to make clear predictions that can be tested by observations with current and future missions such as TESS, NGTS, CHEOPS and PLATO. We are now entering a golden age where tidal theory can confront observation.

\section*{Acknowledgements}
I would like to thank the referee for suggestions that helped me to clarify several points in the paper. This research was supported by STFC grants ST/R00059X/1 and ST/S000275/1. I would like to thank Jacob Hamer and Simon Albrecht for forwarding their latest pre-prints.

\section*{Data availability}
The data underlying this article will be shared on reasonable request to the corresponding author.

\bibliography{tid}
\bibliographystyle{mnras}

\appendix
\section{Equilibrium tide displacements}
\label{EQMCompsection}

\begin{figure*}
  \begin{center}
    \subfigure[$M/M_{\odot}=0.2,\,\text{age/yr}=2.93\times10^9$]{\includegraphics[trim=3cm 0cm 4.5cm 0cm,clip=true,width=0.34\textwidth]{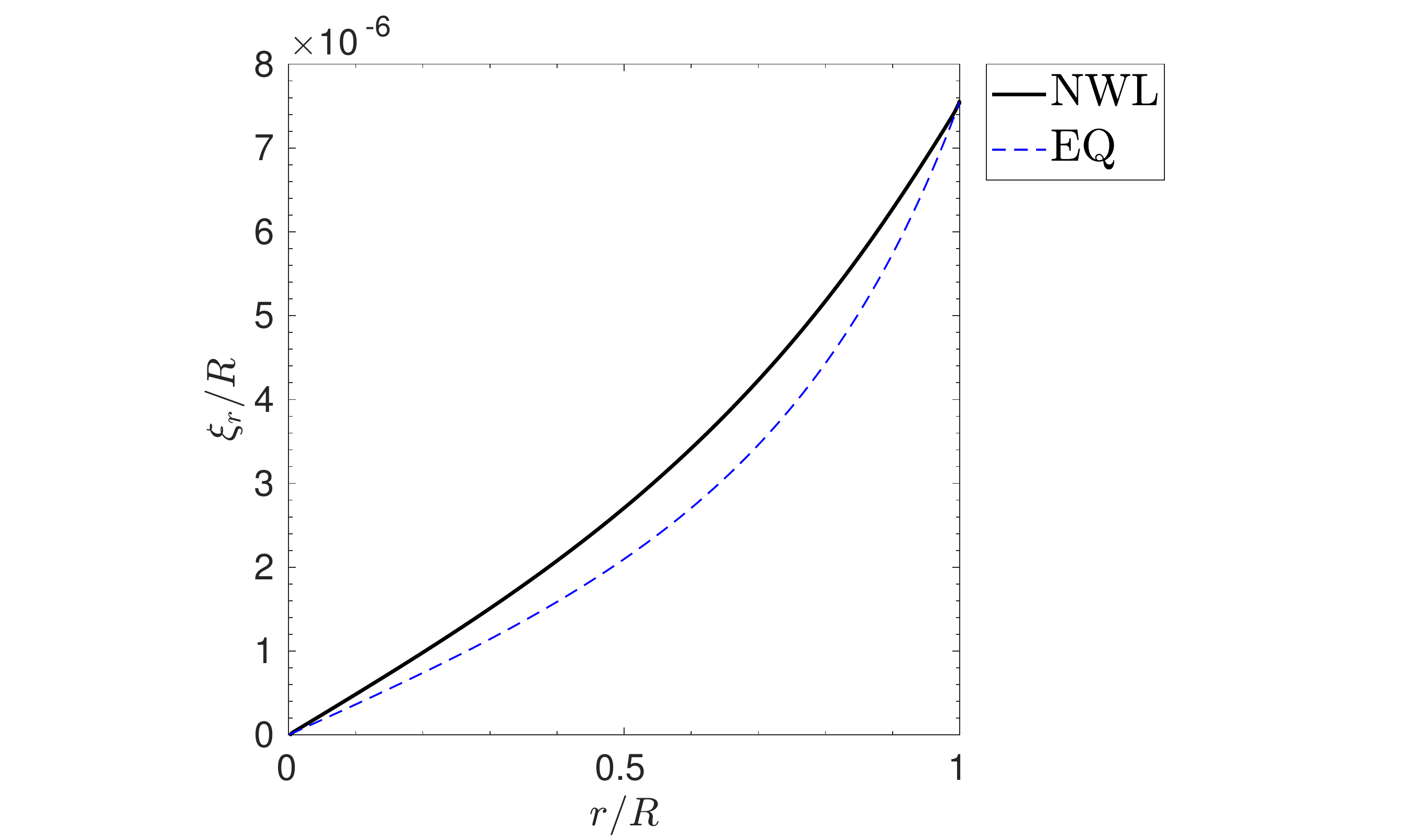}}
    \subfigure[$M/M_{\odot}=0.2,\,\text{age/yr}=2.93\times10^9$]{\includegraphics[trim=3cm 0cm 4.5cm 0cm,clip=true,width=0.34\textwidth]{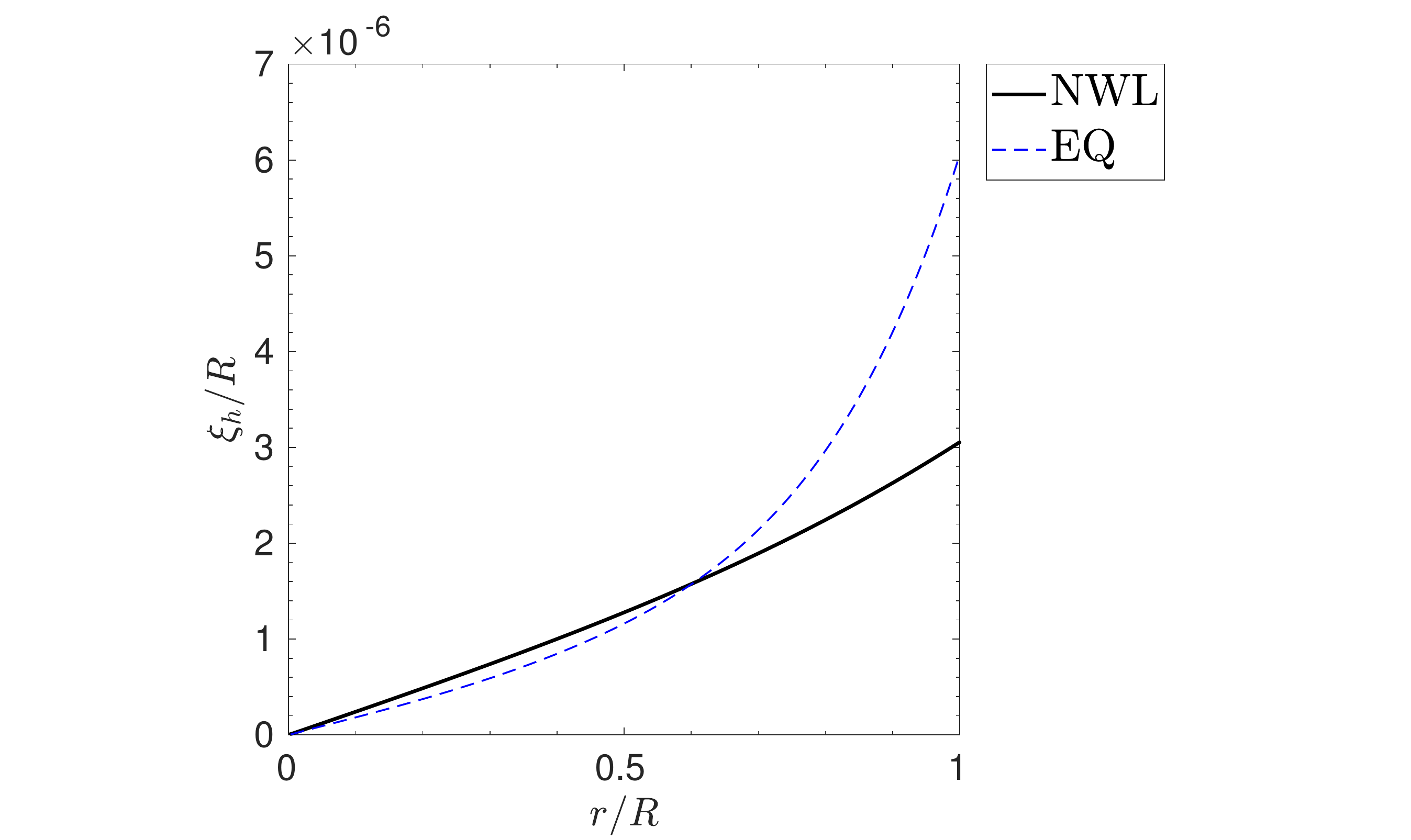}}
   \subfigure[$M/M_{\odot}=1,\,\text{age/yr}=4.70\times10^9$]{\includegraphics[trim=3cm 0cm 4.5cm 0cm,clip=true,width=0.34\textwidth]{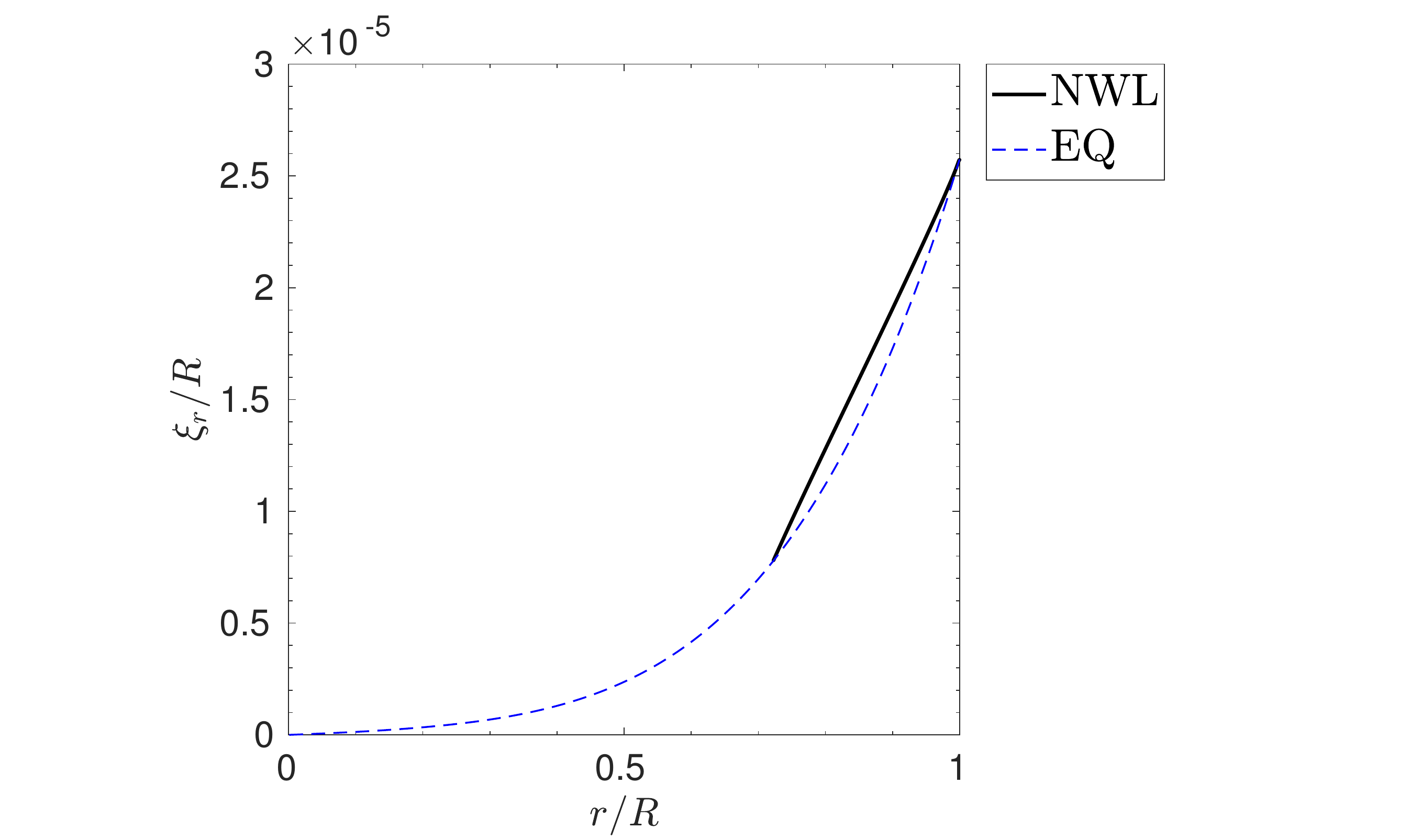}}
    \subfigure[$M/M_{\odot}=1,\,\text{age/yr}=4.70\times10^9$]{\includegraphics[trim=3cm 0cm 4.5cm 0cm,clip=true,width=0.34\textwidth]{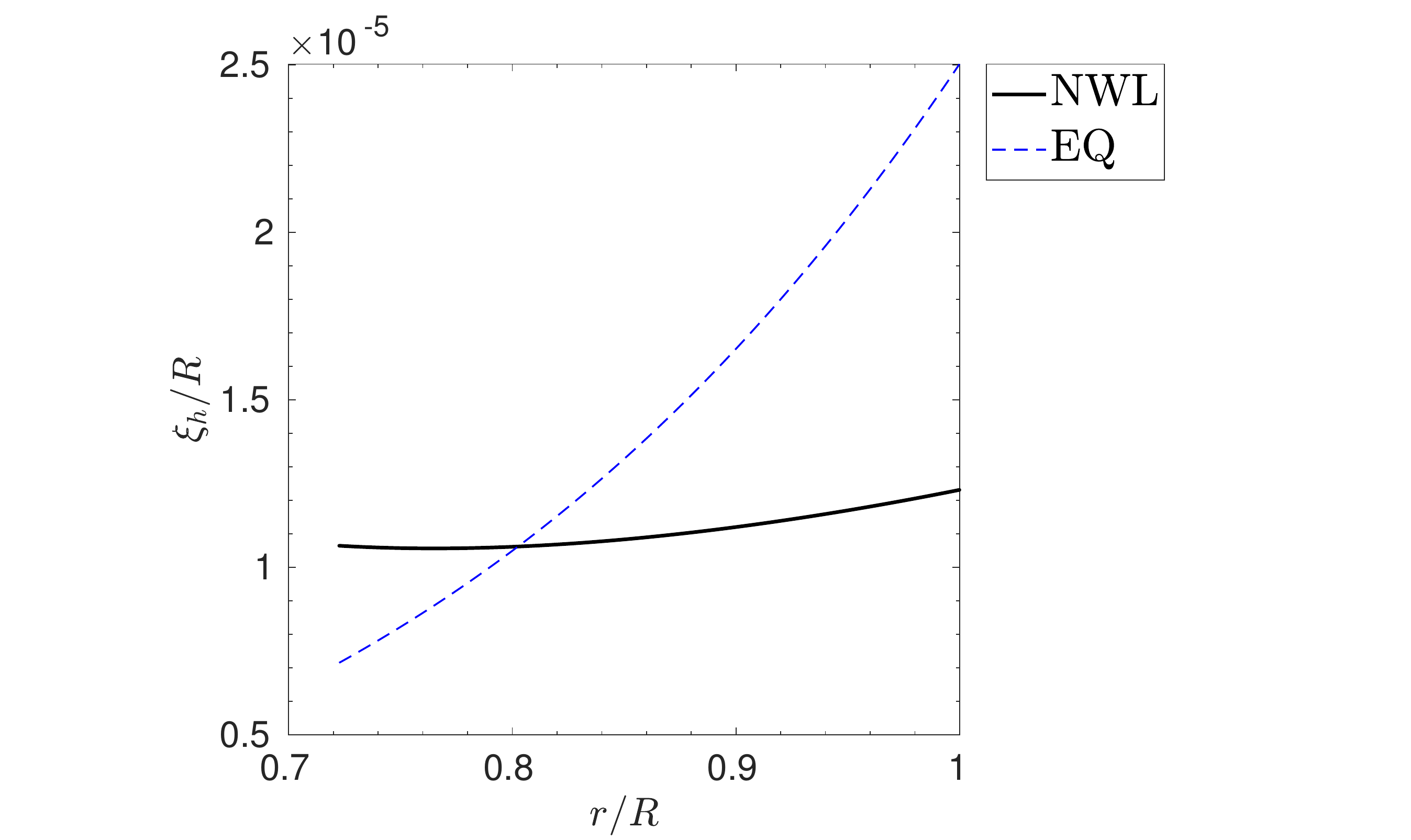}}
    \subfigure[$M/M_{\odot}=1.4,\,\text{age/yr}=1.29\times10^9$]{\includegraphics[trim=3cm 0cm 4.5cm 0cm,clip=true,width=0.34\textwidth]{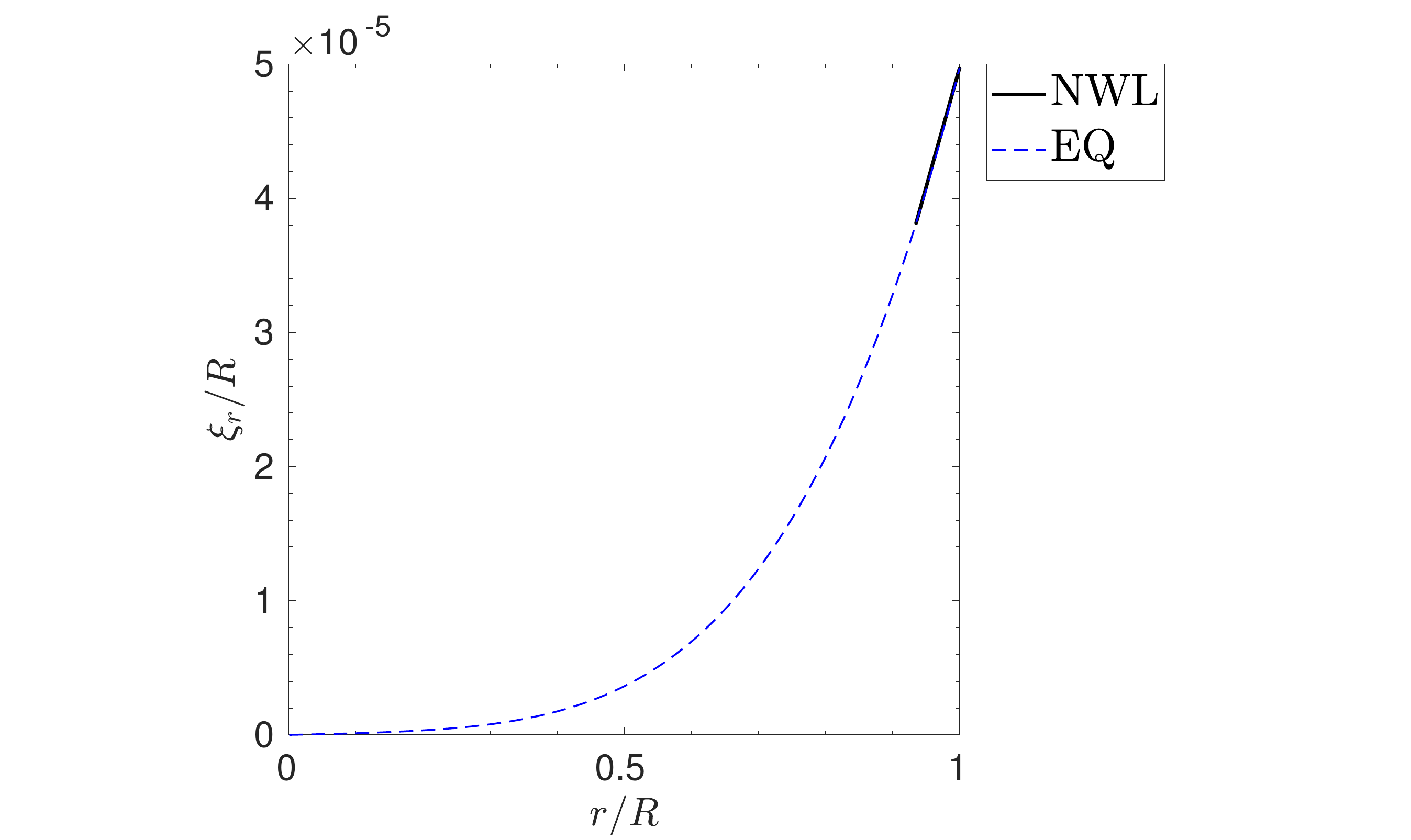}}
    \subfigure[$M/M_{\odot}=1.4,\,\text{age/yr}=1.29\times10^9$]{\includegraphics[trim=3cm 0cm 4.5cm 0cm,clip=true,width=0.34\textwidth]{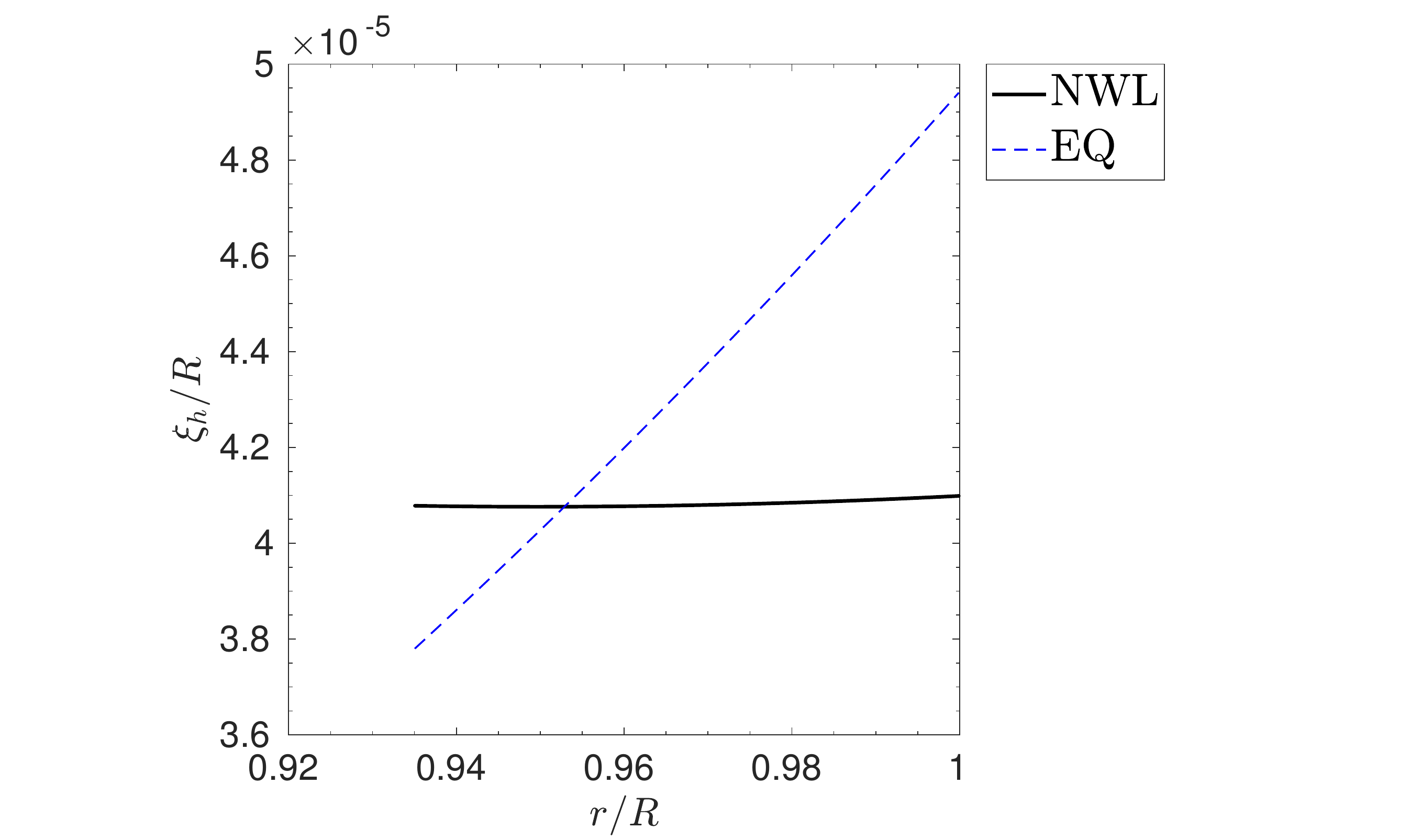}}
    \end{center}
  \caption{Comparison of the radial (left) and horizontal (right) displacements in the correct equilibrium tide (NWL, black solid lines) and the conventional equilibrium tide (EQ, blue dashed lines) in various representative (slowly rotating) main-sequence stellar models. The conventional equilibrium tide (EQ) is strictly invalid in convection zones and typically has larger radial gradients than the correct equilibrium tide (NWL). As a result, it is likely to over-predict the resulting tidal dissipation.}
  \label{EQMNWLcomp}
\end{figure*}

In this appendix we compare the tidal displacements due to the correct equilibrium tide (NWL) with the standard equilibrium tide of \cite{Zahn1966,Zahn1989} (EQ), as defined by the solutions of Eq.~\ref{EQMtide}, to complement the results of \S~\ref{EQMCOMP}. It is important to compare them because many authors have assumed the latter to apply throughout the star, even though it is strictly only valid in radiation zones.

In Fig.~\ref{EQMNWLcomp}, we compare the radial (left panels) and horizontal (right panels) components of the tidal displacements of NWL and EQ as a function of normalised radius in a range of stellar models representing low-mass and solar-type stars on the main-sequence. The masses and ages of each star are given in the panel captions. The amplitude of each panel is arbitrary, but we have chosen to represent that due to a $1M_J$ planet on a $1$-day circular equatorial orbit around each star, and we have normalised this by the stellar radius in each case. The tidal amplitudes obtained increase as we move down the figure because the stellar radius increases for larger stellar masses.

The first two panels show the equilibrium tidal components in a fully-convective $M=0.2M_\odot$ star at an age of $2.93$ Gyr. The left panels show the radial displacement of NWL differs from EQ throughout most of the star, only agreeing at the centre and surface (where it must). The right panels show the corresponding horizontal displacements, which disagree most strongly at the surface. We observe that EQ generally has larger radial gradients than NWL throughout the star. The tide in this model (and other similar fully-convective models) is found to be similar to the solutions for an $n=1.5$ polytrope (not shown).

Similar results are shown for stars with $M/M_\odot=1$ and $1.4$ in the remaining rows of this figure. The second row compares these tides in a model that is very similar to our current Sun (such as model S e.g.~\citealt{JCDModelS1996}). We again find that EQ differs from NWL, having larger gradients throughout the bulk of the convection zone. The bottom row shows the same comparison in an F-type stellar model with $M=1.4M_\odot$, which has a thin convective envelope. In this latter model, the radial components of NWL and EQ look similar but the horizontal components still differ substantially. Similar results have been found in a range of different stellar models at various ages on the pre-main sequence and main-sequence. 

In this section, we have demonstrated that the displacements in the equilibrium tides predicted by NWL and EQ have different radial profiles, and that the radial gradients of EQ are generally larger. This suggests that the dissipation of EQ is likely to be larger than that of NWL, which qualitatively explains our observations in \S~\ref{NWLdissipation}.

\section{MESA Code Parameters}
\label{MESA}

We use MESA version 12778 \citep{Paxton2011, Paxton2013, Paxton2015, Paxton2018, Paxton2019}. The inlist file that we use is given below. We alter {\verb initial_mass } and {\verb initial_z } as required to generate a given stellar model, and the code is stopped manually at a chosen time, usually when the star has left the main sequence.
\begin{verbatim}
&star_job
  create_pre_main_sequence_model = .true.
/ !End of star_job namelist
&controls
! starting specifications
    initial_mass = 1.0
    initial_z = 0.02d0
    MLT_option = 'Henyey'
    max_age = 5.0d10
    max_years_for_timestep = 1.0d8      
    use_dedt_form_of_energy_eqn = .true.
    use_gold_tolerances = .true.
    mesh_delta_coeff = 0.3
    when_to_stop_rtol = 1d-6
    when_to_stop_atol = 1d-6
/ ! end of controls namelist
\end{verbatim}

\section{Table of observed hot Jupiters with new theoretical predictions}
\begin{landscape}
\begin{table}
\begin{tabular}{cccccccccc|ccccc}
\hline
Name & $M_p/M_J$ & $P_\mathrm{orb}/\mathrm{d}$ & $M/M_\odot$ & $R/R_\odot$ & $Z_{init}$ & $T_{eff}$ & $P_\mathrm{rot}/\mathrm{d}$ & $\mathrm{age}/\mathrm{Gyr}$ & $Q'_{\mathrm{obs}}$ & $Q'_\mathrm{IGW}$ & $\tau_a/\mathrm{Myr}$ & $T_\mathrm{shift}/s$ & $M_\mathrm{crit}/M_J$ & Breaking? \\
\hline 
WASP-4b & 1.2 & 1.34 & 0.93 & 0.9 & 0.02 & 5542 & 23 & 5.8 & $4.5-8.5\times10^4$& $3.3-3.8\times 10^5$ & $14-17$ & 13-15 & $2-7$ & No? \\
WASP-12b (MS1) & 1.47 & 1.09 & 1.43 & 1.68 & 0.03 & 6376 & 38 & 1.62 & $2\times 10^5$ & $4.5\times 10^6$ & 11 & 20.5 & $>5000$ & No \\
WASP-12b (MS2) & 1.47 & 1.09 & 1.32 & 1.69 & 0.025 & 6072 & 38 & 3.1 & $2\times 10^5$ & $2.2\times 10^5$ & 0.42 & 522 & 330 & No \\
WASP-12b (SG) & 1.47 & 1.09 & 1.24 & 1.69 & 0.023 & 6126 & 38 & 4.1 & $2\times 10^5$ & $2.7\times 10^5$ & 0.58 & 384 & 0.3 & Yes \\
WASP-18b & 11.4 & 0.94 & 1.24 & 1.29 & 0.02 & 6306 & $6$ & 1.37 & $>1.3\times 10^6$ & $2.6\times 10^6$ & 1.2 & 200 & 207 & No \\
WASP-19b & 1.14 & 0.79 & 0.94 & 1.01 & 0.02 & 5624 & 13 & 9.28 & $3.5-7.5\times10^5$& $0.6-0.8\times 10^5$ & $0.13-0.3$ & 675-1000 & 0.8 & Yes? \\
WASP-43b & 2.03 & 0.81 & 0.72 & 0.67 & 0.02 & 4462 & $6$ & 5.03 & $>0.7-3.5\times10^5$& $1.3\times 10^5$ & 0.98 & 230 & $40$ & No \\
WASP-72b & 1.55 & 2.22 & 1.39 & 2.01 & 0.01 & 6876? & $17$ & $2.3$ & $>2.1\times10^3$& $>10^{12}$ &  &  & 0.01 & Yes \\
WASP-103b & 1.51 & 0.93 & 1.21 & 1.43 & 0.02 & 6115 & $7$ & 3.48 & $>1.1\times10^5$& $4\times 10^5$ & $0.68$ & 322 & 550 & No \\
WASP-114b & 1.77 & 1.55 & 1.29 & 1.42 & 0.03 & 6206 & 12 & 2.12 & & $3\times 10^6$ & 48 & 4.7 & 3300 & No \\
WASP-121b & 1.18 & 1.27 & 1.353 & 1.49 & 0.025 & 6429 & $5.5$ & 1.42 & & $2.4\times 10^7$ & 225 & 1 & $10^4$ & No \\
WASP-122b & 1.284 & 1.71 & 1.24 & 1.50 & 0.04 & 5895 & $23$ & 4.2 & & $3.5\times 10^5$ & 8.4 & 27 & 1000 & No \\
WASP-128b & 37.2 & 2.21 & 1.16 & 1.16 & 0.02 & 6108 & $3$ & 1.57 & & $1.7\times 10^8$ & 1400 & 0.2 & 24.8 & Yes \\
NGTS-6b & 1.33 & 0.88 & 0.787 & 0.74 & 0.025 & 4774 &  & 9.01 &  & $>0.99\times 10^5$ & 1.2 & 182 & 18 & No \\
NGTS-7Ab & 62.0 & 0.676 & 0.48? & 0.645 & 0.02 & 3736 & sync? & $0.0055$ &  & $>0.9\times 10^5$ & & & 357 & No \\
NGTS-10b & 2.16 & 0.77 & 0.696 & 0.68 & 0.02 & 4428 & $8.8$ & 10.06 &  & $0.99\times 10^5$ & 0.5 & 440 & 15 & No \\
HAT-P-23b & 2.09 & 1.21 & 1.13 & 1.22 & 0.03 & 5916 & $7.5$ & 4.3 & $>4.5\times 10^5$ & $3.5\times 10^5$ & 2.5 & 91 & 400 & No \\
HATS-18b & 1.98 & 0.84 & 1.04 & 1.03 & 0.03 & 5735 & $8.3$ & 4.26 &  & $1.1\times 10^5$ & 0.33 & 686 & 1.75 & Yes \\
KELT-16b & 2.75 & 0.97 & 1.21 & 1.38 & 0.02 & 6180 & $9$ & $2.97$ & $>0.7\times 10^5$ & $7\times 10^5$ & 0.98 & 228 & 450 & No \\
TRES-3b & 1.91 & 1.306 & 0.924 & 0.845 & 0.009 & 5699 & 27 & 1.23 & $1.1\times 10^5$ & $6.1\times 10^5$ & $23.6$ & $9.5$ & $65$ & No \\
OGLE-TR-56b & 1.39 & 1.21 & 1.23 & 1.38 & 0.02 & 6235 & $23$ & 2.53 & $>5\times 10^5$ & $1.8\times 10^6$ & 14 & 16.5 & 1920 & No \\
WTS-2b & 1.12 & 1.018 & 0.82 & 0.74 & 0.02 & 4761 & $17$ & 0.43  &  & $1.9\times 10^5$ & 6.4 & 35 & 1600 & No \\
\hline
\end{tabular}
\caption{Table of short-period hot Jupiters reporting the strongest constraints on $Q'$ that are available $Q'_\mathrm{obs}$, along with our theoretical predictions due to internal gravity wave dissipation $Q'_\mathrm{IGW}$ (assuming these waves are fully damped) and the critical planetary mass for wave breaking. The rotation period $P_\mathrm{rot}$ is based on the reported $V\sin i$ \citep[e.g][]{Patra2020} by simply assuming $\sin i=1$, and the values for WASP-4 and WASP-19 from \citet{Maxted2015}. The predictions are based on similar stellar models obtained with MESA by adopting the initial metallicity $Z_{init}$ and mass are given in the table at the age reported. We caution that the predictions for $M_\mathrm{crit}$ are strongly age, mass and also somewhat metallicity dependent, and $Q'_\mathrm{IGW}$ also depends but to a lesser extent on these parameters. Data is taken from \citet{Patra2020} and The Extrasolar Planets Encyclopaedia (\url{http://exoplanet.eu}).}
\label{TableHJ}
\end{table}
\label{lastpage}
\end{landscape}
\end{document}